# An empirical study of testing machine learning in the wild

MOSES OPENJA, Polytechnique Montréal, Canada
FOUTSE KHOMH, Polytechnique Montréal, Canada
ARMSTRONG FOUNDJEM, Polytechnique Montréal, Canada
ZHEN MING (JACK) JIANG, York University, Canada
MOUNA ABIDI, Polytechnique Montréal, Canada
AHMED E. HASSAN, Queen's University, Canada

**Background:** Recently, machine and deep learning (ML/DL) algorithms have been increasingly adopted in many software systems. Due to their inductive nature, ensuring the quality of these systems remains a significant challenge for the research community. Unlike traditional software built deductively by writing explicit rules, ML/DL systems infer rules from training data. Recent research in ML/DL quality assurance has adapted concepts from traditional software testing, such as mutation testing, to improve reliability. However, it is unclear if these proposed testing techniques are adopted in practice, or if new testing strategies have emerged from real-world ML deployments. There is little empirical evidence about the testing strategies.

**Aims:** To fill this gap, we perform the first fine-grained empirical study on ML testing in the wild to identify the ML properties being tested, the testing strategies, and their implementation throughout the ML workflow.

**Method:** We conducted a mixed-methods study to understand ML software testing practices. We analyzed test files and cases from 11 open-source ML/DL projects on GitHub. Using open coding, we manually examined the testing strategies, tested ML properties, and implemented testing methods to understand their practical application in building and releasing ML/DL software systems.

**Results:** Our findings reveal several key insights: 1.) The most common testing strategies, accounting for less than 40%, are Grey-box and White-box methods, such as *Negative Testing*, *Oracle Approximation*, and *Statistical Testing*. 2.) A wide range of 17 ML properties are tested, out of which only 20% to 30% are frequently tested, including *Consistency*, *Correctness*, and *Efficiency*. 3.) *Bias and Fairness* is more tested in Recommendation (6%) and CV (3.9%) systems, while *Security & Privacy* is tested in CV (2%), Application Platforms (0.9%), and NLP (0.5%). 4.) We identified 13 types of testing methods, such as *Unit Testing*, *Input Testing*, and *Model Testing*.

**Conclusions:** This study sheds light on the current adoption of software testing techniques and highlights gaps and limitations in existing ML testing practices.

CCS Concepts: • **Machine learning and Deep Learning → Testing practice**; *Testing strategy*; • **Test types**;

Additional Key Words and Phrases: Machine learning, Deep learning, Software Testing, Machine learning workflow, Testing strategies, Testing methods, ML properties, Test types/ Types of testing

## 1 Introduction

The increasing adoption of Machine Learning (ML) and Deep Learning (DL) in many software systems, including safety-critical software systems (e.g., autonomous driving [31, 104], medical software systems [84]) raises concerns about their reliability and trustworthiness. Ensuring the

Authors' addresses: Moses Openja, openja.moses@polymtl.ca, Polytechnique Montréal, Montreal, Québec, Canada; Foutse Khomh, Polytechnique Montréal, Montreal, Québec, Canada, foutse.khomh@polymtl.ca; Armstrong Foundjem, Polytechnique Montréal, Montreal, Québec, Canada; Zhen Ming (Jack) Jiang, York University, Toronto, Canada; Mouna Abidi, Polytechnique Montréal, Montreal, Québec, Canada; Ahmed E. Hassan, Queen's University, Kingston, Canada.

quality of these software systems is yet an open challenge for the research community. The main reason behind the difficulty in ensuring the quality of ML software systems is the shift in the development paradigm as induced by ML. Contrary to traditional software systems, where the engineers have to manually formulate the rules that govern the behavior of the software system as program code, in ML the algorithm automatically formulates the rules from the data. This paradigm makes it difficult to reason about the behavior of software systems with ML components, resulting in software systems that are intrinsically challenging to test and verify. A defect in an ML software system may come from its training data, the program code, execution environment, or even third-party frameworks. A few research advances in the quality assurance on ML systems recently have adapted different concepts from traditional (i.e., software not using ML) software testing, such as evaluating their effectiveness (e.g., mutation testing [34, 88]) and introduced new techniques to verify the security or privacy of an ML system (e.g., Spoofing attacks in autonomous systems [25, 131]), to help improve the reliability of ML based software systems.

With the support of programming languages like Python and C/C++, ML engineers can effectively perform both white-box and black-box testing on ML models and software systems. Notably, black-box tests can be written in Python even if the system under test (SUT) is not written in Python. ML engineers can develop *test cases*, which are collections of unit tests that verify a function's performance under various conditions. These test cases consider all possible inputs that the function is expected to handle, ensuring comprehensive coverage. By analyzing the test code of an ML software system, we can gain insights into the testing strategies employed by ML engineers, the ML properties they test, and the types of tests or testing types they utilize in their workflow.

We define testing strategies as the various approaches and resources employed during testing activities to identify defects in software or ML systems. It specifies the methods and techniques to be used, the types of testing to be conducted, the test environments to be employed, and the criteria for test completion. These strategies are crucial to ensuring that testing activities meet software quality assurance objectives and that the most suitable methods are used to evaluate the behavior of ML systems and the effectiveness of existing test cases. In ML/DL, factors such as randomness and floating-point arithmetic often lead to significant discrepancies between computed results and expected outcomes across different test runs, making ML systems challenging to test. ML engineers often resort to strategies like Oracle Approximation [98], which allows for slight differences between computed results and the oracle, rather than specifying exact values for tests. For example, to test the correctness of an ML component, engineers might use a rounding tolerance approximation API provided by the `unittest` framework in Python. This approach enables a specified number of decimal places (`dp`) for rounding the actual value of the code under test (`a_val`) and its expected value (`oracle`), as shown in the assertion: `assertAlmostEqual(a_val, oracle, places=dp)`. Similarly, to determine the optimal threshold value for classification, ML engineers often implement appropriate Thresholding strategies [111, 123] that maximize a balance among different performance metrics (e.g., Recall, Precision, Accuracy) instead of relying on the default classification threshold (e.g., 0.5).

In the context of an ML software system, ensuring that the software behaves adequately also involves guaranteeing that critical "ML properties" are not violated during the development and evolution of the system [42]. These "ML properties" refer to underspecification issues and requirements that must be satisfied to ensure that a model behaves as expected. Examples of such properties include functional correctness, consistency, efficiency, bias and fairness, robustness, among others. These requirements are essential to ensure that models generalize as expected in deployment scenarios [18, 132, 150]. Violating these properties often results in instability and poor model performance in practice. Testing these properties is crucial not only for producing high-quality software systems but also for regulatory and auditing purposes [48]. Therefore, ML properties should be tested throughout the ML workflow to ensure that the generated models meet expected requirements, by following some "testing strategy (s)" and "test types". Software test types encompass various activities used to validate the System Under Test (SUT) for defined test objectives. These methods are categorized into types such as black-box or white-box tests. Similar to traditional software systems, ML engineers employ numerous testing types, including unit tests, integration tests, and manual tests, throughout the development process [1]. This comprehensive testing ensures that the ML software system can operate successfully under multiple conditions and across different platforms.

However, little is known about the current practices for testing ML software systems, and it remains unclear if the testing techniques proposed in research are being adopted in practice. Additionally, the emergence of new ML testing techniques from real-world deployments is largely unknown. To fill this gap, we conduct the first fine-grained empirical study on ML testing practices in the wild, aiming to identify the ML properties being tested, the testing strategies employed, and their implementation throughout the ML workflow. We manually analyzed the test code contents of 11 ML software systems from four (4) different application domains; Computer Vision (CV) systems (e.g., Autonomous driving), NLP systems (e.g., Voice recognition), Recommendation systems (e.g., Lightfm), and Application Platform (e.g., Medical system, Distributed ML, Music and Art Generation with Machine Intelligence), by following an open coding procedure to derive a taxonomy of the ML testing strategies used by ML engineers. Second, we examine the specific ML properties that engineers commonly test throughout the ML workflow. Third, We examine the software types of tests/testing types that are used to test the ML software systems in the ML development phases. Moreover, we compare the usage of the different testing strategies, the tested ML properties and the tests methods across the studied ML software system's and the ML domains.

Overall, our findings provide actionable implications for different groups of audiences: (1) ML practitioners can use our presented taxonomy to learn about the existing ML testing strategies that they can implement in their ML workflow, especially the most frequently used testing strategies, Grey-box and White-box testing, such as: *Exception and Error Condition*, *Statistical Analysis, Decision & Logical Condition*, and *Oracle Approximation*. Our results can also guide them on how to verify specific ML properties, such as functional correctness and consistency, efficiency and Robustness. (2) Researchers can build on our work to further investigate the effectiveness of different ML testing strategies and investigating the underlying reasons for the non-uniform application of these methods across different ML software systems with the end goal to developing standardized best practices for ML testing. (3) Designers of testing tools can develop better tooling support to the help testing for ML properties (e.g., Security and privacy, Bias and fairness, Certainty, Compatibility and Portability), across the different ML domains such as CV, NLP, and Recommendation Systems.

In summary, we make the following main contributions:

- To the best of our knowledge, this is the first empirical study on the adoption of research advances in software testing of ML software systems by the industry, specifically by examining the testing strategies, the ML properties, and the testing types adopted in the field. Specifically, we highlighted 15 major categories of testing strategies, 17 different ML properties, and 13 different types of tests that the ML engineers commonly test.
- We compared how the tests are distributed across the ML workflow, ad observed that a significant portion of testing activities occurs during model training and feature engineering.
- We provided a comparison of the testing strategies, the testing properties and testing types used across the studied systems to identify common or inconsistent practices in their existing testing practices. We observed an overall uneven distribution in the testing practice, with a significant number of studied ML software systems relying on a limited number of testing strategies and testing types, which could potentially lead to gaps in coverage.

- We highlighted some challenges and proposed new research directions for the research community. Moreover, we discussed the potential applications of the identified testing strategy to help drive future research. ML practitioners can also leverage our findings to learn about different testing strategies, properties to test, and testing types that can be implemented to improve the reliability of their next ML software system.

**Paper organization:** The remainder of the paper is organized as follows: in Section 2, we provide background information on ML workflows and discuss related works. Section 3 describes the eight major steps of our methodology. Section 4 presents the results of our analysis, addressing three research questions. Section 5 discusses more our findings, the critical analysis, the challenges and lessons learnt and the implications of our findings. Section 6 outlines potential threats to the validity of this work. Finally, Section 7 concludes the paper.

## 2  Background on Machine Learning workflow

In this study, we want to assess the practice of testing ML software systems across the entire ML workflow. Figure 1 highlights nine activities of a machine learning workflow to build and deploy ML software system adopted from Microsoft [5]. This workflow is comparable to those used by other major companies, such as Google [58], and IBM [72]. Below are the nine activities of this ML workflow:

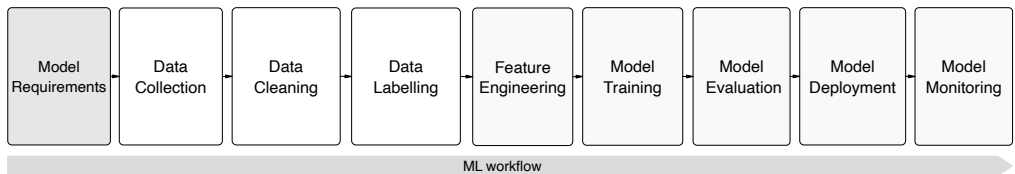

Fig. 1: An overview of the machine learning workflow adopted from [5]

(1) *Model Requirements:* The primary activity in constructing a machine learning model is specifying the model's requirements. Based on the product and the problem statement, the decision is made on the appropriate types of models for the problem and features to implement with machine learning. (2) *Data collection:* This activity follows after the model requirements, where the data from different sources are identified and collected. This data may come from batch storage software components such as databases and file storage or generated from external components such as detection components (e.g., camera or LiDAR sensors). Indeed, most ML training would require a considerably large amount of data to train from and make meaningful inferences on the new data, making the data collection challenging. (3) *Data cleaning:* Once the data is gathered, the next activity (*Data cleaning*) is to process or clean the data to remove anomalies that are likely to hinder the training phase. Most activities common to data science are performed during this step, such as generating descriptive statistics about the features in the data and the distribution of the number of values per example. (4) *Data labeling:* Assigns ground truth labels to each data record, for example, assigning labels to a set of images with the objects present in the image. This step is required in most supervised learning techniques to induce a model. (5) *Feature engineering:* This activity is where the features are being prepared and validated for training. This activity also includes data validation, which ensures that properties such as data schema (e.g., features present in the data, expected type of each feature) are correct. This step further ensures the quality of the training data. As an ML platform scales to larger data and runs continuously, there is a strong need for an efficient component that allows a rigorous inspection of data quality. (6) *Model Training:* The training activity uses the prepared features on different implementations of algorithms to train ML models. Also, the implemented algorithms are subjected to hyperparameter tuning to get the best-performing ML model. Tests such as verifying the validity of model input parameters or memory leak error detection may be performed during this step. The result from this step is a trained ML model. (7) *Model Evaluation:* The trained model is then evaluated on a holdout data set, commonly referred to as a validation set. This evaluation activity is required to confirm the model adequacy for deployment; *i.e.,* that it can generalize beyond training data. (8) *Model Deployment:* Once the model is evaluated, the ML Model and entire workflow are then exported (i.e., the meta-data information of the model is written in a file or data store.) to be reused (imported) in other platforms for scoring and—or making predictions. ML engineers may test the exported model against the original model's performance for consistency. Moreover, the portability of the model may be tested at this activity to allow its deployment on different platforms (e.g., a model designed in Python can be utilized inside an Android app). (9) *Model Monitoring:* The final activity is monitoring to ensure that the ML model is doing what is expected from it in a production environment.

Behold the standard workflow structure; in most cases, there is a need for shared *configuration and utilities* to allow for integrating these components in a single platform, ensuring consistency across the workflow. For example, transformations at a serving time may utilize statistics generated by the data analysis component using a shared utility.

Throughout this paper, we will refer to the different activities of the ML workflow that are described in Figure 1 (including the *utility, frameworks, and system*) as ML workflow activities.

## 2.1 Related works on testing ML software systems:

Tests in machine learning software systems are applied to the following three high-level components: data, model, and ML code. Each of these components contains multiple functions. Braiek and Khomh [19] described the main sources of faults occurring during the development process of ML-based systems, from data preparation to the deployment of models in production. Then, they reviewed testing techniques proposed in the literature to help detect these faults both at the implementation and model levels, describing the context in which they can be applied and their anticipated outcome. In the following, we present the most relevant literature on software testing broken down into three: 1) the ML testing strategy, 2) ML tested properties, and 3) testing methods:

### 2.1.1 *ML testing strategy*

This study uses the term *testing strategy* to refer to the technique used to verify that an application or function behaves according to its specification (i.e., the property being tested). By adopting a given testing strategy, ML engineers can detect and fix incorrect behavior of ML components for the corner case inputs before using the ML software system. Among the ML testing strategies discussed in the literature, *Oracle Approximation* is the strategy that recently received attention [98].

● **Oracle Approximation (OA):** ML engineers often resort to Oracle approximations in ML software systems for various reasons. For instance, the inherent randomness in ML model outcomes can lead to slight variations in results across different test runs. Moreover, specifying exact values for tests involving floating-point numbers can be challenging, resulting in significant discrepancies between computed results and expected outcomes. In the *OA* testing strategy, the ML component's output is permitted to fall within a predetermined range of the oracle, diverging from the conventional equality checks prevalent in traditional software testing methodologies. Nejadgholi et al. [98] conducted a comprehensive investigation into *OA* practices within compute-intensive software systems, particularly focusing on DL libraries. The study examined the prevalence of oracle approximations in test cases within DL libraries and identified that up to 25% of all assertions utilized oracle approximations. The identified *OA* testing strategies encompassed techniques such as *Absolute Relative Tolerance*, *Absolute Tolerance*, *Rounding Tolerance*, and *Error Bounding*. Furthermore, the research delved into the diversity of utilized test oracles and thresholds in *OA*, revealing that computation-derived oracles predominantly featured. Finally, the study scrutinized the motivations behind code changes and elucidated maintenance challenges encountered by DL engineers when employing oracle approximation techniques.

Next, we discuss some of the examples of testing strategies that have not received much attention both from the research and the practices, but was observed in the studied ML software systems:

● **Error and Exception Condition (EEC):** This testing strategy employs the concept of exception and error handling [23] to derive the test cases and help detect or signal the errors (or potential errors) that can occur during the running of the ML software system. It is used to detect problems that may arise in the sequence of statements through the exception-related code or by executing a specific statement to trigger an exception, usually using the *throw* or *raise* statement. The steps in *EEC* include understanding limitations in the ML artifacts – data/features, model, or code across the ML workflow and trying to make these aspects as robust as possible. Unfortunately, there is no one-size-fits-all solution for addressing ML challenges. Depending on the nature of the ML problem, various error and exception conditions may need to be considered. For example, in the context of computer vision tasks, it is imperative to ensure that the input data satisfies the necessary quality criteria, including validity. Furthermore, the features need to conform to the expectations of the model. Moreover, DL models often exhibit complexity and opacity, rendering the features utilized in the model difficult to interpret (explainable) due to their non-trivial nature.

● **Decision and Logical Condition:** This structural testing strategy [35] uses conditional expressions to test the possible outcomes of a program's decisions and ensure that the program's different points of entry or subroutines are verified at least once. The test strategy helps to validate the branches in the SUT, ensuring that no branch leads to abnormal system behavior. Previous works [128, 140] proposed coverage criterion for deep neural networks, inspired by the modified condition/decision coverage criterion capturing the logical structures of the deep neural network and their semantics.

● **Thresholding:** Limited studies [111, 123, 151] have focused on finding the classification threshold for testing or probabilistic forecasts calibration. This testing strategy is used to determine the best or optimal threshold value of a decision system (classification) by maximizing a balanced performance measurement (for example, F1 score, Precision, Recall, Accuracy). The step to convert the prediction probability or a given scoring measure into a class label is guided by a parameter called "decision threshold," "discrimination threshold," or simply "threshold." In some cases, the default threshold value may not represent an optimal interpretation of the predicted probabilities or the scoring for the following reasons: 1) Uncalibrated prediction probabilities [111, 123]; 2) Imbalance classification problem [151] (i.e., the distribution of class is severely skewed); 3) The metrics for training and evaluating the model are not the same.

● **Back Testing (Back-testing):** This testing strategy is a concept in finance/business to evaluate the performance of an investment/trading model based on historical data [68]. It examines how well the implemented strategy can perform in the future based on a simulation with the previous data. This includes testing the idea on different samples of historical dataset drawn in/ out of sample (or simulating various possible outcomes) and comparing the consistency of the strategy's performance. Different Back Testing strategies are discussed in [24] and [37] to assess the correctness of a Value-at-Risk model at several quantiles/ intervals.

● **Swarming testing:** The aim of the testing strategy is to improve the diversity of test cases to enhance test coverage and fault detection during random testing [62]. Swarming testing does not follow the general practice of potentially including all features in every test case. Rather, a large "swarm" of randomly created configurations containing only specific features (omitting some features) is used, with the configurations accepting similar resources. Gheisari et al. proposed a score-based learning algorithm in [54], employing Particle Swarm Optimization (PSO) principles. PSO, inspired by social behaviors like fish schooling and bird flocking, is a stochastic meta-heuristic optimization algorithm. It treats potential solutions as particles navigating problem space, updating their positions and velocities based on local and global information. The algorithm introduces novel velocity and position update formulas, integrating stochastic mutation and crossover operations from Genetic Algorithms.

### 2.1.2 ML testing properties

 A testing *property* is a software requirement (functional and non-functional) that should be met to ensure that the software system behaves as expected [18, 132]. In the context of ML-based software systems, ML properties are essential to ensure that the models generalize as expected in deployment scenarios [18]. Zhang et al. [150] presented a comprehensive review of ML testing research, covering testing properties (such as correctness, robustness, and fairness). Among the ML testing properties discussed in the literature, *Bias and Fairness* [51], and *Robustness* [15, 78, 115, 122] are the ML properties that received attention and was studied empirically in details.

● Consistency and Correctness: A specification is consistent when its provisions align harmoniously with each other or with the overarching specs or objectives.[16]. In data labeling, consistency measures the agreement among annotators/ labelers (machine or human) [47]. Properties influencing consistency include non-interferential behavior, boundedness, reversibility, liveness, etc. Specifically, these entail (1) identification of operations negatively affecting other operations, (2) ensuring finite capacity of resources, (3) maintaining initial state integrity post-operation, and (4) enabling every operation during execution. Notably, one critical aspect contributing to an AI system's trustworthiness is its ability to consistently and reliably produce correct results with the same input [6, 45]. Correctness, on the other hand, is a measure of the probability that an ML software system yields the correct outcome [150]. In the case of model prediction, correctness involves evaluating the likelihood that the model's predicted outcome aligns with the actual result. Additionally, ensuring a consistent and comprehensive set of scenarios contributes significantly to correctness [56].

● **Bias and Fairness:** ML software system, in addition to being correct and robust, should be fair and without bias. Galhotra et al. [51] proposed a metric-based testing approach for measuring fairness in ML software systems. The approach measures the fraction of inputs for which changing specific characteristics causes the output to change by relying on causality-based discrimination measures. Using these metrics, automatic test suites are generated to allow for the detection of any form of discrimination in the ML models under test. The test cases modify training data inputs that are related to a sensitive attribute, aiming to verify if the modifications cause a change in the outcome. Also, for a given sensitive attribute (SA), a fairness test validates the model under test by altering the value of SA for any input and verifying that the output remains unchanged. Using the proposed fairness test, they examined the fairness of 20 real-world ML software systems (12 of these ML software systems were designed with fairness as the primary objective) and observed that even when fairness is a design goal, developers can easily introduce discrimination in software. Also, different bias mitigation algorithms are broadly categorized into pre-processing, in-processing, and post-processing [4, 28, 41, 100, 105] that try to improve the fairness metrics by modifying the training data, the learning algorithm, or the results of the predictions.

● **Robustness:** Robustness measures how well a system (SUT) can accurately function even when subjected to stressful environmental conditions or specific input streams [150]. The input streams may be either a valid input, an invalid input, or a random input stream. Therefore, a robust system should maintain its performance even when there is a change in input stream or introduction of noise (e.g., through perturbation) [141, 150]. Prior works have evaluated the stability of the model due to the adversarial changes [78, 115] or model robustness to noise in the data [15, 122]. For instance, Belinkov and Bisk [15] studied the robustness of the Neural Machine Translation (NMT) models when presented with both natural and synthetic kinds of noise. Then, they investigated different techniques for increasing model robustness for the NMT model, i.e., robust training on noisy texts and structure-invariant word representations. They reported that a model based on a character convolutions neural network could simultaneously learn representations robust to multiple kinds of noise during the model training. Also, they observed rich aspects of natural human errors that the current models cannot efficiently capture.

● *Compatibility and Portability*: Compatibility is the ability of two or more ML components or systems to perform their respective functions within a shared environment. In contrast, Portability concerns the ease of moving/ porting ML components or systems between environments (software or hardware environments).

● *Data Uniqueness*: This ML property measures unnecessary duplication in or across the ML software system within a particular field, record, or data set [14], by discrete measures of repeatable data items within or comparison with the counterpart in different data set that complies with the exact business rules or information specifications.

● *Data Timeliness*: This property measures the degree to which the information/data is up-to-date and made available within the acceptable timeline, time frame, or duration. The main dimensions for measuring the Data timeliness proposed in the literature are currency (measures how up-to-date data is at a given moment.), volatility (measures the degree of fluctuation in data over time; high volatility indicates frequent changes while low volatility shows signs of stability.), and Timeliness [14, 17, 85, 125] (measures how quickly and efficiently data is delivered, including data freshness, processing and transmission delays.). This paper investigates the practical adoption of testing properties and strategies outlined in the literature review. We explore how these properties and strategies manifest in real-world scenarios. The comprehensive list of ML properties identified in the studied ML software projects is detailed in Section 4.2.

### 2.1.3   *Test types (Types of testing)*

Sato et al. [1] describe the type of testing (Test types) as Test Pyramid introduced by Mike Cohn [93] to help engineers of ML software systems visualize the different layers of testing and estimate the amount of testing effort or frequency for each layer (*i.e.,* the higher is the level of a type of test on the Pyramid, the lower should be the number of its instances). The main components of the Test Pyramid are:

● **Unit tests:** Are in the first level of testing at the bottom of the test pyramid. They aim to verify the correctness of each functionality once the implementation is complete.

● **User Interface (UI) tests:** Are used to verify the aspects of any software components that an end-user will interact with. This involves testing any visual elements of the software system, verifying that they are functioning according to requirements (in terms of functionality and performance and ensuring they are bug-free).

● **Contract tests:** Are a type of testing used to check the interaction between two separate systems, such as two microservices for an API provider and a client. It captures and stores the communication exchanged between each service in a contract, and the stored interaction is used to verify that both parties adhere to it. Compared to other closely related approaches (such as Schema testing, Compatibility testing [135]), in the Contract test, the two systems can be tested independently from each other and the contract is autogenerated using code.

● **Integration tests:** Aim to ensure that small combinations of different units (usually two units) behave as expected and testing that they coherently work together.

● **End-to-end tests:** Represent a form of black-box testing aimed at evaluating the entire software product's functionality from initiation to completion. By simulating real user scenarios, this type of testing verifies the seamless interaction and interoperability of all integrated components within the system, ensuring they collectively meet expected behavior.

● **Exploratory tests:** Are tests performed on the fly (i.e., test cases are not created in advance), whereby the test cases are designed and executed simultaneously, and the results observed are used to design the next test. The Test methods described in the Pyramid proposed by Sato et al. [1], and Mike Cohn [93], above can be considered as a reference to derive the different types of tests from the practitioner's point of view. To better capture both the research and practitioners' point of view on testing ML systems, we complement the above types of tests with the "test level" proposed by Riccio, Vincenzo et al. [116], —by merging the: *Input Tests*, *Model Test*, and *System Test*.

● **Input Test:** This type of testing analyzes the training, validation or test dataset to identify the potential cause for unsuccessful training on the data, therefore minimizing the risk of faults due to prediction uncertainty [116].

● **Model Test:** This type of testing focuses on testing the ML model in isolation, e.g., without considering other parts of the ML software system. The *Model performance tests* and *Model Bias and Fairness tests* discussed above can be considered as part of these tests. Other tests include checking the model architecture and the model training process.

● **System Test:** Is performed on a complete system to evaluate its behavior or compliance against the functional requirements [116]. Typically, this test is conducted after integration tests to assess the entire ML software system in a production-like environment. The goal is to identify any bugs or defects that may have been overlooked during earlier testing stages. It is carried out before end-to-end tests. We will use "types of testing" and "test types" interchangeably for convenience.

## 3 Methodology of the Study

This section describes the methodology used in this study, aiming to investigate the following research questions (***RQs***):

***RQ1 What are the common testing strategies used in an ML workflow?***
*Are they consistently tested across the ML software systems?* Through a manual analysis of ML test cases following an open coding procedure, we want to derive a taxonomy of testing strategies used in the ML software system. Specifically, we want to examine the highly dominated testing strategies across the ML workflow and the consistency of the usage across the ML software projects.

***RQ2 What are the specific ML properties that are tested in an ML workflow?***
In this ***RQ***, we will manually classify the tested ML properties, following an open coding procedure, to identify the commonly tested ML properties in ML workflow and across the different ML software domain.

***RQ3 What are the software types of testing used in an ML workflow?***

We want to examine different types of testing used in the studied ML software systems. Moreover, we aim to map the various types of testing to their corresponding testing strategies (already studied in **RQ1**) to gain insights into how ML engineers operationalize these tests. This mapping will provide a valuable guideline for ML engineers aspiring to implement diverse testing types within their ML software delivery pipeline.

To answer the above research questions, we used a sequential mixed-methods [77] comprising of both qualitative and quantitative analysis to answer our proposed research questions **RQ1** through **RQ3**. Figure 2 shows an overview of our methodology.

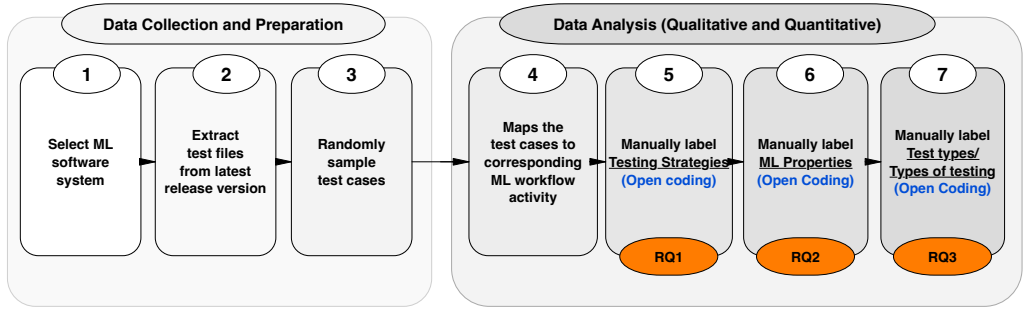

Fig. 2: An overview of our study design/ methodology

① *Select ML software systems*:

To select ML software systems for our study, we first generated a set of relevant keywords for searching GitHub. We queried GitHub using the keywords, and filtered the resulting set of repositories using the inclusion/ exclusion criteria described bellow. Specifically, we proceed as follows.

(i) *Generate Search Keywords ($T_m l$)*:

This step is focused on identifying a comprehensive set of keywords (topics) to encompass a wide range of domains for machine learning (ML) software systems hosted on GitHub. To generate the search keywords, we utilized the search API [55] provided by GitHub. Initially, we searched GitHub topics using keywords such as: "machine-learning" and "deep-learning". GitHub topics are labels assigned to repositories, which facilitate searching and exploring GitHub repositories based on technology, type, or category. The initial keywords returned a set of repositories, which we then manually summarized into twelve (12) major tag categories ($T_{ml}$) by examining co-occurring tags such as 'machine-learning,' 'deep-learning,' 'deep-neural-network,' 'reinforcement-learning,' 'artificial-intelligence,' 'computer-vision,' 'image-processing,' 'neural-network,' 'image-classification,' 'convolutional-neural-networks,' 'object-detection,' and 'machine-intelligence.'

(ii) *Extract Machine Learning Repositories Using $T_{ml}$*: Using the list of keywords obtained in the previous step, we queried the GitHub API [2], searching for repositories that: 1) contain at least one of the keywords in $T_{ml}$ (case insensitive) either in the repository name, descriptions, or system README file; 2) are mainline software systems (i.e., not a forked repository); 3) the README or description is written in English (for easy understanding to allow us to obtain more details about the system); 4) are not archived repositories. This initial search returned a total of $29, 972$ unique repositories.

(iii) *Apply inclusion/exclusion criteria*: We applied the following inclusion/exclusion criteria to retain only mature ML software systems for our study: 1) We sorted the repositories by the number of stars and selected the top 500 most-starred repositories. This resulted in a list of repositories with stars ranging from 2,563 to 160,398. Specifically, we used star counts as a proxy for project popularity, although it may correlate with other project metrics such as commits, forks, issues, and pull requests. For instance, all the top 500 most starred projects had recorded at least 10 commits and had been forked at least 95 times, reducing the chance of selecting experimental systems such as students' class projects, etc. This criteria follows best practices established by previous studies [21, 22, 96]. 2) Our goal is to understand the testing practices in ML software systems by ensuring that the analyzed ML repositories contain at least one ML model. To this end, we manually read through the top 500 repositories to ensure that they are indeed related to ML systems (i.e., contain ML models). Notably, we filtered only the repositories that use machine learning or deep learning to perform specific tasks for the end-user (i.e., containing at least one ML model integrated with other components such as user interface, databases, or detection system) and for the repositories that host ML models. Additionally, we manually analyzed ML repositories that contain at least five test files and have been active for at least one year (i.e., the active period is computed as the difference between the date of the last update and the creation date of the system) and have been released at least once. This step removed 362 ML repositories that are either ML toolkits, DL frameworks, documentation, or for which we could not identify any ML model, leaving us with 138 ML repositories. 3) We manually filtered systems to retain only those where tests are written in either Python or C/C++ programming languages, which are considered the main programming languages [36] for building ML model software systems. This criterion left us with 11 ML software systems.

(iv) *Final List of Machine Learning Repositories*: Table 1 provides descriptive statistics about the nine retained ML software systems. These ML software systems are of varying sizes covering different domains (i.e., Recommendation systems, CV applications/ model, medical application platform, and natural language processing (NLP)), has been stared more than 4, 000 times (the popularity of the selected ML software system [138]), and containing at least one model.

② *Extraction of the test files from the latest release of software systems*:

This study aims to understand the testing practices applied in ML software systems by analyzing the test-related code. Hence, in this step, we identified all tests that are related to source files from the latest release of each ML software system as follows: First, we identified the latest release version ($R_n$) of each selected ML software system (from which we need to extract each tests' related code) using Git tags API of the form: `https://api.github.com/repos/{owner}/{repo}/tags` (e.g., `https://api.github.com/repos/numenta/nupic/tags` for Nupic software system). Git releases represent the different snapshots of a system by marking a specific point in time of the repository's history using Git tags [1]. In this study, we considered only the latest stable release version (i.e., no pre-release versions) of each studied ML software system, to ensure that we analyze fully tested versions. Hence, whenever the latest release version $R_n$ of a system is a pre-release (i.e., contain the word 'alpha' in the release name indicating an alpha version), we select the previous stable version of the system. We do not consider the alpha release because usually, an alpha release may indicate that the current version is unstable and may not contain all of the planned features for the final version. The latest release for the studied ML software systems (at the time of this study) are listed in column 'Release' of Table 1.

---

[1]https://git-scm.com/book/en/v2/Git-Basics-Tagging

Table 1: **List of studied Machine learning software systems and their characteristics.** *Com*: Number of commits, *Stars*: Number of Stars, *Language*: the most used programming language and their composition, *Sample*: The random sample size of the test cases based on 95% confidence and a 5% confidence interval (refer to following step), *Release*: The release version studied, *Testing Framework*: The unit testing framework

| | ML System | Com | Stars | Language | Test files | Test cases | Sample | Release | Domain | Testing Framework | system Descriptions |
|---|---|---|---|---|---|---|---|---|---|---|---|
| 1 | apollo | 17,536 | 17,650 | C/C++(83.7%), Python(5.1%) | 150 | 439 | 207 | v6.0.0 | CV System/ Autonomous driving. | gTest, unittest | A high performance, flexible architecture autonomous driving platform for the development, testing, and deployment of Autonomous Vehicles. Apollo contains 28 ML models that are trained based on various deep learning frameworks (e.g., Caffe, Paddle, and PyTorch). |
| 2 | lightfm | 428 | 3,831 | Python(99.1%) | 7 | 63 | 55 | 1.16.0 | Recommender | pytest | lightFM [81] is a recommendation system based in Python, developed by Lyst, a Fashion e-Commerce site based in London. lightFM implements multiple algorithms such as Bayesian Personalized Ranking from Implicit Feedback and explicit feedback utilizing Weighted Approximate Rank Pairwise (WARP) ranking losses. |
| 3 | qlib | 1,541 | 6,976 | Python (98.5%) | 20 | 51 | 46 | v0.8.5 | Recommender | numpy, unittest | Qlib is an AI-oriented Qualitative investment platform [147], developed by Microsoft. Qlib covers the entire chain of quantitative investment: alpha seeking, risk modeling, portfolio optimization, and order execution and also helps with the data processing, training model, and back-testing. |
| 4 | paperless-ng | 4,061 | 2,785 | Python (53.3%) | 34 | 398 | 196 | ng-1.5.0 | CV System and NLP System | unittest, pytest, django | Paperless is an ML application that indexes scanned documents and allows for searching for documents and storing the document's metadata. When given a document file (e.g., Pdf, images, plain text, office documents) paperless uses machine learning to can learn and automatically assign tags, mark the document as read, flag them as important, and employs integrated sanity checks before storage. |
| 5 | mmf | 977 | 4,657 | Python (98.9%) | 14 | 23 | 23 | v0.3.1 | NLP | unittest | MMF is a modular framework powered by Pytorch implementing the state-of-the-art vision and language models utilized by multiple research projects at Facebook AI Research (FAIR) [133]. |
| 6 | mycroft-core | 4,848 | 5,431 | Python (92.7%) | 71 | 425 | 202 | v21.2.2 | NLP/ Speech System | unittest | Microsoft-core, a Mycroft Artificial Intelligence platform, is a voice assistant software system that runs on multiple devices such as a desktop computer, inside an automobile, Raspberry Pi. |
| 7 | nanodet | | | Python (76.4%), C/C++ (18.5%) | 32 | 40 | 40 | v1.0.0 | CV Model | unittest | An object detection model on mobile devices. It uses the Generalized Focal Loss [85] for regression and classification loss. |
| 8 | deepchem | 7,280 | 3,250 | Python (99.7%) | 158 | 828 | 263 | 2.6.1 | Application Platform/ Medical | numpy, unittest | DeepChem is an open-source software based on deep learning used in drug discovery, quantum chemistry, materials science, and biology [112]. The use-case for DeepChem includes analyzing protein structures and extracting helpful descriptors, predicting the solubility of molecules, predicting binding affinity for a small molecule to protein targets, predicting physical properties of simple materials, and counting the number of cells in a microscopy image. |
| 9 | magenta | 1,393 | 17,142 | Python (99.2%) | 51 | 256 | 164 | 1.1 | Application Platform/ Art generation | numpy, unittest, absltest | Magenta is a research project from Google Brain exploring the use of deep learning and reinforcement learning algorithms for generating songs, images, drawings, and other materials in the music and art generation. |
| 10 | nupic | 6,625 | 6,187 | Python (97.7%) | 130 | | 270 | 1.0.5 | Application Platform/ Intelligent Computing | numpy, unittest, gTest, BOOST | Numenta Platform for Intelligent Computing is an implementation of Hierarchical Temporal Memory (HTM), a theory of intelligence based strictly on the neuroscience of the neocortex. The HTM uses time-based continuous learning algorithms to store and recall spatial and temporal patterns. |
| 11 | DeepSpeech | 3,329 | 16,165 | C/C++ (68.1%) Python (21.4%) | 121 | | | v0.9.3 | 124 | NLP/Voice recognition | Boost, gLag | An embedded (offline, on-device) open source speech recognition engine which can run in real time on devices ranging from a Raspberry Pi 4 to high power GPU servers. At its core, DeepSpeech uses a trained recurrent neural network (RNN) consisting of 5 layers of hidden units to ingests speech spectrograms and generates English text transcriptions. |

After identifying the tags of the latest release $R_n$ and the corresponding tagged source code in the target systems, we manually downloaded this source code, for further processing locally. We followed prior studies (e.g., [106, 149]) and utilize the naming convention to identify test files. Specifically, we extracted all the files of the local copy of the target systems versions $R_n$ that contain the word 'test' either at the beginning or end of their file names. We used a spreadsheet software to store source code file information such as the file name, file path, and other related meta-data. We further assigned each of the file a sequential unique identifier ($F_{id}$), starting from the value 1, to help us easily refer to each file later during the manual labeling and the rest of the analysis.

Next, we detailed the terminology used during the next steps of data extracted and throughout this paper (i.e., test cases, tests, testing strategies, and ML properties) using examples. Specifically, in this study we are focusing on the testing practices of the Python and C/C++ programming languages, since they are main languages [36] used in ML models and ML software systems. Python provides a rich support of test packages or frameworks to handle both white-box and black-box testing. Moreover, it is also possible to write any black-box test in Python, even if the SUT is not written in Python. Depending on the types of tests, different test frameworks may be chosen.

For example, a basic unit test could be handled by the `unittest` framework (i.e., a build-in test framework into Python standard library), but other test frameworks like `pytest` may work better for higher-level testing such as using test fixtures [110].

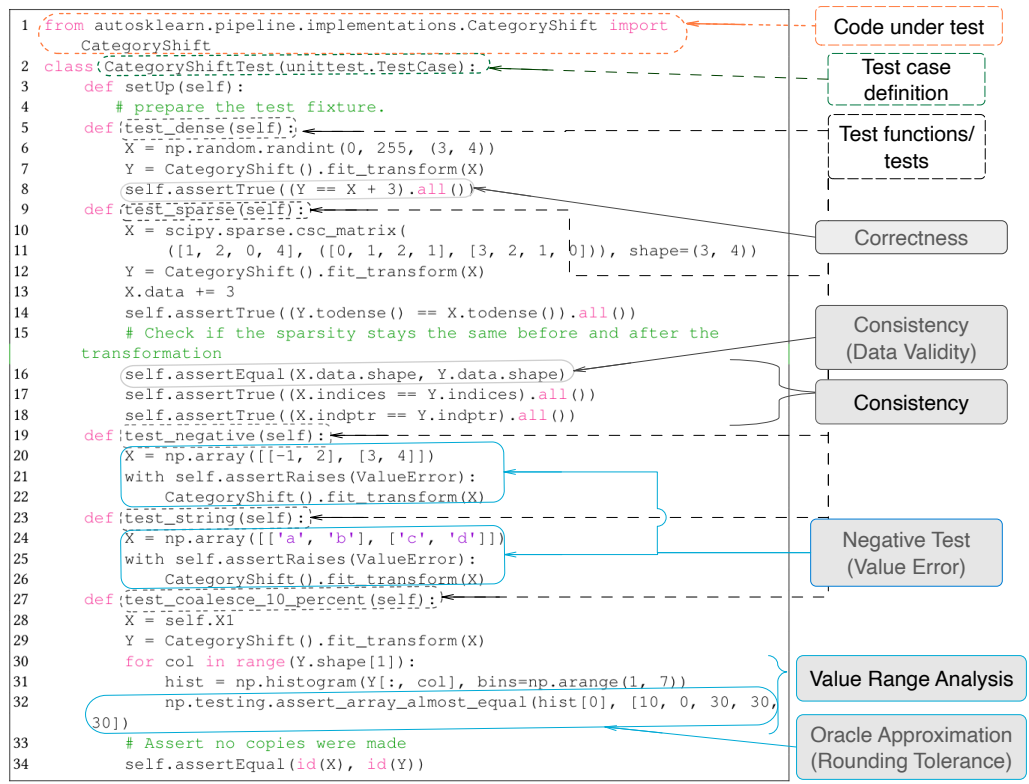

Fig. 3: Example of a test file in Python that uses the `unittest` testing framework. We use different color codes defined as follows: The *dotted orange* line is the code under test, *dotted green* line indicates the test case definition, *dotted gray* lines indicates the test functions/tests, **bold gray** lines indicates the ML properties being tested, and the bold light blue color indicates the testing strategies.

Figure 3 shows the example code of a test case in Python that uses the build-in unit test framework (`unittest`). A *test case* is a collection of unit tests that demonstrates that a function works as expected under a broad range of conditions (in which that function may find itself and is expected to work successfully). Usually, a test case considers all possible kinds of input a function under test is expected to receive from users and therefore contains the tests (test functions/methods) representing each expected input situation. According to Figure 3, the name of the test case is 'CategoryShiftTest' and it contains five (5) test functions (i.e., methods/functions containing the word 'test') such as 'test_dense', 'test_sparse()', 'test_negative' 'test_string()', and 'test_coalesce_10_percent()'. For simplicity, we will refer to these functions/methods defined inside the test cases containing the word 'test' as *Test functions* or *Tests* interchangeably throughout this paper. A test file may contain one or multiple test cases, for example, the following reference [8] contains a test file with three test cases corresponding to three functions. Also, for the situations where the test functions/methods are independent of each other (no class defined), e.g., manual testing or integration testing, such as in [7], each test function is defined as a separate test case. C/C++ supports multiple unit testing frameworks. The most popular testing frameworks are *Google C++* (gTest) [59] and *Boost.Test* (Boost) [53]. Both frameworks have closely related features that allow the creation of test cases and their organization as test suites that can be registered automatically or manually. Other features supported by both frameworks include the addition of test fixtures for each test case, test suite, or globally for all test cases. Test fixtures allow for a consistent initialization and cleaning of resources during the testing process. They also reduce code duplication between test cases. Other unit testing frameworks used for C/C++ based software systems include CppUTest [40], Unity [90].

Figure 3 present an example of test cases implementing different testing strategies for testing different properties. In this paper, we identify ML properties and testing strategies from such a test. In Figure 3, we further highlighted the different testing strategies and ML properties that are being tested using the color codes *Bold light blue* lines and *Bold gray* lines, respectively. We will describe in detail the labeling procedure that we followed to derive the different testing strategies and ML properties in Step ⑤ and Step ⑥.

③ *Randomly sample the test cases*:

In this step, first, we manually extracted all test cases and their corresponding test functions defined inside the test files while referring to the system documentation. For example, for the test using python unit testing frameworks, the test functions contain the word 'test' appended to the name of the function (for automatic test runner), while tests in google test use the macro TEST() to define the test or test fixtures. Then, for each system, we randomly selected a sample of test cases using a 95% confidence level and a 5% confidence interval. This resulted into a total sample of 1,588 test cases ($T_s$). The distribution of the sample sizes of test cases obtained for each of the 9 studied systems is shown in column '*Samples*' of Table 1.

④ *Map the test cases to the corresponding ML workflow activity*:

As part of the goal of this study is to understand how the different testing practices expand across the ML workflow, the first two authors manually mapped all the test cases ($T_s$) to the corresponding ML workflow activities (e.g., *Data Cleaning*, *Feature Engineering*) illustrated in Figure 1. To map a test case to an activity of the ML workflow, the authors used information such as the name of the test file, the incode comments (i.e., comments left inside the test source code describing what the test is about) and official documentation related to the code under test (to get familiar with what the test cases are about). They then assigned the ML workflow activity name (i.e., mapping) to the test case manually. Each decision was discussed between the two authors until a consensus was reached before a label was assigned.

⑤ *Label the testing strategies implemented to test different ML properties*:

Tests strategies are the various techniques or approaches that are used to test a system, in order to ensure that its behavior matches the specification. In this part, we specifically focused on forming a taxonomy of categories and sub-categories of testing strategies that are used in ML software systems. Six individuals consisting of the first four authors of this paper and two additional researchers (a graduate student and a research professional with a PhD degree) participated in the initial labeling and construction of the taxonomy. All these participants have strong knowledge of machine learning and software testing. Furthermore, participants leveraged the research literature on testing strategies, as outlined in Section 2, throughout the preparatory stages and while inductively assigning labels at both the statement and function levels, as depicted in Figure 3. To generate the taxonomy, first, the documents, each containing randomly sampled test cases selected from the total samples $T_s(1,588)$ of the test cases, were distributed to the labeling team. During this first iteration, the first author labeled 800 test cases, and the remaining test cases were distributed among the five other participants. Each individual then constructed a taxonomy of categories and sub-categories of testing strategies, following the open coding procedure [49, 126], by analyzing the test file contents in a bottom-up process (i.e., from labels, sub-categories to the main/core category) as described below. The participants were provided with the cloned source codes of the studied ML software systems release version $R_n$ together with a spreadsheet document containing the name of the sampled test cases, the corresponding test file name, $F_{id}$, and the complete path of the file corresponding to the source code files in $R_n$. They then read through the test source codes of the target ML software systems and refer to the official documentation to be familiar with code implementation before assigning short sentences as initial labels, to indicate the test strategy. Specifically, to derive the complete set of test strategies, we proceeds by following these two steps:

(i) Examine the structure and content of the test code; the algorithm being tested, control statements, logical statements, loop and error-handling techniques, and other run-time options. For instance, the test `test_coalesce_10_percent` (line 27 to 34) in Figure 3 compares each element of the transformed target data $Y$ using a looping statement and approximates them with the expected value [10, 0, 30, 30, 30].

(ii) Identify and extract the assertions API used within the test cases; to help identify categories and sub-categories of testing techniques based on the extracted assertions as follows:

Test assertions are statements within the test functions through which desired program specifications are checked against actual program behavior. As such, assertions represent the core part of test functions that evaluate the program's internal state. Most unit testing frameworks provide multiple assertion functions to compare different types of actual-vs-expected variables [82]. Assessing the most used assertions can help understand what is being tested. In this step of our analysis, we categorized the commonly used assertions to derive the tests strategies in ML software systems.

First, we identify the test functions that are used in the studied ML software systems to express assertions. For example the test functions `test_dense()`, `test_sparse()`, `test_negative()`, `test_string()`, and `test_coalesce_10_percent()` in Figure 3. We only considered the test functions within the randomly selected samples described in Step ③ above. Then, we extracted assertions by finding the usages within the identified test functions. In particular, for every test functions, we manually extracted all the lines representing the assertion statement. In Python, assertions are expressed in two ways: 1) assert keyword, which is then followed by a Boolean expression; and 2) using customized assertion APIs, e.g., internally defined by each Python system, the Python *unittest* built-in functions, and assertion APIs provided by NumPy, i.e., a commonly used library in Python that supports computations on arrays and matrices. The most popular unit testing frameworks for C++ are *Boost.Test* (Boost), and *Google C++* (gTest). Both frameworks have similar features; for example, in Boost, engineers can organize test cases into test suites that could be registered automatically or manually. Moreover, they both allow a broader number of checkers/assertions such as Exceptions: Throws/Not throws, Equal, Not Equal, Greater, Less, the Equality checking for collections & bits, explicit Fail/Success, Floating-point numbers comparison, including control of Closeness/Approximation of numbers. To extract the assertions statement, we followed a combined approach of both manual and automated techniques that include first reading the official documentation of the target ML software systems and understanding the code structure to initiate regular expression queries such as assert*, EXPECT*, BOOST* or CHECK*/ ACHECK* while continuously examining the query results. Prior to manually labeling the test cases and the assertion APIs—as described in the next steps, the six participants confirmed that all the assertions are extracted and correctly mapped to the respective test functions. The manual analysis allowed us to identify rarely used assertion APIs that could not be identified using an automatic process. For example, some systems define a custom assertion API instead of using the standard assertion API provided by the testing framework (e.g., `SLOPPY_CHECK_CLOSE` defined in DeepSpeech system instead of using the standard `BOOST_CHECK_CLOSE` provided by the Boost framework, due to reasons such as types matching problem). Our approach to extracting the assertion APIs is similar to the following previous work [98]. Leveraging the extracted information, the team proceeded to build the categories of the related assertions statements separately before the general group discussion.

Figure 3 is an example to illustrate the labeling steps described above. We identified testing strategies from such test by proceeding as follows. First, we followed the path to the code under test specified in Line 1, to understand the content of the code being tested. Next, we examine the structure and content of the test code; test data generation, control statement, logical statement, loop, error-handling technique, run-time options, and assertion API choices, following a top-down or bottom-up procedure. For instance, the test `test_coalesce_10_percent` (line 27 to 34) compares each element of the transformed target data $Y$ using a looping statement and approximates them with the expected value $[10, 0, 30, 30, 30]$. The assertion API `assert_array_almost_equal`, provided by NumPy [99], at line 32 approximates the two values with the default of 7 decimal places [2], an example of *Oracle Approximation* of sub-category *Rounding Tolerance* testing strategy. Note that some approximation assertions API (e.g., `EXPECT_NEAR`, in gTest [3]) instead approximates the result using the absolute range (i.e., $1e - 4$). We referred to this sub-category of *Oracle Approximation* as *Absolute Tolerance*.

Also, the test (i.e., `test_coalesce_10_percent`) at the same time analyses the transformed target $Y$ elements expecting the test to pass within the range of value using the loop statement at line 30. We therefore referred to this kind of testing strategy as *Value Range Analysis*. Following the initial labeling process, the team consolidated all the labeled items and collectively reviewed each. Subsequently, they embarked on generating a hierarchical taxonomy of test strategies by clustering related labels into cohesive categories (see Figure 4). This categorization process is iterative, with team members continuously revisiting and refining the taxonomy by iteratively grouping related categories and labels. Throughout the categorization process, we observed a spectrum of complexity among testing strategies. While some exhibited hierarchical structures with four levels of intricacy, others presented simpler frameworks with only two levels. This diversity underscores the multifaceted nature of testing activities intersecting ML and software engineering. Three domain experts with several years of industrial and academic experience in ML and software testing were engaged to ensure a systematic and rigorous classification approach. Drawing upon the collective expertise of these experts (co-authors), we meticulously categorized the testing strategies into three overarching categories: White-box, Black-box, and Grey-box. This classification framework was informed by expert opinions and supported by pertinent literature where applicable.

In particular, the experts differentiated among White-box, Black-box, and Grey-box testing strategies by examining the characteristics and structure of the test code, with a focus on the level of knowledge about the system's internal workings and the approach taken in testing ML systems. White-box testing is characterized by direct manipulation and assertions on internal code components, detailed path testing, and the use of code coverage tools. In contrast, Black-box testing centers on inputs and outputs that adhere to specifications and tests through external interfaces without referencing internal code. Grey-box testing incorporates a mix of internal and external knowledge, tests integration points, and utilizes high-level internal APIs or modules. The experts effectively annotated and differentiated the testing strategies by analyzing these aspects. For instance, after collaborative deliberation among the experts, swarm testing was classified under heuristic-based strategies within the grey-box category. Consequently, our taxonomy of testing strategies and their respective prevalence provides a clear delineation into White-box (23.44%), Grey-box (69.38%), and Black-box (16.10%) categories. This taxonomy offers valuable insights into the nuanced nature of testing approaches used in ML systems.

---

[2]https://het.as.utexas.edu/HET/Software/Numpy/reference/generated/numpy.testing.assert_almost_equal.html
[3]https://google.github.io/googletest/reference/assertions.html

Furthermore, it is important to note that a test function may align with one or multiple testing strategies simultaneously. Any conflicts arising from this categorization were thoroughly deliberated and resolved through the involvement of an additional expert—a practitioner with extensive research experience in machine learning and software testing. This collaborative approach to conflict resolution mirrors practices documented in prior literature [89, 98, 101]. For a comprehensive overview, Tables 6 and 7 in the appendix provide a detailed representation of the selected testing strategies derived from this meticulous analysis. We discuss this analysis (**RQ1**) in Section 4.1.

⑥ *Label the ML Properties being tested*:

ML properties are requirements that should be satisfied to ensure that a model behaves as expected. To categorize ML properties being tested, we followed the same labeling procedure discussed in step ⑤, only now focusing on analyzing the test properties and classifying the requirements being tested into functional and non-functional requirements. The labeling team used the test properties reviewed by Zhang et al. [150] (i.e., *Correctness, Model Relevance, Robustness, Security, Efficiency, Fairness, Interpretability, and Privacy*), during the preparation steps and when assigning labels. To derive the labels, the labeling team focused on understanding the code under test based on the following three dimensions:

(1) Understanding the problem domain, by analyzing the test input data, the process, and the output, by continuously referring to the official documentation related to the code under test. (2) Examining the algorithm being tested and scrutinizing it for any imprecision. This helped understand how the testing data set are constructed. These two initial steps were important to understand the algorithms being tested. For example, they allowed understanding the kinds of inputs the algorithms expect and the different outputs that are expected from the algorithms. (3) Examining the oracle comparison and the run-time options. This step helped to understand the specific point in the test cases at which the engineers verify the test results and how they permute the order of the input data.

Using one or a combination of the above three dimensions, the labeling team members were able to understand the ML properties being tested. Then, leveraging the ISO 25010 Software and Data Quality standard[4] and the ISO/IEC TR 29119-11:2020 Software and systems engineering — Software testing standard (Part 11: Guidelines on the testing of AI-based systems)[5], they assigned initial labels to each identified property.

For instance, in Figure 3, to understand what ML properties are being tested, in the test function `test_dense` (line 5 to 8), which verifies the correctness of the implemented function `fit_transform` by checking that the returned transformed data $Y$ has all the components of input $X$ successfully incremented by 3. The property being tested is *Correctness*. Test `test_sparse()`, in addition to functional correctness (line 14) also verifies the consistency of the transformed data (line 16 to 18) against the original data. It also verifies the validity of data

---

format (line 16). We referred to these two ML properties being tested as *Consistency* and *Validity*. Other test scenario related to *Consistency* could be an test oracle executed inside a loop statement without changing the test input or the oracle value.

After this initial labeling step, they grouped the labels into categories to form the final taxonomy of ML test properties. The team followed an iterative process by going back and forth between categories while refining the taxonomy. The team discussed and resolved all the conflicts by introducing a new practitioner with extensive research experience in software engineering and machine learning. The results of this step is discussed in Section 4.2 to answers our **RQ2**.

⑦ *Categorization of Tests types (Types of testing) implemented in the studied systems*:

Software testing types are sets of testing activities categorized into three basic levels of transparency: opaque, translucent, and transparent. These activities aim to validate a System Under Test (SUT) for a defined set of test objectives. Following the open coding procedure, the authors classified the different testing activities used in the studied ML systems as follows: For each sampled test file in the spreadsheet, the authors first read through the corresponding source code, using the official documentation as a reference to familiarize themselves with the test activities. Specifically, the authors focused on three main dimensions: 1) understanding the overall goal and achievement of the test file when executed (test objective), 2) identifying the different test strategies used, as described in Step ⑤, and 3) recognizing the test deliverables. In many cases, test types such as integration, unit, regression, and swarming were already labeled by the developers via folder and file names (e.g., in ML software systems like `Nupic` and `mycroft-core`). In these instances, the authors directly used these names as labels. When the test type was not specified in the folder name, the labeling team examined the tests based on the three dimensions mentioned above to ensure correct labeling. For example, a test containing multiple print statements instead of assertions would likely be labeled as a manual/static test. All disagreements during the labeling process were discussed and resolved before final labels were assigned. A practitioner with extensive research experience in software testing participated in these discussions. A consensus was reached on all labels. The results of this analysis answer our **RQ3** and are discussed in Section 4.3.

We share the replication package of this study in [3].

## 4 Results

In this section, we present the results of our analysis, answering the proposed research questions.

### 4.1 *RQ1: What are the common testing strategies used in an ML workflow?*

This subsection reports the results for testing strategies, divided into three parts: (1) A taxonomy of the types of testing strategies and their breakdowns, (2) Mapping the testing strategies to the ML workflow, and (3) Comparing the testing strategies across the studied ML systems and domains.

*4.1.1 Taxonomy of the Testing Strategies:* Figure 4 presents the high-level and low-level categories of test strategies identified in the studied ML systems (at Step ⑤ of our analysis methodology). For each strategy, we computed and presented the average percentage of tests corresponding to each testing strategy across all the studied ML software systems.

The example code for each testing strategy is highlighted in the column 'Code Example/Assertion API' of Table 3. In the following sections, we describe each category and sub-category of the obtained taxonomy in detail, providing examples of test scenarios.

**Grey-Box Testing:** The test cases are designed with partial knowledge of the internal structure of the component. This category of the testing strategy takes the highest percentage (69.39%) of the studied test cases. Table 2 shows the composition breakdown of the testing strategies across the ML workflow activities. We discuss the sub-categories of *Grey-Box* testing identified below:

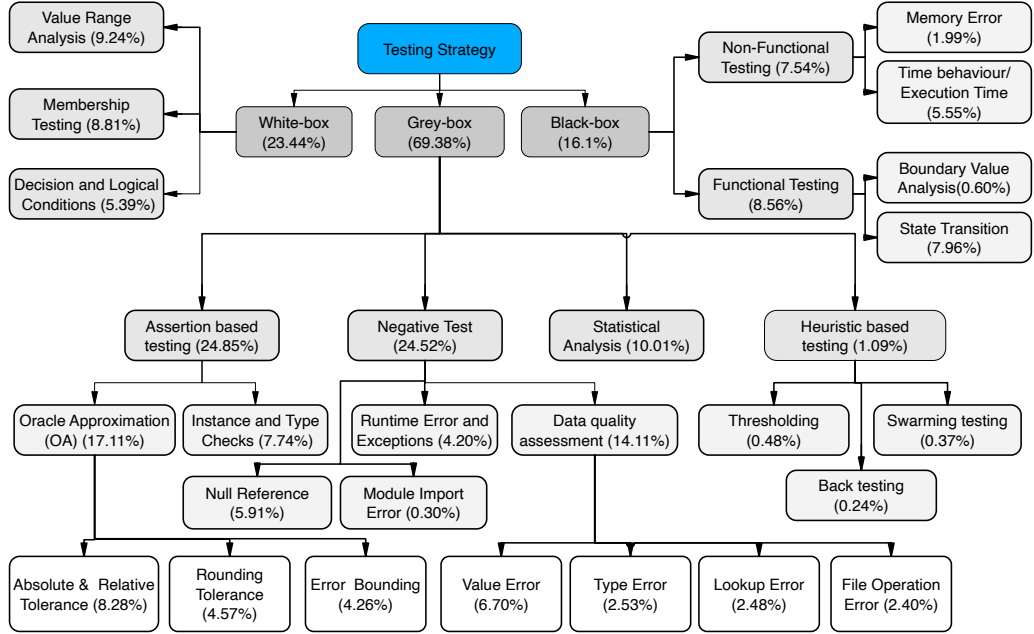

Fig. 4: Overview of the common testing strategies found in the studied ML Systems. The nine (9) high-level categories of the testing strategies are highlighted in light gray color code, while the sub-categories are shown in white boxes.

• *Negative Testing:* This category focuses on how the System Under Test (SUT) can gracefully handle invalid inputs. *Negative test* checks whether the software system behaves as expected with negative or invalid data inputs. An example code of a negative assertion test is shown in Table 7a from the Appendix. Negative testing ensures that the software does not crash but continues running normally when given invalid data inputs. In negative testing, exceptions are expected, which indicates that the software handles improper data input behavior correctly. Moreover, negative test cases can detect more defects in the software system compared to positive tests [27]. This testing strategy dominates in four major ML workflow activities, i.e., Model Training (17.48%), Model monitoring (14.71%), Data labeling (12.42%), and Data cleaning (10.5%), according to our results presented in Table 2.

Table 2: The list of common Testing Strategies and their percentage composition across the ML Workflow. The highlighted cells in **bold-face** indicate the highest value. Due to space constraints, the following terms are abbreviated: *Data*: Data Collection, *Clean*: Data Cleaning, *Label*: Data Labelling, *Feat*: Feature Engineering, *Train*: Model training-related activities, including model fit, prediction, hyper-parameter tuning, *Eval*: Model Evaluation and Post Processing, *Deploy*: Model Deployment activities, including Model inspection, model update, pickling and pipeline export, *Moni*: Monitoring, including Model Monitoring and Inspection, *Util*: shared configurations and Utility file, *Infras*: Infrastructures or frameworks used, *Syst*: tests related to system.

| | Category | Test Properties | Data | Clean | Label | Feat | Train | Eval | Deploy | Moni | Util | Infras | Syst |
|---|---|---|---|---|---|---|---|---|---|---|---|---|---|
| 1 | Exceptions and Error Condition | Data Quality Assessment | 13.86 | 8.77 | 10.9 | 10.41 | **28.55** | 11.79 | 3.89 | 4.75 | 7.08 | 0 | 0 |

Grey-Box

| # | Category | | Test | | | | | | | | | | | |
|---|---|---|---|---|---|---|---|---|---|---|---|---|---|---|
| 2 | | | Negative Test | 9.24 | 10.5 | 12.42 | 8.65 | **17.48** | 8.13 | 7.77 | 14.71 | 11.12 | 0 | 0 |
| 3 | Assertion based Testing | | Oracle Approximation (OA) | 4.05 | 18.09 | 3.88 | 13.84 | **27.09** | 19.54 | 4.14 | 4.85 | 4.52 | 0 | 0 |
| 4 | | | Instance & Type Checks | 8.9 | 12.01 | 13.96 | **24.53** | 18.13 | 7.71 | 5.93 | 4.19 | 4.65 | 0 | 0 |
| 5 | Heuristic based | | Swarming Testing | 4.98 | 0 | 0 | 29.24 | **51.16** | 7.31 | 0 | 7.31 | 0 | 0 | 0 |
| 6 | | | Back testing | 50 | 0 | 0 | 0 | 0 | 0 | 0 | 0 | 50 | 0 | 0 |
| 7 | | | Thresholding | 0 | 0 | 27.23 | 0 | 25.75 | **29.87** | 0 | 0 | 17.15 | 0 | 0 |
| 8 | | | Statistical Testing | 8.71 | 10.37 | 4.55 | 22.98 | **27.23** | 12.7 | 4.5 | 4.68 | 4.27 | 0 | 0 |
| 9 | Black-Box | Functional | State Transition | 11.45 | 11.31 | 2.59 | 12.24 | **14.86** | 10.87 | 7.34 | 6.88 | 11.79 | 6.13 | 4.55 |
| 10 | | | Boundary Value Analysis | 8.67 | **45.34** | 0 | 13.36 | 13.77 | 18.86 | 0 | 0 | 3.65 | 0 | 0 |
| 11 | | Non-functional | Memory Error | 15.59 | 7.96 | 0 | 16 | 9.65 | 8.6 | 6.59 | **17.65** | 1.95 | 8 | 8 |
| 12 | | | Time behaviour | 7.77 | 9.87 | 4.17 | 18.14 | 4.57 | 3.27 | 2.78 | 5 | **23.78** | 16.89 | 3.76 |
| 13 | White-Box | | Value Range Analysis | 10.45 | 8.67 | 4.65 | **27.91** | 20.92 | 13.45 | 5.61 | 4.7 | 3.65 | 0 | 0 |
| 14 | | | Membership Testing | 7.16 | 12.17 | 4.23 | 13.18 | **26.8** | 8.36 | 4.13 | 8.28 | 7.64 | 4.36 | 3.67 |
| 15 | | | Decision & Logical Condition | 8.85 | 10.19 | 9.56 | **22.23** | 20.45 | 11.79 | 4.13 | 3.79 | 2.58 | 6.44 | 0 |
| **AVERAGE** | | | | **11.31** | **11.02** | **6.54** | **15.51** | **20.43** | **11.48** | **3.79** | **5.79** | **10.01** | **2.79** | **1.33** |

(1) ***Runtime Error and Exceptions:*** This category employs the concept of exception and error handling [23] to derive the test cases and help detect or signal the errors (or potential errors) that can occur during the running of the ML software system. The idea behind exception and error handling is to detect problems that may arise in the sequence of statements through the exception-related code or by executing a specific statement to trigger an exception, usually using the throw or raise statement. This implies that developers and ML engineers can provide an efficient mechanism to notify users in case of errors in the ML software system, handle abnormal situations, and implement better recovery strategies, typically within the exception class or objects. An example of this category is testing for timeout errors, such as a socket timeout when a network client hangs while trying to request a server. Additionally, a prediction or model training function can be set to throw a Runtime error whenever required parameters for pipeline optimization are not specified before calling the training or prediction function. Furthermore, this category also represents a specific *negative testing* technique used to identify the incorrect operation of a program or code under test. Specifically, it addresses: (a) Code logic errors that handle program statement sequence mistakes, such as using the wrong formula or function. Errors in this category may compile successfully but still produce inaccurate results. (b) Syntax errors are mistakes in the programming language rules or are violations of the rules governing a language's structure. Moreover, this category includes checking for assertion errors to indicate that something in the code under test should never have happened. This negative testing strategy allows the creation of test cases to ensure that the program's state, inputs, and outputs are correct.

```
1  # e.g., x=range(0,10,2)
2  def value_check(x):
3    for i in x:
4      try:
5        return i%2 == 0
6      except ValueError: pass
```

Listing 1: original code/ function under test

```
1  # Testing incorrect value.
2  def test_value_check():
3    x=range(1,10,2)  #1,3,5,7,9
4    with
       pytest.raises(ValueError):
5      value_check(x)
```

Listing 2: modified input, line 3

(2) **Data Quality Assessment:** This sub-category represents the testing strategies used to assess data quality by employing various data quality dimensions, such as validity, accessibility, correctness, and other potential errors that may occur during data operations. Table 2 indicates that this category of testing strategy is most frequently performed during model training activities (28.55%), followed by data collection (13.86%), model evaluation (11.79%), data labeling (10.9%), and feature engineering (10.41%). We identified four testing strategies related to *Data Quality Assessment*, which include *Value Error*, *Type Error*, *Lookup Error*, and *File Operation Error*, as shown in Figure 4:

```
1   # Checking for Value Error
2   def test_text_input_with_illegal_dim():
3       x = utils.generate_data(shape=(32,))
4       input_node = input_adapter.TextInputAdapter()
5       with pytest.raises(ValueError) as info:
6           x = input_node.transform(x)
7   # Checking for Type Error
8   def test_image_input_numerical():
9       x = np.array([[['unknown']]])
10      input_node = input_adapter.ImageInputAdapter()
11      with pytest.raises(TypeError) as info:
12          x = input_node.transform(x)
```

Listing 3: Example test cases of value error vs type error in Python

(a) **Value Error**: In this test, a modification is made by introducing an incorrect value along the test code path of a given function under test. An example of a value error is shown in Listing 1 and Listing 2. Considering the code under test in Listing 1, which expects the input list $x$ to contain all even numbers, the corresponding test cases was created in Listing 2 by testing that calling the function "`value_check`" with incorrect input that starts with value of $x$ from 0 to 1 (i.e., $x = range(1, 10, 2)$). This modification (mutated test code) will cause the function to fail to execute code inside try, instead throw ValueError due to the value error introduced in the input $x$. (b) **Type Error**: Unlike the *Value Error*, the target program under test is injected with an inappropriate type to invoke a *Type Error*, indicating that the operation on the attempted object is not supported. An example of this test includes passing an incorrect argument to a function (such as a transformation function) expecting a matrix array but receiving unsupported data of type dictionary or set, resulting in a *Type Error*. Conversely, passing arguments with the wrong value (e.g., a one-dimensional array when a two-dimensional array is expected) results in a *Value Error*.

(c) **File Operation Error:** This category of testing strategy assesses issues related to file operations, such as reading or creating files. Some of the issues tested during file operations include invalid filename, invalid directory name specification, permission rights for creating files, disk errors, end-of-file errors, or file schema errors (i.e., problems with the file structure, order, or content, such as an invalid character preventing the file from being read).

(d) **Lookup Error:** This category concerns faults arising from an invalid or incorrect index or key specification in a given sequence or dictionary.

(3) **Null Reference**: This error targets operations on a specific object considered 'null' or a method call on the null object (i.e., a pointer or reference attempting to access an invalid object). Usually, 'null' is used (in most high-level languages, e.g., Java, C/C++) to indicate a problem in the program that ML engineers should be aware of. In Python, the singleton 'None' represents the NULL object. Examples of assertion APIs for testing the None reference in Python are `assertIsNone(indices)` (unittest), `assert statement is None` (pytest), or checking that a given object is not a NULL reference: `assertIsNotNone(x)`.

(4) **Module Import Error**: This category concerns testing for faults raised when the code under test cannot successfully import the specified module, typically due to a problem such as an invalid or incorrect path.

• **Assertion-based Testing:** This category of testing strategy specifies the test conditions using the assertion statements to evaluate a given point in the program under test against the desired actual program behavior. In the following, we detail the identified sub-categories:

(1) **Oracle Approximation (OA):** Here, the output is allowed to accept a value within a specified range, unlike the equality checks commonly used in traditional software testing. As shown in Figure 4, 17.11% of the test cases, and each of the studied ML software systems uses at least one OA test case. The test cases belonging to this category were observed more in Model training related activity (27.09%), followed by Model evaluation (19.54%), Data Cleaning (18.09%), and Feature engineering (13.84%), as shown in Table 2. We further grouped the OA into three main categories adopted from the related work [98]. (a) **Absolute and Relative Tolerance**: this test expresses the range of accepted oracles by using absolute and/or relative thresholds. Whether the assertions using these APIs pass or not depends on the result of $assert\ abs(res - oracle) < aVal + rVal * abs(oracle)$, where the variable $res$ is the result from the code under test, $aVal$ is absolute tolerance, and $rVal$ is relative tolerance. Some assertions API approximate the code under test by considering only the absolute range/ tolerance instead of using the combination of absolute and relative tolerance, represented as $assert\ abs(res - oracle) < aVal$. We, therefore, use the category name *Absolute and Relative Tolerance* to refer to both types of Oracle Approximation throughout this paper. (b) **Rounding Tolerance**: This oracle approximation uses a significant decimal digit to round between the resulting value of code under test and its closeness with the expected described as: $assert\ abs(oracle - res) < 1.5 * 10 * * (-dp)$. The variables $res$ is value of the code under test, $oracle$ is the expected oracle, and $dp$ variable is used to specify the number of significant digits in $res$. (c) **Error Bounding**: In this category of Oracle Approximation, the first step is to calculate the difference between the two values (i.e., the code under test and expected value) instead of directly comparing variables $res$ and $oracle$, then asserts whether the difference (i.e., error) is smaller than a threshold as follows: $assert\ error < threshold$.
Table 7b in the Appendix highlights commonly used assertion APIs for expressing Oracle Approximations extracted from the studied ML software system.
Examples of such test cases include using the Absolute and Relative Tolerance assertion API to compare the closest in model prediction precision when performing cross-validation with different sizes of training and test sets. The model performance is expected to remain consistent, while slight variability at different test runs may be allowed (e.g., due to randomness, floating-number representation, etc.).

(2) **Instance and Type Checks:** This test strategy verifies if the object argument under test is an instance or a subclass, or if the type of the argument matches with the specified class argument. An example of this category includes verifying if a transformation function returned data of a valid type. The tests were observed in all activities of ML workflow in the following order: Feature engineering (24.53%), Model training (18.13%), and Data Labeling (13.96%).

● **Statistical Analysis:** This category of testing strategy employs statistical methods in data analysis to uncover trends, patterns, relationships, and other meaningful insights. Additionally, it provides mechanisms to determine whether an assertion (e.g., a hypothesis) about a quantitative feature of a population should be accepted or rejected [75]. This category includes statistical methods such as the mean, median, sample size analysis, statistical hypothesis testing, and inter-rater reliability (consistency), among others.

● **Heuristic-based Testing:** This category represents testing strategies that utilize rules, estimation, or educated guessing to solve problems, discover information, and learn about specific (often unknown or new) issues. We discuss three subcategories of testing strategies under this category: Backtesting, Swarming Testing, and Thresholding Testing.

(1) **Backtesting :** This testing strategy assesses the performance and effectiveness of a trading strategy or predictive model on historical data [68]. Backtesting involves simulating a trading strategy by sampling data from relevant periods that reflect various market conditions [24] (e.g., bear markets characterized by prolonged price declines or average-gain/loss statistics). The premise is that a trading strategy that performed well will likely perform well in the future. In contrast, historically, a poorly performing strategy will likely continue to perform poorly. Our analysis indicates this is the least used testing strategy (0.24%, as shown in Figure 4). Backtesting is primarily used during data collection and model monitoring.

(2) **Swarming Testing:** This strategy involves the random generation of test cases [62], allowing for the creation of a more diverse and focused set of inputs. Unlike tests conducted under an "optimal" configuration, swarming tests only the necessary features, omitting others. This approach is beneficial when certain program sections are suspected to be more prone to faults. We observed swarming tests exclusively in the `nupic` (application platform domain) and `paperless-ng` (CV System), with the highest percentage (51.16%) of swarming tests occurring during the model training phase of the ML workflow, as shown in Table 2.
In the `Nupic` system[6], the swarming algorithm is employed to determine the best model for a given dataset automatically. The search space includes identifying the model's optimal components (e.g., encoders, spatial pooler, temporal memory, classifier) and their corresponding parameter values. Due to the high computational complexity of running the swarming algorithm, `Nupic` employs concurrency, using multiple workers. Swarming attempts various models on the dataset and ultimately selects the model that produced the lowest error score. The test cases include running a worker on a model for a while, then having it exit before the model finishes, and starting another worker, which should detect the orphaned model. Additionally, they test the worker's correct behavior when the model's connection differs from what is stored in the database.

(3) **Thresholding:** This testing strategy is used to find the best or optimal threshold value of a decision system (or classification). This strategy accounts for, on average, only 0.48% of all the test cases we analyzed (as shown in Figure 4). The testing strategy was observed during the three major phases of the ML workflow: Data labeling, Model training, and Model evaluation, as well as in test cases for utility functions. Examples of such test cases, highlighted in Table 3, include testing the classification decision point for binary and multiclass classification problems, testing the confidence score threshold to filter out false positives, and determining the minimum score when predicting the bounding box corresponding to classes in object detection—observed in the `Apollo` autonomous system.

---

[6]https://github.com/numenta/nupic/tree/master/tests/swarming

**Black-Box Testing:** This testing strategy focuses on ML components that can only be evaluated using input-output data pairs. These components are not visible to developers or users and can only be assessed by observing their behavior in response to variations in input and output data.

• ***Functional (Black-Box) Testing:*** This type of black-box testing examines the ML system's functionalities against defined requirements. The test involves providing inputs and verifying the outputs against expected outcomes.

(1) ***State Transition:*** This testing strategy analyzes the behavior of a software system by changing input conditions that cause state or output changes in the code under test [74]. The program states refer to the different conditions of the program based on the inputs. In state transition testing, various input conditions (both valid and invalid) are passed in sequence, and the system's behavior is observed. This is the most commonly used black-box functional testing strategy, accounting for an average of 7.96% of the test cases in the studied ML systems, as shown in Figure 4. Additionally, this strategy is observed in all activities of the ML workflow, according to Table 2. Examples of such tests in the studied ML systems include cases where an upload API is called with valid versus invalid data (e.g., in the `paperless-ng` ML software system). The tests verify whether the loaded ML model is already trained and whether there is a change in model parameter values from default to non-default (e.g., losses, embeddings) during model training or retraining. This testing strategy allows ML engineers to understand the behavior of the ML system under varying input conditions, making it effortless to cover multiple conditions and design tests effectively and efficiently.

(2) ***Boundary Value Analysis:*** This testing strategy is designed to test for the values at the boundaries, i.e., the values near the extreme ends or limit where the system's behavior usually changes. For instance, at the lower-upper, minimum-maximum, start-end, and Just inside-just outside values, usually between partitions of inputs. Specifically, this testing strategy was observed more in `deepchem` and `apollo` ML software systems (see Figure 5a), covering only 0.6% of the studied test cases. In `apollo`, i.e., CV (Computer Vision) system domain, the Boundary Value Analysis test cases are used when checking the input point cloud coordinate in perception (from the LinDar sensor) is contained within some boundary. The `deepchem` application platform uses deep learning for Drug Discovery, Quantum Chemistry, Materials Science, and Biology, and the test cases related to Boundary Value Analysis are used to test for the data points within the neighborhood of the collection of atomic coordinates.

• ***Non-Functional (Black-Box):*** This category of testing is used to verify the non-functional requirements of ML systems, focusing on how the system should perform tasks [118, 136]. Two sub-categories of non-functional testing strategies are identified: memory errors and time behavior.

(1) ***Memory Error:*** This testing strategy targets memory usage-related errors that can affect system stability and correctness. These are memory access errors (occur when a read/write instruction references an unallocated or deallocated memory), memory leaks (occurs when an allocated memory is not released), and memory corruptions (occur when a random area of memory is altered unintentionally by a process or program ). Other examples of memory errors in C/C++ are broadly classified as *Heap Memory Errors* and *Stack Memory Errors* [65] include missing allocation (freeing memory which has been freed already), mismatched allocation/deallocation, cross stack access (a thread try to access stack memory of a different thread), uninitialized memory access. On the other hand, Python adopts a memory management architecture similar to the C language (i.e., `malloc()` function). In rare cases,

Python may raise `OutofMemoryError` when the interpreter entirely runs out of memory, allowing the raised error to be caught and enabling recovery from it. The ML engineers, therefore, may catch out-of-memory errors within their script (as `except MemoryError` or `assertRaises(MemoryError)`); however, in some cases, MemoryError will result in an unrecoverable crash. Moreover, Python provides a cross-platform library called Psutil [7] (process and system utilities) that allows for retrieving the information on running processes and system utilization (memory, disks, CPU, network). Solutions that depend on the Psutil module, such as a memory profiler (the example code is shown in Table 3), have been proposed to help monitor the memory usage of a process and perform a line-by-line analysis of memory consumption for Python programs.

(2) ***Time Behaviour/ Execution time:*** This testing strategy is a measure of how long the ML system takes to execute the given tasks. Examples of such test cases include the testing for the amount of time the ML software system takes to make inferences (inference time), load time, and training/learning machine learning algorithms given different input data sizes, etc. A longer inference/response time of the software system can lead to poor system quality, impact user satisfaction, and may lead to discontinuing using the system [102]. In Table 2, we show that the majority of the test cases related to Time Behaviour are executed in the Utility functions (23.78%) (e.g., testing the performance of the code used across multiple phases of ML Workflow), Infrastructure (16.89%) and during Feature engineering (18.14%). The test cases are highlighted in Table 3. We observed that time monitoring functionality across the ML workflow is commonly done using the utility function (i.e., functions/ classes to perform some common functionally and are often reusable across the activities of the ML workflow [89]). Subsequently, the high percentage (23.78%) of the test cases related to *Time behavior* is done on the utility function as shown in Table 3. Other cases include testing the performance when fetching single and multiple data from different data sources.

---

Table 3: The list of common Testing Strategies, the examples of the observed use case scenarios from the studied ML systems, and the test code examples.

| Category | | Test Properties | Example test cases scenarios | Example Source code |
|---|---|---|---|---|
| Grey-Box | Exceptions and Error Condition | Data Quality Assessment (1) | Checking the data quality based on some quality attributes (e.g., correctness, validity, accessibility, consistency). Tests that check that the correct data is sent as JSON format during the API call; Do not allow division by zero, and check the input data to the model is of valid shape and types (float, int, string, array, tensor, etc.), invalid model graphs; Checking the type of document; valid range or out of range; throw value error when trying to shuffle an empty array. | ```\n1  # raises on invalid range\n2  scaler=MinMaxScaler(feature_range=(2,1))\n3  assertRaises(ValueError,scaler.fit,data)\n``` ```\n1  x = np.array([[['unknown']]])\n2  with pytest.raises(TypeError) as info:\n3      x = input_node.transform(x) #unsupported input x\n``` ```\n1  EXPECT_FALSE(JsonUtil::GetString(json_obj,"int",\n      value))// Value is not string\n``` ```\n1  def testGetOutOfBounds:\n2      assertRaises(IndexError, patternMachine.get, *\n      args)\n``` |
| | | Negative Test (2) | Tests to handle the unwanted input and user behavior, usually to prevent the ML system from crashing. Throwing runtime errors and exceptions when unexpected errors occur during model training, timeout error when API takes long, e.g., due to network issue, assertion errors due to some programming error or logical error in the code, Null data referencing | ```\n1  assertTrue(a != b)\n2  EXPECT_FALSE(edges_[3] < edges_[4])\n``` ```\n1  pool.submit(lambda a, v: a.f.remote(v), 0)\n2  with pytest.raises(TimeoutError):\n3      pool.get_next_unordered(timeout=0.1)\n4  def test_predict_proba_3():\n5      #Assert that the model predict function raises a\n      RuntimeError when no optimized pipeline\n      exists.\n``` ```\n1  assertIsNone(fs.getNextRecordDict())\n2\n3  EXPECT_EQ(nullptr, obstacle_ptr103);\n``` |
| | Assertion based Testing | Oracle Approximation (OA) (3) | Accepting non-negligible difference (numerical error) between the actual value and the value that is calculated by the code implementation (e.g., model precision during multiple run of cross-validation, due to factors like randomness). These include Machine representation errors due to limited representation and floating-number computation and approximation errors. | ```\n1  assertAllClose(out1[i - 3], out2[i], rtol=1e-4,\n      atol=1e-4)\n2  approx assert np.allclose(precision.mean(),\n      expected_mean_precision)\n``` ```\n1  EXPECT_LE(box.area(), min_area + 1e-5)\n``` ```\n1  assertAlmostEqual(result['actualValues'][0], 34.7,\n      places=5)\n``` |
| | | Instance & Type Checks (4) | Checking Object or variable is an instance of the specified class type or data type, e.g., after data transformation or converting Numpy array to PyTorch tensor and vice-versa, verifying if the loaded model class if it is of valid, check the validity of an object type return by API. | ```\n1  assertIsInstance(transform, np.ndarray)\n2  self.assertTrue(isinstance(cnn_lstm, torch.nn.\n      Module))\n3  assert scale.scale.dtype == torch.float\n``` |
| | | Swarming Testing (5) | Used in random testing to improve test cases' diversity to enhance test coverage and fault detection. The example test is found in the link [8] | |

---

[8] https://github.com/numenta/nupic/blob/master/tests/swarming/nupic/swarming/swarming_test.py

| | | |  |  |
|---|---|---|---|---|
| 6 | **Heuristic based** | **Back testing** | Testing a trading strategy using historical transactional data to predict its value for future investment decisions. For example, testing cash limit for buying/ selling, the minimum cost for buying/ selling, testing the held stock limit for selling, compute measures like return, net profit or loss, market exposure, and volatility. The steps could be to choose a sample of historical orders within a relevant period of time reflecting different market conditions, run the trading strategy, and verify the correctness of the outcome. | ```python
def test_trading(self):
    # 1) get order information like price, factor,
         price per unit
    # 2) generate orders orders = [...]
    # 3) run the strategy
    strategy_config = {...}
    backtest_config = {"freq": "day",
                       "start_time": "2020-01-01",
                       "end_time": "2020-01-16",
                       "account": 150000, ...}
    report_dict, indicator_dict = backtest(strategy=
        strategy_config, **backtest_config)
    print(report_dict, indicator_dict)
``` |
| 7 | | **Thresholding** | Classification threshold, including binary prediction, multiclass prediction, proper thresholding in one-hot-encoded data. threshold estimation for anomaly detection (i.e., anomaly threshold); Confidence score threshold to filter out false positives in object detection. | ```python
def test_threshold_predictions_binary():
    """Test thresholding of binary predictions."""
    # Get a random prediction matrix
    y = np.random.rand(10, 2)
    y = y / np.sum(y, axis=1)[:, np.newaxis]
    y_thresh = threshold_predictions(y, 0.5)
    assert y_thresh.shape == (10,)
    assert (y_thresh == np.argmax(y, axis=1)).all()
``` |
| 8 | | **Statistical Testing** | Making quantitative decisions about a population or population sample. Hypothesis tests such as computing the population mean, median, standard deviation comparing the mean of two population, among others. Reliability measure between two samples, e.g., by using reliability coefficients measurement like Kappa statistic [91], Peason r [38]. | ```python
def test_pearsonr():
    """Test the Pearson correlation coefficient is
       correct."""
    metric = dc.metrics.Metric(dc.metrics.pearsonr)
    r = metric.compute_metric(
        np.array([1.0, 2.0, 3.0]), np.array([2.0,
        3.0, 4.0]))
    assert_almost_equal(1.0, r)
```<br>```python
self.assertEqual(sd1.mean(), 2.5)
counter = statistics.Counter('name_123')
self.assertEqual(counter.count, 0)
counter.increment(10)
self.assertEqual(counter.count, 10)
``` |
| 9 | **Functional (Black-Box)** | **State Transition** | Testing the ML system with different input conditions, e.g., valid and invalid, and expecting different system statuses. Examples include API calls with valid vs. invalid data, comparing the status of trained ML model and untrained ML model. Other cases include resetting the model status by discarding the old results before making a new prediction so that the results are not inherited. | ```python
EXPECT_FALSE(environment_features.has_ego_lane())
environment_features.SetEgoLane("1-1", 1.0)
EXPECT_TRUE(environment_features.has_ego_lane())
``` |
| 10 | | **Boundary Value Analysis** | Tests the values near the extreme ends or limit where the system's behavior usually changes. For instance, at the Lower- Upper, Minimum-maximum, start- end, Just inside-just outside values, usually between partition of inputs. | |
| 11 | **Non-functional (Black-Box)** | **Memory Error** | Testing memory related errors that can occur when using ML system, such as memory access errors or improper management of memory when a read/ write management references unallocated or deallocated memory, memory leak, and memory corruptions. Data type overflow error, space complexity, performing ML task like featurization often generates large arrays which quickly eat up available memory. | ```python
def test_memory:
    device = "CUDA_VISIBLE_DEVICES=0 "
    script = "setsid python -m memory_profiler
             benchmark_test.py "
    for name in LIBSVM_DATA:
        command = device + script + "--
        pipeline_name " + pipeline_name + " --name "
        + name + " --object " + test_object + " >" +
        log_name + " 2>&1 &"
        os.system(command)
``` |

(The row 9-11 left spanning column reads **Black-Box**)





| | | | |
|---|---|---|---|
| 12 | Time behaviour | Inspect the time cost to fetch data from different storage type such as Dataframe, Hash storage structure for different test scenarios, e.g., random single and multiple fetching; Testing the performance of the code, training and testing time; Thread execution time; Converting the raw audio to MIDI events for training and testing the model. | <pre>1 def test_cal_sam_minute(self):<br>2     # test the correctness of the code<br>3     ....<br>4     # test the performance of the code<br>5     args_l = list(gen_args())<br>6     with TimeInspector.logt():<br>7         for args in args_l:<br>8             cal_sam_minute(*args)</pre> |
| 13 | White-Box / Value Range Analysis | Bounds and range testing (i.e., checking if a variable is within some bounds before using it and checking if the value is within a specific range, respectively), e.g., Array bounds analysis, finding data dependence, testing Data partitioning and data sharding, data distributions, restricting the range of values a domain can take | <pre>1 results = [c for c in clock_range]<br>2 assertTrue(all(map(lambda x: expected_min <= x <=<br>        expected_max, results)))<br>3 assertTrue(0 <= b1 and b1 < 1000)</pre> |
| 14 | Membership Testing | Finding out if the collection items (e.g., dictionary, list, tuple, set) contain specific items. Example include the use the 'in' and 'not in' to check if an object is part of the object's collection, in Python. Indexing and data selection, e.g., checking data (meta-data) using a known indicator, checking the subset of features in the dataset, slicing and dicing during data analysis. | <pre>1 def testNeighbors0Radius(self):<br>2     neighbors = self.encoder._neighbors(np.array<br>        ([100, 200, 300]), 0).tolist()<br>3     self.assertIn([100, 200, 300], neighbors)</pre> |
| 15 | Decision & Logical Condition | exercising the logical conditions and expression within the test cases to test all (or most important) branches of the ML component. Testing the correctness of the intermediate results in the data processing pipeline, and intermediate connection in Directed Acyclic Graph, using the logical expression to assert many decision options including positive and negative outcome, e.g., exporting model graph gives either checkpoints, meta graph, bundle or both/ or returns a runtime error if none is set, or checking correct handling of different hyper-parameter using conditional branching. | <pre>1 # Assert that pick_two_individuals_<br>        eligible_for_crossover<br>2 # picks the correct pair of nodes<br>3 # to perform crossover with<br>4 def test_pick_two_individuals_<br>        eligible_for_crossover:<br>5     assert ((str(pick1) == str(ind1) and str(pick2)<br>        == str(ind2)) or str(pick1) == str(ind2) and<br>        str(pick2) == str(ind1))</pre><br><pre>1 assertTrue(all(map(lambda x: isinstance(x, int_type<br>        if is_int else float_type), results)))</pre> |

**White-Box Testing:** This category tests ML components accessible to ML engineer that could be tested using additional test code or manual interaction with these ML components.

• *Value Range Analysis:* Is a testing strategy similar to an interval analysis [67, 95], are static analysis techniques used to infer the set of values that a variable may take at a given point during the execution of a program. This category of testing strategies includes array and loop bounds checking, array-based data dependence analysis, pointer analysis, feasible path analysis, asserting for values between a given minimum and maximum, for example: `self.assertTrue(all(map(lambda x: expected_min <= x <= expected_max, results)))`, and loop timing analysis, among others, presented in the appendix' Table 6. This category represents (9.24%) of test cases in this study and represents the most used testing strategy in the White-Box testing category based on our results shown in Figure 4. Furthermore, Table 2 shows that the *Value Range analysis* testing strategy is used in all the activities of ML workflow, with the highest usage happens during feature engineering activity (27.91%). Examples of such tests include: Checking the distribution of training data given some range of values; counting the sample size of the historical data given two-time slots; checking the size of data shards which partition the large databases into smaller ones for faster execution and easy management.





• ***Membership Testing:*** This testing strategy checks if a specific value is contained within a collection of items or sequences (i.e., a list, a set, a dictionary, or a tuple). Usually, the time complexity when checking for a specific value depends on the type of the target collection (or data type) [73, 127]. This implies the choice of appropriate target collection is crucial for implementing algorithms that handle operations in large volumes of data. According to Figure 4, Membership testing represents 8.81% of all the test cases we analyzed. Specifically, we observed tests corresponding to *Membership testing* happens in the Model training (26.8%) related source-code, followed by the Feature Engineering (13.18%), Data Cleaning (12.17%), and Model Evaluation (8.36%) activities of ML workflow. In Table 3 and in Appendix, Table 7, we highlighted the examples of the *Membership Testing*, including the assertion API in Python, for example, which uses the build-in Python-based membership testing operator called `'in'` [127].

• ***Decision and Logical Condition***: This structural testing strategy employs conditional expressions to cover the different branches or subroutines of the decision system. As shown in Figure 4, 5.4% of the test cases analyzed belong to this category, and the test is predominantly used during Feature engineering (22.23%) and model training (20.45%) of ML workflow.

### 4.1.2 Mapping of the testing strategies to the ML workflow.

Here, we examine the uses of the testing strategies across the ML workflow. We mapped the identified testing strategies across the ML workflow, as follows: For each testing strategy, we identified and counted the number of unique test cases corresponding to each ML workflow activity for every ML software system in which the test exists at least once. Next, we computed the percentage of tests for each testing strategy in a single ML software system.

The results of our analysis are shown in Table 2, representing the proportion of testing strategies for each ML workflow activity. For example, the test *Statistical Testing* dominates (i.e., highest percentage proportion) within the ML model training activity with 27.23% compared to its percentage proportion in other ML workflow activities, shown in Table 2. Also, in Table 3 provides more details about the test case example scenarios for each of the major testing strategies and the corresponding test code example. In the Appendix Table 7, and Table 6, we also provide the assertion API example code expressing some of the selected testing strategies identified.

> We derived a taxonomy of testing strategies (from three major categories of Grey-box, White-box, and Black-box testing) that ML engineers use to identify software bugs. On average, Grey-box testing is the most dominating (*Assertion-based* (24.85%), *Negative testing* (24.52%), *Statistical Analysis* (10.01%)). Among these testing strategies, *Oracle Approximation* (i.e., Assertion-based category) has been empirically studied in the previous work [98]. We encourage further research to evaluate the effectiveness of these identified testing strategies.

Overall, we observe a non-negligible proportion of testing strategies along the ML workflow. The majority of testing is happening during ML training (20.43%), followed by the Feature engineering (15.51%), Model Evaluation (11.48%). In Table 2, we highlighted in **bold-face** the testing strategies that demonstrate the highest percentage of test cases in the ML workflow activity. For example, *Thresholding* test cases are found more in Model training (29.87%) related activity, *Time behavior* observed more in test cases related to utility code (23.78%).





ML engineers employ different testing strategies throughout the development and deployment of ML models. A high proportion of testing activities happens during model training (20.43%) followed by feature engineering (15.51%). Grey-box testing (e.g., *Data Quality Assessment*, *Negative test*, Statistical analysis) and White-box (e.g., *Value Range Analysis*, *Membership testing*, *Decision & Logical condition*), and Non-function testing (*Memory error*, and *Time behavior*) are used across at least 50% of ML workflow activities. ML engineers can enhance their testing strategies, leading to more robust and reliable ML models.

### 4.1.3 Comparing the Use of Test Strategies Across Systems

In alignment with the investigation into our **RQ1**, this section explores the percentage composition of testing strategies employed in various ML software systems. Consequently, analyzing these testing strategies enables us to gain insights into developers' decision-making processes. Also, by examining the prevalence of different testing strategies, we aim to understand the factors influencing developers' choices and practices in implementing tests. This exploration can help identify trends, challenges, and best practices in the testing landscape of ML software development.

Figure 5a presents a visual comparison of test cases distributions for the identified testing strategies, across the studied ML software systems. Our analysis shows a non-uniform proportional use of testing strategies, i.e., there is a general high deviation between the most dominating test strategies and the least dominating test strategies, within the studied ML software systems. Testing strategies such as *Oracle Approximation (i.e., Absolute Relative Tolerance, and Error bounding)*, *Statistical testing*, *State Transition*, and *Value Range Analysis*, and *Negative testing* are among the most commonly used testing strategies; these testing strategies are consistently used in at least 80% of the studied ML software systems. For example, the proportion of the *State Transition* tests ranges from 2% to 27% of all the test cases across the studied ML software system. The proportion of *Value Range Analysis* tests ranges from 5% to 21% across the studied ML software systems. For instance, in `lightfm`, a recommendation system, we observed the use of State Transition to tests for consistency during model training/ retraining (or to help with scheduling training) by checking the learned recommender model parameter (embeddings) values after resetting (discarding the old result) and retraining the model under different epochs. In `paperless-ng`, the status of a file in a given directory is checked before and after applying an archive function (except to return true and false, respectively), indicating the file is no longer of type file and hence was successfully archived. Other cases in `paperless-ng`, a CV system, include checking if the given document classifier is already trained or not and uploading valid vs. invalid documents, among others. For the `nanodet` system, the State Transition testing was used to check the change in the value of the loss function to change from zero (initial value) to none-zero when the model weight is learned.

Regarding the *Value Range Analysis*, we observed a diverse range of test cases scenarios, including data operations such as testing of database sharding, training, and test set data split functionality, and usage in computing the sample size of the data given two date differences (e.g., in `deepchem` and `lightfm` ML systems). Moreover, this testing strategy was observed as part of test cases for the loss function when computing weight distribution, e.g., testing the Generalized Focal Loss [83], implemented in `nanodet` ML system. In the `nupic`, an application platform category, *Value Range analysis* tests the anomaly detection algorithm by generating an estimate using fake distribution (with fake data) of anomaly scores and verifying the results within some range, among others. In `DeepSpeech`, an NLP system, various augmentation techniques are implemented on raw audio data that can be used while training the DeepSpeech model. For example, the corresponding Value Range Analysis test cases include checking a range of uniformly random values between two float





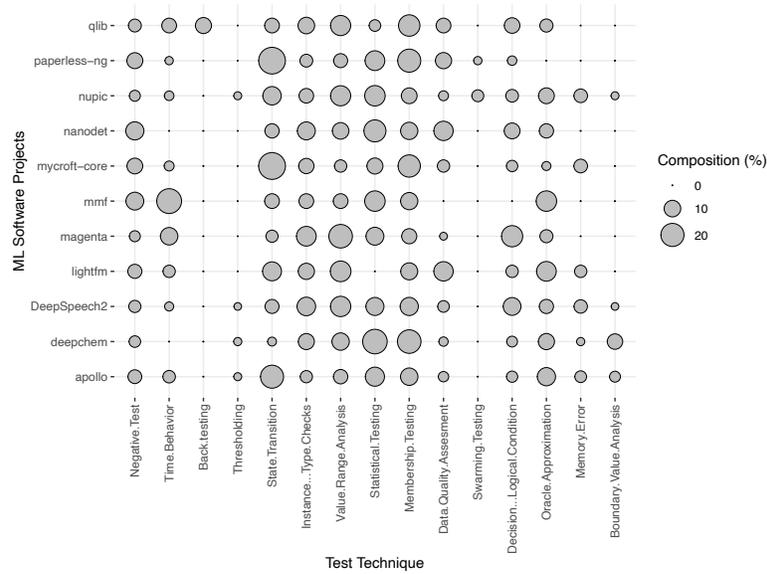

(a) The composition of the testing strategies across the studied ML software systems.

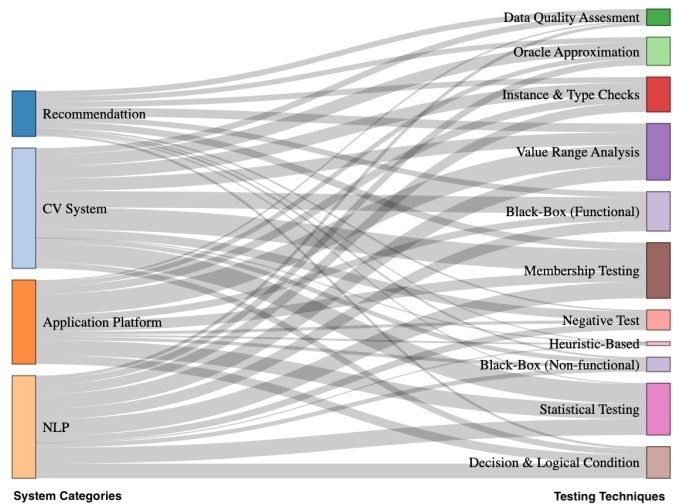

(b) Sankey diagram summarising the testing strategies across ML system category

Fig. 5: Comparing the testing strategies across the ML Software System and their categories

values on each sample augmentation or two sets of values specifying the argumentation sample at the beginning of the training to the argumentation at the end of the training, respectively.

In addition, *Oracle Approximation* (specifically *Absolute and Relative tolerance*, and *Rounding Tolerance*) is adopted by 90% of the studied ML software system with the test cases ranging from 0% to 16% of the total tests across the studied ML software systems.





*Decision and Logical Condition* testing is one of the testing strategy used in at least 90% of the studied ML software systems, and `magenta` application, implements the highest percentage of test cases. In `magenta`, there are multiple ML models which require different types of inputs. Some models train on melodies, and some on raw audio or Musical Instrument Digital Interface (MIDI) data. A data processing pipeline that transforms input data types to output data types is implemented to help convert easily between these different data types. The multiple pipelines are then connected using Directed Acyclic Graph (DAG) to generate a new pipeline used to build a new model. Multiple test cases are designed using Decision and Logical Condition strategy to test that the different intermediate output (usually with different length) of the data transformation and DAG generations, are handled correctly. Testing compatibility issues, we observed a case in `DeepSpeech`, an NLP system where the *Decision and Logical condition* are used to check the compatibility of the deployment platform, a simple logical condition was used to verify the current platform (such as the operating system application programming interface) where the system is running on; before creating or running process/ multitasking operating systems.

On the other hand, *Heuristic based* testing strategies (i.e., *Back Testing, Thresholding, Swarming test*), *Boundary Value Analysis*, and non-function testing (*Memory error* and *Time behavior*) are not observed in some of the studied ML software systems, suggesting that test strategies are currently being used inconsistently in the field where they applied. Indeed, this is not surprising given the nature and the goal of these testing strategies. For instance, *Back testing* can be more important specifically in the trading systems and subsequently was observed only in one ML system (i.e., `qlib`) to help in optimizing the trading strategies. `qlib` is a business platform used in quantitative investment management covering the entire chain of quantitative investment strategies, including risk modeling, portfolio optimization, and order execution. Similarly, *Swarming test* strategy was observed only in two ML software systems (i.e., `nupic` and `paperless-ng`).

Testing for the *Memory Error and Memory complexity* was observed in about 50% of the studied ML software systems, including `apollo` —CV system (5%), `DeepSpeech` —NLP (6%), `deepchem` —Medical application platform (2%), `lightfm` —Recommendation system (5%), `mycroft-core` (7%) and `nupic` (7%). For instance, in `mycroft-core` —NLP, the tests case for checking amount of memory allocated during the data operations, model deployment and monitoring. Ideally, the expected memory consumed should be low needs to run in terms of the input size.

Figure 6 shows the summary comparison of the test strategy across the ML systems domains, presented in the descending order of test case composition. The horizontal green line in Figure 6 illustrates the percentage of test cases contributing to at least 80% of the test cases across the studied ML software domains. From this summary, we can see that there is a non-uniformly on which test strategies are highly dominate (high percentage) or least dominate across the different studied ML system's domain. Overall, 80% of the identified test strategies are used across the studied ML system's domain. Moreover, in each of the ML system domain, only four (accounting for the 26%) of the unique testing strategies contributes to more than half (> 50%) of the tests. The testing strategy with a high percentage of usage across all the studied ML system domains are: *Membership Testing*, and *Value Range Analysis*. On the other hand, some testing strategies (e.g., *Heuristic-based* testing) are consistently only used in a small percentage across the ML studied domain. Moreover, there are more test cases related to *Statistical Testing* in CV systems (16% of all test cases), Application platforms, and NLP, yet is least used in Recommendation Systems. Also, the usage of *Data Quality Assessments* varies across studied ML system domain, —it is least used in the domain of NLP and Application Platform and yet one of the most used testing strategy in the Recommendation System domain. Overall, there is an average assessment of *Data Quality* in the CV system. This result suggests that ML engineers pay more attention in assessing the quality of data when building recommendation and CV systems when compared to the studied NLP systems ('mycroft-core',





'mmf', 'DeepSpeech') or ML Application Platform like 'magenta', 'deepchem', 'nupic'; instead NLP and Application Platform focuses on the *Value Range Analysis*, *Statistical*, and *Membership* testing.

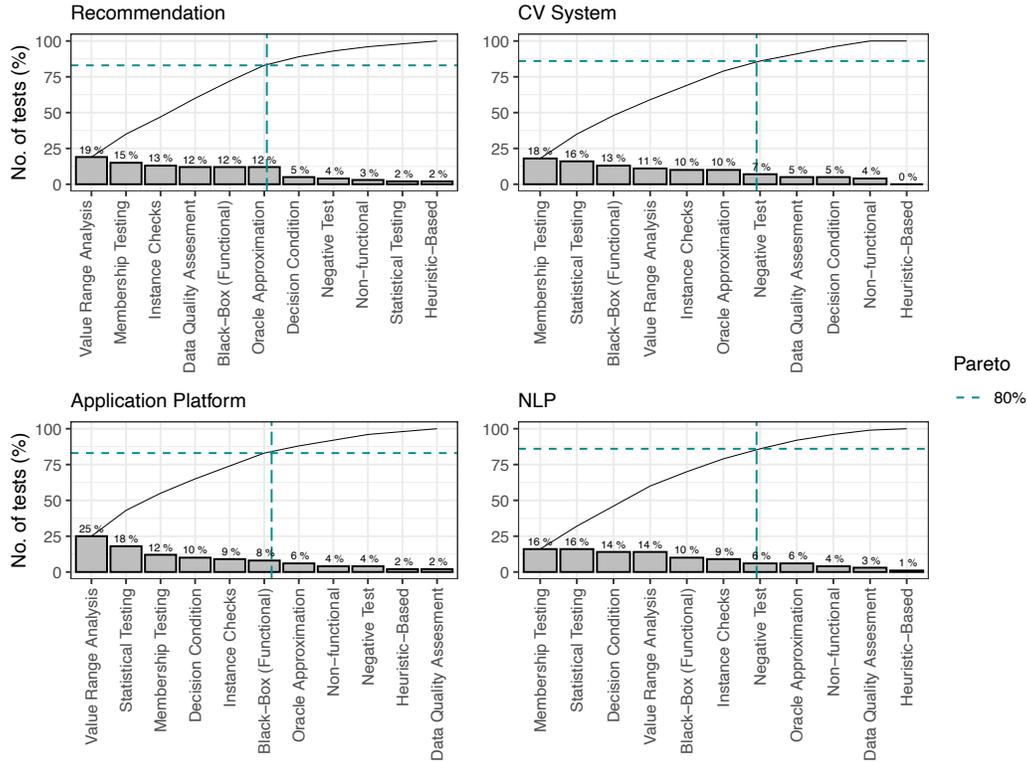

Fig. 6: The Test strategies used in each ML system domain, and Pareto analysis using the cumulative sum (%)

On the one hand, there is a non-uniform use testing strategy across the studied ML software systems—out of which more than half (> 50%) of the tests are implemented using only four (26%) unique testing strategies. On the other hand, at least 80% of the testing strategies are implemented by all the studied ML system domains.

Comparing usage across ML domains; there are more *Statistical Testing* in Application platforms (18%), CV (16%), and NLP systems (16%), yet is least used in Recommendation Systems. Moreover, *Data Quality Assessments* varies across studied ML system's domain;—it is least used in the domain of NLP and Application Platform, yet one of the most used testing strategy in the Recommendation System. Overall, there is an average assessment of *Data Quality* in the CV system:— implying that more attention is paid on assessing the data quality, when building Recommendation and CV systems. Similar trends observed in the used of Oracle Approximation with averagely used more in Recommendation and CV systems— suggesting these systems are more prone to computation variations in results across different test runs. Heuristic based testing strategies are the least used across the studied ML software systems (< 40%), suggesting that they have a more unique test objectives.





## 4.2 RQ2: What are the specific ML properties that are tested in an ML workflow?

This Section details the results of ML properties (derived in Step ⑥ of our methodology) broken down into: (1) identification of ML properties, and their respective mapping to the ML workflow activities, and (2) Comparison of the ML properties across the studied ML software system's domain.

### 4.2.1 The commonly tested ML properties in the ML workflow.

Table 4 presents the ML properties identified during our data analysis, classified as both high-level and low-level functional and non-functional requirements. Table 4 also shows the mean composition of properties in the ML workflow. We also provide examples of test scenarios to highlight how the different ML properties were tested. Note that, the ML properties presented are not strictly independent of each other, when given the features that facilitate their measurement. Yet, their violations result in different manifestations of the behaviors of an ML software system.

● **Consistency and Correctness**: This property measure the degree of agreement within or across the ML components. We highlighted in Table 4 the areas where consistency was observed in the ML test cases, including the model prediction, read and write operations, and data migration. Consistency during the model prediction is the ability of the successive model to make consistent output (i.e., either correct or incorrect, however, more desirably in producing consistent correct results) for the same inputs. In Listing 26 we illustrate three test case scenarios related to testing for consistency during data preparation and model training. They are extracted from the studied ML software systems (`lightfm` and `Nupic`). The first scenario (i.e., function `test_consistency_1()`) checks for consistency during prediction, where similar code was executed two times without any modification, and the results are expected to remain consistent. The second scenario (i.e., `test_model_api()`), tests the consistency in setting and getting the model parameters api. In the third test scenario (i.e., `test_pca_default()`), the transformation function was executed ten (10) times, and the two results are compared against each other for consistency. Finally, in the fourth test scenario, read/ write tests the code for encoding data using adaptive scaler encoder in `nupic`. A proto encoder was written into a temporary file and then read back into a new proto. The resulting proto was then used to encode the new data, which was then compared with the default encoder to verify consistency.

An example of functional correctness and completeness test (extracted from `apollo` software system) is shown in Listing 27 where the return results of the polynomial function were checked against its specification of the form $f(x) = 1 + 2 * x^2 + 3 * x^3 + 5 * x^5$ (i.e., all the four different input values $x = \{0, 2, 3, 5\}$ to the function should output the correct values $\{1.0, 2.0, 3.0, 5.0\}$, or inputting invalid values $x = \{1, 4\}$ should return correct default value 0.0). Also, reversing the input order should return the correct results values of $x$.

● **Compatibility and Portability**: This ML property is among the least tested properties in the studied ML software system, taking on average 1.6% of the studied test cases. As shown in Table 4, we observed the tests related to *Compatibility and Portability* mainly in verifying the infrastructures used across the ML workflow (41.94%, observed mainly when deploying `DeepSpeech` NLP system), followed by Model training (8.01%), utility (6.42), model evaluation (5.26%), Model deployment (3.53%), monitoring (3.21%) activities, and in the configuration and utility related test files. The example of a compatibility test includes checking for identical functionality of the same algorithm implemented in different programming languages (e.g., python and C++) by running similar tests side by side, iteratively with random input data and monitoring the consistency of the outcome (in `Nupic` application platform). Similarly, we observed Backward compatibility [11] tests performed by checking that users can effectively use either the legacy code implementation or the related updated version shown in Listing 28. Testing for the model's portability was observed during





Table 4: **The list of common ML Tests Properties and their percentage composition across the ML Workflow.** *Data*: Data Collection, *Clean*: Data Cleaning, *Label*: Data Labelling, *Feat*: Feature Engineering, *Train*: Model training activities including model fit, prediction, hyper-parameter tuning, *Eval*: Model Evaluation and Post Processing, *Deploy*: Model Deployment activities including Model inspection, model update, pickling and pipeline export, *Moni*: Monitoring including Model Monitoring and Inspection, *Util*: share Utility file *Infras*: test cases on ML infrastructures and frameworks used, *Syst*: System testing, *C&R*: Category and Related works

| | C&R | Test Properties | Data | Clean | Label | Feat | Train | Eval | Deploy | Moni | Util | Infras | Syst | Example of test scenarios and/or related dimension from studied systems separated by commas |
|---|---|---|---|---|---|---|---|---|---|---|---|---|---|---|
| 1 | Correctness [16, 56, 150] | Consistency | 9.3 | 9.1 | 2.1 | 18.8 | 31.4 | 11.5 | 8.2 | 3.4 | 3.6 | 1.8 | 0.8 | The representations of the same information across multiple datasets, multiple prediction, read/write, data migration, and the consistent result across programming languages (compatibility). |
| 2 | | Correctness | 9.3 | 11 | 3.5 | 16.1 | 25.2 | 13.6 | 4.2 | 3.8 | 6.4 | 3.8 | 3 | Functional correctness (input and expected output), Data syntax conformance (i.e., format, range, type), accuracy and precision, consistency and completeness |
| 3 | [51, 109, 150] | **Bias & Fairness | 0 | 0 | 6.5 | 1.7 | 56.5 | 20.5 | 14.8 | 0 | 0 | 0 | 0 | Model evaluation, mask pruning, probability thresholds (remove unlikely predictions likelihood thresholds), classification fairness |
| 4 | [63, 66, 97, 150] | **Explainability & Interpretability | 0 | 0 | 0 | 38.82 | 24.08 | 11.88 | 0 | 0 | 0 | 0 | 0 | Post-hoc methods to understand trained model mechanisms or predictions, visual and/or interactive artifacts to describe a model's behavior, features selection and neural network layers Transformation, Features importance estimation for Neural Network Pruning [94] |
| 5 | [150] | Efficiency | 10.1 | 7.4 | 4.1 | 19.4 | 9.6 | 4.5 | 3.6 | 6.3 | 17.3 | 14.6 | 3.2 | Performance, Training efficiency, data storage, Time behavior, API calls, computing resource usage, Neural network slimming and level pruning |
| 6 | | Scalability | 9.5 | 2.4 | 0 | 36 | 11.8 | 6.2 | 6 | 12.6 | 4.3 | 7.6 | 3.7 | Hyper-parameter search, model training efficiency, batch processing, Deep neural network morphism, parallel computing and distributed machine learning. |
| 7 | [11, 137] | Compatibility & Portability | 0 | 0 | 0 | 53.1 | 10.1 | 6.7 | 4.5 | 4.1 | 8.1 | 8.8 | 4.6 | On-device machine learning inference, Backwards compatibility test, environment variables and different computing environment, identical functionality of python and C++ implementation of same algorithm. |
| 8 | | Concurrency & Parallelism | 4.4 | 11.7 | 4.3 | 12.3 | 7.3 | 4.4 | 3.9 | 34.5 | 4.7 | 8.9 | 3.7 | Producer/consumer queue (pcqueue), concurrent queue, thread pool, parallel training, and parallel prediction |
| 9 | | Uniqueness | 12.1 | 12 | 12.4 | 8.2 | 13.7 | 14.3 | 6.7 | 9.8 | 10.8 | 0 | 0 | Only new records accepted, item measured against itself or its counterpart, scenario model prediction (e.g., estimate the lane change), and sensor fusion |
| 10 | | Timeliness | 15.8 | 12.1 | 7.7 | 19.4 | 7.6 | 15.4 | 3.8 | 5.7 | 4.8 | 3.9 | 3.8 | The time difference between data captured and the event being captured, Object detection, catch speed, extraction of point cloud feature |
| 11 | | Data Relation | 5.3 | 17 | 5 | 13.8 | 23.8 | 17.3 | 3.7 | 4.6 | 7.2 | 2.3 | 0 | Attribute relation File Format (ARFF), Feature meta-learning and feature set selection, performing crossover of two parent nodes to generate new offspring in genetic algorithms, data encoding like scalar encoder and geospatial coordinate (e.g checking closeness) in Nupic system, and data segmentation |
| 12 | | Uncertainty | 0 | 12.7 | 0 | 25.4 | 41.3 | 20.6 | 0 | 0 | 0 | 0 | 0 | Evidence theory or Dempster–Shafer theory (DST), information filter |
| 13 | | **Anomaly | 11.8 | 11.8 | 0 | 29.4 | 35.3 | 11.8 | 0 | 0 | 0 | 0 | 0 | Training anomaly (i.e., unreliable training behaviors), prediction and classification region, spike frequency, distribution estimation and anomaly likelihoods |
| 14 | Security & Robustness [15, 150] | Data Integration & Integrity | 26 | 26 | 8.7 | 6 | 0 | 0 | 0 | 17.3 | 10.7 | 0 | 5.3 | Data aggregation, data and feature fusion, efficient object detection, and spatial dimension reduction |
| 16 | | **Data Poisoning | 6.9 | 8.9 | 3.5 | 16.3 | 45 | 8.5 | 8.3 | 2.5 | 0 | 0 | 0 | Robust to errors introduced to training data, robustness in feature processing, classification algorithm, |
| 17 | | **Adversarial Perturbation | 0 | 0 | 0 | 0 | 40.6 | 51.3 | 0 | 8.1 | 0 | 0 | 0 | Robustness to given conditions valid input, invalid input and random input data, noise robustness in HTM spatial pooler (Nupic system) |
| 16 | | Security & Privacy | 23.9 | 21.7 | 7.8 | 18 | 0 | 0 | 0 | 8 | 9.4 | 0 | 11.3 | Data visitation, data migration/ transmission, secure matrix, data encapsulation |
| | | **AVERAGE** | 8.5 | 9.6 | 3.9 | 19.6 | 22.5 | 12.9 | 4 | 7.1 | 5.1 | 3 | 2.3 | |

** Properties that could collectively enhance the generalization of ML models against robustness of distribution shifts.

model exports; by ensuring that the exported model is consumed from different platforms in the ML-based production system (e.g., freezing a model designed in Python and general framework





```
1    # consistency in prediction
2    def test_consistency_1:
3        for i in range(2):
4            assert expected_score == result
5    # test consistency in loading model params
6    def test_model_api():
7        model = LightFM()
8        params = model.get_params()
9        model2 = LightFM(**params)
10       params2 = model2.get_params()
11       assert params == params2
12   # consistency in feature preprocessing
13   def test_pca_default(self):
14       t = []
15       for i in range(10):
16           transform, original = _test_prep(PCA)
17           t.append(transform)
18       np.assert_allclose(t[-1],t[-2],rtol=1e-4)
19   # consistency in data preprocessing
20   def testReadWrite(self):
21       originalValue = self._1.encode(1)
22       p1 = AdaptiveScalarEncoderProto.new_message()
23       self._1.write(p1)
24       # write proto and read back into new proto
25       with tempfile.TemporaryFile() as f:
26           proto1.write(f)
27           f.seek(0)
28           p2 = AdaptiveScalarEncoderProto.read(f)
29       encoder = AdaptiveScalarEncoder.read(p2)
30       # ensure the encodings match on new value
31       r1 = self._1.encode(7)
32       r2 = encoder.encode(7)
33       self.assertTrue(numpy.array_equal(r1, r2))
```

Listing 26: Tests examples for consistency extracted from *Autokeras* and *Nupic* ML system

```
1    TEST(BaseTest, polynomial_test) {
2        // f(x) = 1 + 2 * x^2 + 3 * x^3 + 5 * x^5
3        Polynomial poly;
4        poly[0] = 1.0;
5        poly[2] = 2.0;
6        poly[3] = 3.0;
7        poly[5] = 5.0;
8        EXPECT_NEAR(poly[0], 1.0, 1e-8);
9        //unknown input 1
10       EXPECT_NEAR(poly[1], 0.0, 1e-8);
11       EXPECT_NEAR(poly[2], 2.0, 1e-8);
12       EXPECT_NEAR(poly[3], 3.0, 1e-8);
13       //unknown input 4
14       EXPECT_NEAR(poly[4], 0.0, 1e-8);
15       EXPECT_NEAR(poly[5], 5.0, 1e-8);
16       EXPECT_NEAR(poly(0.0), 1.0, 1e-6);//reverse
17       EXPECT_NEAR(poly(1.0), 11.0, 1e-6);//reverse }
```

Listing 27: Tests examples for correctness and completeness

into a portable format to be utilized inside mobile apps). `DeepSpeech` an NLP system make use of the KenLM Language Model Toolkit [69] in performing many of its tasks and subsequently tests its compatibility when performing tasks such as selecting macros, model inference, data processing and generation when dealing with big data sets.

```
1    def test_object_id_backward_compatibility(ray_start_shared_local_modes):
2        # We've renamed Python's `ObjectID` to `ObjectRef`, and added a type
3        # alias for backward compatibility.
4        # This test is to make sure legacy code can still use `ObjectID`.
5        # TODO(hchen): once we completely remove Python's `ObjectID`,
6        # this test can be removed as well.
7
8        # Check that these 2 types are the same.
9        assert ray.ObjectID == ray.ObjectRef
10       object_ref = ray.put(1)
11       # Check that users can use either type in `isinstance`
12       assert isinstance(object_ref, ray.ObjectID)
13       assert isinstance(object_ref, ray.ObjectRef)
```

Listing 28: Tests examples for backwards compatibility





● **Bias and Fairness**: A bias can be defined as a systematic error introduced into sampling or testing of data by choosing or promoting one outcome over others. An unfair algorithm is one which decisions are skewed toward a specific sub-group, as defined by sensitive attribute [150]; on the one hand, a model is considered fair if errors are shared similarly across protected attributes or sensitive attributes. The example of the recognised sensitive attributes include sex, race, colour, religion, national origin, citizenship, age, pregnancy [150].

While the mitigation of model or data bias and fairness are done in various ways throughout the machine learning life-cycle such as by addressing the problem of over-fitting/ under-fitting (inductive bias [70]) and feature selection (e.g., removing or imposing some constraints towards the sensitive attributes like sex, race from dataset to achieve fairness). We identified tests about bias or fairness in the studied ML software system during ML model training, and model evaluation activities of ML workflow as shown in Table 4. An example of such tests was observed in the code pruning filters and weights when compressing Convolution Neural Network (CNN) model [86]. Also, testing for the inductive bias [70] was observed more frequent in `deepchem` ML software system. Further details will be provided in Section 4.2.2.

● **Data Uniqueness**: This data quality dimension measures unnecessary duplication in or across the ML software system within a particular field, record, or data set [14]. An example of a Data uniqueness test case is shown in Listing 29. The test ensures that there is no duplicate entry into the object table by first checking if the object already exists before adding the object with a similar identity to the object table.

```
1  def testInvalidObjectTableAdd(self):
2      # Check that Redis returns an error when RAY.OBJECT_TABLE_ADD
3      # adds an object ID that is already present.
4      self.redis.execute_command("RAY.OBJECT_TABLE_ADD", "object_id1", 1, "
         hash1", "manager_id1")
5      response = self.redis.execute_command("RAY.OBJECT_TABLE_LOOKUP", "
         object_id1")
6      self.assertEqual(set(response), {b"manager_id1"})
7      with self.assertRaises(redis.ResponseError):
8          self.redis.execute_command("RAY.OBJECT_TABLE_ADD","object_id1",
            1, "hash2", "manager_id2")
```
Listing 29: Tests examples for data uniqueness

● **Data Timeliness**: This ML property measure the degree to which the information/data is up-to-date and made available within the acceptable timeline, time frame, or duration. Listing 30 illustrate the example of test scenario that checks for the outdated object detected using Rada sensor extracted from `apollo` autonomous system.

● **Scalability**: The scalability of a machine learning system is its capability to handle varying amounts of the dataset and able to perform multiple computations in a cost-effective and time-saving way instantly for different the numbers of users [39]. While testing for scalability property, the performance of the ML system is monitored, such as the response time, network throughput, and resource consumptions (e.g., memory usage, CPU usage, Network usage) under varying application requirements. Testing for scalability was observed more during the Feature engineering phase, followed by model training, model monitoring, and data collection according to our analysis result.

● **Data Relations**: Defines the relationship between a given variable and the predicting variable of a ML model for a given dataset. It also defines the association between two or more variables and entities of a dataset and the predicted ones for those variables.

● **Uncertainty**: Different types of uncertainty can be observed in machine learning due to in-complete information [129]. Factors such as noise, errors, outliers in the input data or incomplete domain coverage, and an imperfect model of the problem can impact the generalization of the ML component. Therefore, ML model output should be accompanied by a measure that allows





evaluating the certainty or belief of the decision. Specifically, this is crucial in safety-critical domains such as aerospace field, medical systems, autonomous systems like self-driven cars, etc.

• **Security & Privacy**: The security [150] of an ML model is defined as the system's durability against the illegal entry of any of the components or illegal data manipulation, which degrades ML models' prediction performance or reverses the actual decision. There lies a symmetric relation between model's robustness and security [26]. A less robust ML system may be insecure; for instance, a system with low robustness may easily suffer from adversarial attacks when provided with perturbation examples to predict. On the other hand, the Privacy of the ML system is its ability to preserve private data or information [150]. Thus, an ML model is expected to remain consistent in performance even in the presence of noise in the dataset.

• **Explainability & Interpretability**: Explainability in ML/DL involves making the behavior of a machine learning or deep learning model understandable to humans [63]. This is typically achieved through post-hoc techniques that do not require understanding the model's internal workings. Interpretability is the extent to which an observer can comprehend the reasons behind a decision made by a ML/DL system [66, 97, 150]. It encompasses two aspects: transparency (understanding how the model operates) and post-hoc explanations (other information derived from the model). For instance, explaining the feature importance— measures the weight or score of the input features during feature selection of a predictive model, indicating the relative importance of the individual features when making a prediction. This interns helps to reduces the number of input features while providing more insights about the data set by indicating which features are most relevant or least relevant to the target or the model in general. Indeed, there are various techniques and models for measuring the feature importance, such as using model coefficients or permutation testing [107]. We identify the test related to feature importance, such as during structural pruning of neural network that uses the first-order Taylor expansions [142] to estimate the contribution of a neuron (filter) to the final loss, and iteratively removed those with smaller scores [94].

• **Data Poisoning**: A malicious tactic [124] employed to compromise the integrity and performance of ML models. Adversaries strategically inject tainted data instances into the training dataset with the intent of influencing the learning process, leading the model to adopt erroneous patterns or associations. This attack vector may manifest as targeted assaults, focusing on specific vulnerabilities, or indiscriminate campaigns, which cast a wide net to disrupt the model's integrity. The ramifications of data poisoning can be profound, particularly in domains where the accuracy and reliability of model predictions are paramount, such as in safety-critical or autonomous systems. Detecting and mitigating data poisoning attacks represent formidable challenges in ML systems. Effective defenses must be developed to prevent the manipulation of training data and ensure the trustworthiness and robustness of ML models in practical deployments.

• **Adversarial Perturbation/ Inference attack**: This post-training attack [145] involves intentionally manipulating the input data with imperceptible perturbations, to induce targeted erroneous prediction/misclassifications by ML models. These perturbations are crafted with precision to exploit weaknesses, for example, in the decision boundaries of the model, which can lead to incorrect predictions despite the minimal visual changes introduced to the input data. In the context of autonomous vehicles, adversaries could subtly alter traffic signs or road markings in images captured by onboard cameras. This could cause the vehicle's perception system to misinterpret critical information and potentially lead to hazardous driving decisions. Adversarial attacks pose a significant challenge to the robustness and reliability of machine learning models, especially in safety-critical domains like autonomous vehicles, medical diagnosis, and cybersecurity.





```
1   TEST(ContiRadarIDExpansionSkipOutdatedObjectsTest,
      skip_outdated_objects_test) {
2   ContiRadarIDExpansion id_expansion;
3   ContiRadar raw_obstacles;
4   auto *sensor_header = raw_obstacles.mutable_header();
5   sensor_header->set_timestamp_sec(0.0);
6   sensor_header->set_radar_timestamp(0.0 * 1e9);
7   ContiRadarObs *radar_obs = raw_obstacles.add_contiobs();
8   radar_obs->set_meas_state(static_cast<int>(ContiMeasState::CONTI_NEW));
9   auto *header = radar_obs->mutable_header();
10  header->set_timestamp_sec(0.0);
11  header->set_radar_timestamp(0.0 * 1e9);
12  id_expansion.SkipOutdatedObjects(&raw_obstacles);
13  EXPECT_EQ(raw_obstacles.contiobs_size(), 1);
14
15  sensor_header->set_timestamp_sec(0.7);
16  sensor_header->set_radar_timestamp(0.7 * 1e9);
17  ContiRadarObs *radar_obs2 = raw_obstacles.add_contiobs();
18  radar_obs2->set_obstacle_id(0);
19  radar_obs2->set_meas_state(static_cast<int>(ContiMeasState::CONTI_NEW))
20  auto *header2 = radar_obs->mutable_header();
21  header2->set_timestamp_sec(0.7);
22  header2->set_radar_timestamp(0.7 * 1e9);
23  id_expansion.SkipOutdatedObjects(&raw_obstacles);
24  EXPECT_EQ(raw_obstacles.contiobs_size(), 1);
```

Listing 30: Tests examples for Timeliness

> ML Engineers test at least (17) different ML properties in ML workflow, such as: *Correctness, Consistency, Bias & Fairness, Explanability & Intepretability, Compatibility, Efficiency, Security & Privacy, Data poisoning* and *Adversarial Perturbation*.
>
> The ML properties, i.e., *Uncertainty, Security & Privacy, Concurrency & Parallelism, Adversarial Perturbation*, and *Model Bias & Fairness* are tested in less than half ($\leq 50\%$) of ML workflow activities. On the other hand, the ML properties *Consistency, Correctness, Efficiency*, and *Data poisoning* are tested in the majority ($\geq 80\%$) of ML workflow activities.

### 4.2.2 Comparing the Tested ML properties Across Software Systems

This Subsection dive deeper to analyse how the identified ML properties are being tested across the different studied ML software systems, to help us understand if there are any ML properties which are examined in different ML software systems. Notably, we computed the percentage of tests verifying each of the ML property, in a given ML software.

Figure 7a compares the ML properties being tested across the studied ML software systems. According to Figure 7a, only 20% to 30% of the ML properties such as *Correctness, Consistency, Data Poisoning*, and *Efficiency* are consistently tested across at least 90% of the studied ML software systems. In contrast, the ML properties *Bias and Fairness, Compatibility and Portability, Security and Privacy, Data Timeliness* and *Uncertainty* are not tested consistently in about 80% of the studied ML software systems. Figure 7b compares the tested ML properties across the ML system domains. The results show that *Bias and Fairness* is tested more in Recommendation (6%) and CV System (5%) and least tested in NLP (2%) and Application Platform (2%) categories; *Security & Privacy* is tested in CV system (2%), Application Platform (2%), and NLP (1%); *Compatibility and Portability* is tested only in NLP (3%), Application Platform (2%) and CV system (0.2); *Uncertainty* is tested in CV system (2%) and Application Platform (0.3%).

For instance, the test about *Security & Privacy* (*Data Integrity* inclusive) was only found in five ($< 50\%$) of the studied ML software systems: i.e., two CV systems `paperless-ng` (6.5%), and `apollo` (1.6%); two Application Platform: `nupic` (1.4%), and `deepchem` (0.3%), and an NLP system: `mycroft-core` (0.4%). It's not surprising to see the highest number of test cases related





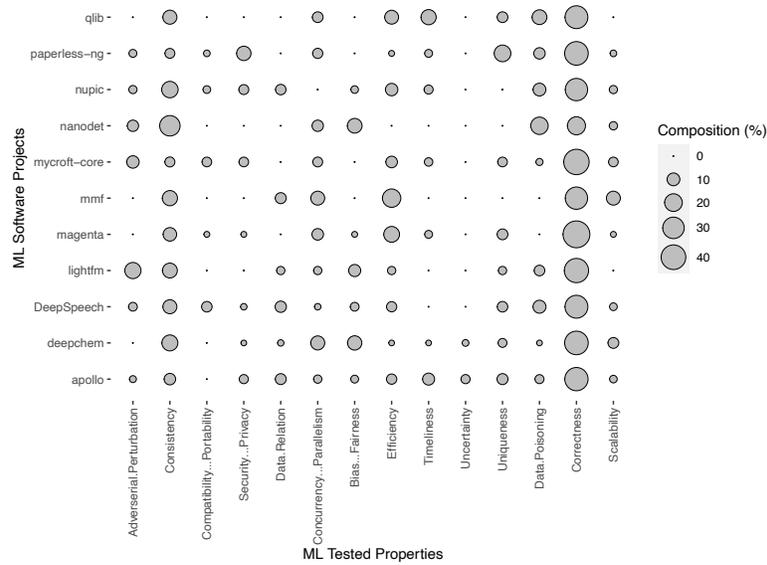

(a) The composition of ML test properties across the studied ML software systems

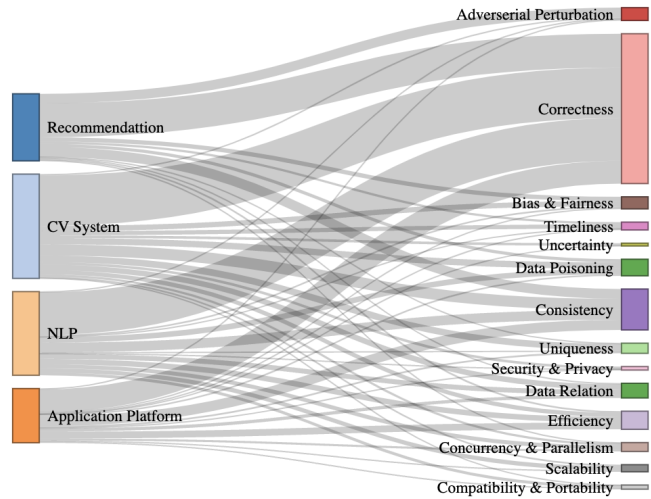

(b) Sankey diagram summarising the tested properties across ML system category

Fig. 7: Comparing the ML test properties across the ML Software System and their categories

to *Security & Privacy* comes from the ML system `paperless-ng`, given the nature of the project where it deals with the indexing, searching and storing of the scan documents meta-data. Testing for these properties is essential in this kind of system since the end users expect stronger data privacy and security safeguards, or other governance and regulations involve in storing the user's data. The example of the test cases in `paperless-ng` include checking for malformed data before storing,





detecting, and handling spam mail/ phishing attacks, data encryption, managing user sessions, and data access control, among others. In the `apollo` system, we observed test cases about *Security & Privacy* that are checking the data access management to protect the privacy and confidentiality of data storage and ensuring that only the data allocated to a given channel is fetched by a user. Other *Security & Privacy* test cases include a protocol test that verifies the correctness of the encryption function and ensures that the data is consistent between the two communication endpoints. Usually, data sets are encrypted to protect against potential interceptions of communications by a third party during a data transfer, which could violate the confidentiality and privacy of the data.

Testing for *Bias and Fairness* was also observed in about six of the studied ML software system, in the order of percentage composition: `deepchem` (6.3%), `nanodet` (5.9%), `lightfm` (3.8%), `DeepSpeech` (1.2%), `nupic` (1.0%), `apollo` (0.5%), and `magenta` (0.3%). As mentioned in Section 4.2, we considered the test cases related to checking for the model overfitting/ underfitting as Bias and Fairness property (also called inductive bias [70]), and the test cases of this kind was observed more frequent in `deepchem` ML software. `DeepSpeech` ML system uses Graph convolutions [52] whereby the convolutional methods are performed on the input graph itself, with structure and features left unchanged, unlike the case of graph embedding methods [60] where the graph is transformed to a lower dimension. This implies graph of the Graph convolutions learning remains closest to its original form in a higher dimension, causing the relational inductive bias much stronger. In `nanodet` (anchor-free object detection model), a novel label assignment strategy is proposed called assign guidance module (AGM) and dynamic soft label assigner (DSLA) [139] to solve the problem of bias label assignment. Similarly, we observed test cases scenarios designed to test for inductive bias when the bias parameter is turned off/on in specific layers of the convolution network; before and after Batch Normalization.

Figure 8 details the comparison of the tested properties across the ML systems domain, the tested ML properties are presented in the descending order of test case composition. We also highlight the top 20% of the tested ML properties with the vertical red dotted line, and the ML properties contributing to at least 80% of the tests (as horizontal green dotted line). From this summary, we can still see that few of the test cases yet dominates across the different domain. Overall, as shown in Figure 8, more than half (50%) of the test cases are used to verify only the top 20% of the ML properties in all the studied ML software domains, while the bottom 20% of the ML properties are verified by a few test cases ($\leq$ 5%). For instance, the properties *Correctness* and *Consistency* dominates across the whole domain. *Efficiency* is more verified in *Application Platform* followed by *NLP* and compared to *Recommendation* and *CV* systems. The least tested ML properties varies in composition of the test cases across the ML system domain.

Overall, more than half (> 50%) of the tests are used to verify the top 20% unique ML properties in all the studied ML software domains. The ML properties such as *Correctness*, *Consistency*, *Data Poisoning*, and *Efficiency* are consistently (i.e., using the same strategy) tested across at least 90% of the studied ML software systems. In contrast, the ML properties *Bias and Fairness, Compatibility and Portability, Security and Privacy, Data Timeliness* and *Uncertainty* are not tested consistently in about 80% of the studied ML software systems. Among the least tested ML properties, *Bias and Fairness* property is more tested in domain of Recommendation system (6%) and CV System (5%); *Security & Privacy* is tested only in CV system (2%), Application Platform (0.9%), and NLP (0.5%); *Compatibility and Portability* is tested in NLP (3.1%), Application Platform (1.5%), and CV system (0.2); *Uncertainty* is tested in CV system (1.9%) and Application Platform (0.3%).





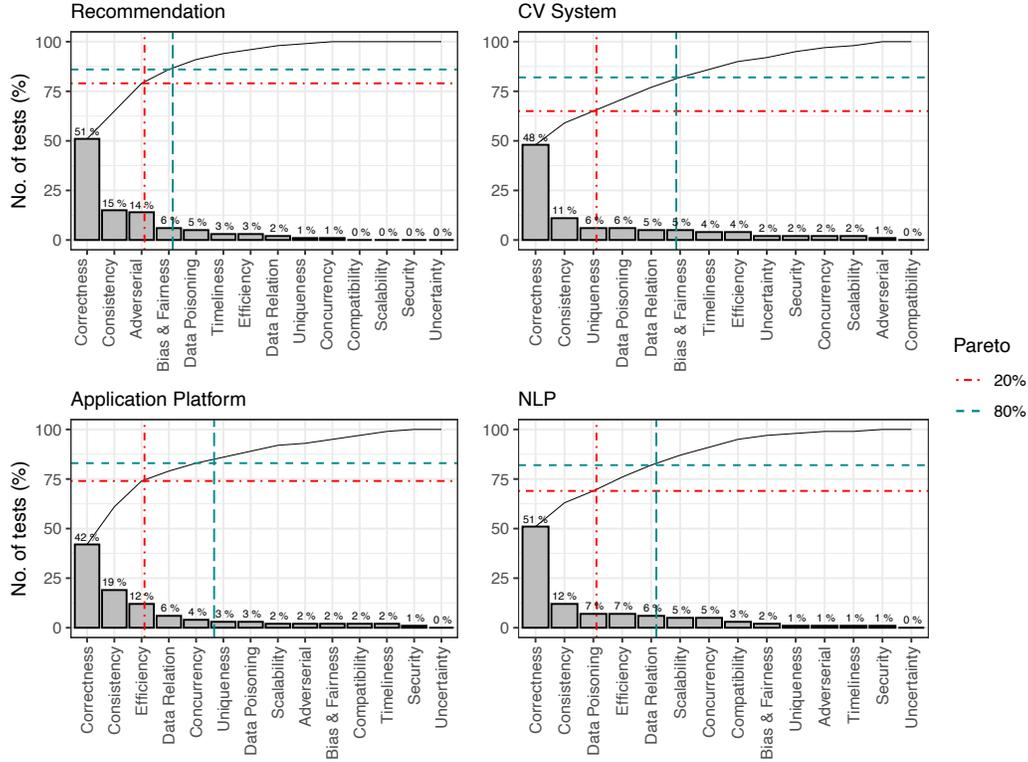

Fig. 8: Comparing ML Properties across the ML system's domains, and Pareto analysis using the cumulative sum (%). The <span style="color:red">dotted red</span> vertical line indicate the top 20% of ML properties, and the <span style="color:teal">dotted green</span> horizontal line show the 80% of the test cases.

### 4.3 RQ3: What are the software types of testing used in an ML workflow?

We have presented the different testing strategies and the ML properties tested in ML workflow. Earlier in Section 2.1.3 we discussed the various types of tests during the delivery process of ML software system, following both the practitioner's point of view in [1] and the research [116]. In this research question, we aim to understand how ML engineers operationalize the tests during the development of their ML software system. Notably, we examine the types of testing (Test types) used in the studied ML software systems. Table 5 shows the list of 13 Test types (types of testing) used by the ML engineers of the studied ML software systems, derived in Step ⑦ of our methodology. The presentation of the tests in Table 5 follows the layers of the Test Pyramid, with *Unit test* at the bottom (i.e., last row) and *Exploratory test* at the top (i.e., first row). We highlighted rows corresponding to tests that we found in the studied ML software systems but which are neither mentioned in the Test Pyramid of ML software system proposed by Sato et al. [1] nor in the literature [116]. The newly observed types of tests are: *Regression testing*, *Sanity testing*, *Periodic Validation and Verification*, *Thread testing*, and *Blob testing*. In the following, we describe each of these newly observed tests in more details.

• **Regression Testing:** This type of testing is used to verify that a previous modification or code change has not adversely affected existing features. The test involves the partial or full selection and





Table 5: **A list of** 11 *Test types* **(Types of testing) and their corresponding test categories, arranged based on the test level from lowest (Cat 13) to highest (Cat 1) during the continuous delivery of the ML software system.** *Data*: Data Collection, *Clean*: Data Cleaning, *Label*: Data Labelling, *Feat*: Feature Engineering, *Train*: Model training related activities including model fit, prediction, hyper-parameter tuning, *Eval*: Model Evaluation and Post Processing, *Deploy*: Model Deployment activities including Model inspection, model update, pickling and pipeline export, *Moni*: Monitoring including Model Monitoring and Inspection, *Config*: Share configurations and Utility file or frameworks used across ML workflow activities, *Cat*: The test category (transparency level), *proj:* Percentage number of systems containing the test types/Types of testing

| Cat* | Types of testing | proj | Data | Clean | Label | Feat | Train | Eval | Deploy | Moni | Config | Infra | System | Example of test scenarios and/or related dimension from studied systems. |
|---|---|---|---|---|---|---|---|---|---|---|---|---|---|---|
| 13 (Opaque) | Exploratory test | 89% | 9.03 | 19.72 | 4.93 | 10.48 | 23.52 | 11.82 | 7.13 | 5.4 | 3.03 | 0 | 4.93 | Explore the application for unexpected behaviors (using the human creativity to hunt possible hidden bugs). The use of print statements instead of assertions API from the automated testing framework to verify the test outcome. |
| 12 | Performance/ Blob Test | 67% | 14.57 | 7.43 | 0 | 14.95 | 9.01 | 8.04 | 6.15 | 18.35 | 1.82 | 8.46 | 11.21 | The system's stability and responsiveness under various workloads, computation overhead during Binary Large Object (BLOB) operations (i.e., reshape, read/write, test header such as Canonical Axis Index or offset or Legacy shape, source pointer CPU/ GPU or mutable CPU/ GPU data), Memory allocation, Time complexity. |
| 11 | Compatibility Test | 47% | 22.55 | 0 | 0 | 37.58 | 8.61 | 5.66 | 3.8 | 3.45 | 6.91 | 7.52 | 3.93 | System's compatibility with the running environment, identical functionality of algorithms in different languages, backward compatibility. |
| 10 (Translucent) | End-to-End Test | 11% | 0 | 0 | 0 | 0 | 100 | 0 | 0 | 0 | 0 | 0 | 0 | Testing for application dependencies ensures that all integrated components can work together. End-to-end testing for the compute function of the HTM spatial pooler algorithm in `Nupic` without any mocking. Testing plumbing: create a random 'dataset', train the network, and then run inference to ensure the correct classification. |
| 9 | Periodic Validation | 22% | 0 | 0 | 0 | 13.53 | 9.52 | 7.42 | 0 | 69.54 | 0 | 0 | 0 | Frequently checking the program requirements, automated model retraining, and continuous learning. |
| 8 | API Test | 33% | 11.11 | 8.1 | 25.93 | 0 | 0 | 2.73 | 0 | 21.61 | 26.88 | 0 | 3.65 | Verifying the results or behaviour produced during the sequence of API calls, observed in the test cases for the studied ML software systems: `Nupic`, `Ray`, `autokeras` |
| 7 | Integration Test | 33% | 5.96 | 10.7 | 5.95 | 17.33 | 13.66 | 10.91 | 5.26 | 12.17 | 3.54 | 13.23 | 1.29 | Testing the combinations of different units or modules to ensures they can work together. |
| 6 | Thread Test | 22% | 2.7 | 2.7 | 0 | 0 | 0 | 0 | 32.13 | 0 | 10.71 | 41.05 | 10.71 | Verifying the key functional capabilities of specific task, concurrent queue Testing, thread pool test, workers threads, performance analysis |
| 5 (Transparent) | Regression Test | 33% | 7.72 | 0 | 3 | 7.72 | 50.93 | 15.67 | 11.98 | 0 | 0 | 3 | 0 | Test that a previous modification or code change has not adversely affected existing feature. For example validating that that the model predictions don't change after modification (i.e., a checkpoint such that changes that affect prediction results will cause the test to fail). |
| 4 | Sanity Test | 0.09 (11%) | 12.68 | 0 | 0 | 18.02 | 30.56 | 12.54 | 6.01 | 16.24 | 3.96 | 0 | 0 | Test the stability of new functionality or code changes in an existing build such as in Finite State Transducer (FST) [119] algorithms implementation in `Deepspeech` system |
| 3 | Model Test | 100% | 0 | 0 | 0 | 4.9 | 55.45 | 28.75 | 7.62 | 3.26 | 0 | 0 | 0 | Tests for an ML model in isolation, e.g., without taking any of the other ML components into account. |
| 2 | Input Test | 100% | 0 | 19.7 | 8.63 | 44.5 | 5.58 | 5.04 | 2.66 | 13.88 | 0 | 0 | 0 | Analysing the training and testing data . |
| 1 | Unit Test | 100% | 17.65 | 22.99 | 2.11 | 10.97 | 14.41 | 5.21 | 5.44 | 5.81 | 10.58 | 3.89 | 0.94 | Tests for individual components such as a newly developed module, mostly contained within a test folder 'unit tests'. |
| | AVERAGE | | 7.07 | 6.46 | 3.89 | 14.77 | 24.55 | 7.81 | 6.02 | 15.07 | 4.52 | 6.57 | 3.27 | |

**\*** *The level of transparency in each testing category indicates the type of test used, unlike the testing strategy in **RQ1**, which is a methodology.*





execution of already executed test cases to ensure that existing functionalities still work as expected after a change. We observed Regression testing in only three (3) of the studied ML systems (33%) (i.e., `DeepSpeech`, `deepchem` and `nupic` software systems) with an average of 1.43% of the total test cases in the target ML software systems. In the `nupic` system, a regression benchmark test [9] is executed every time a change is made to ensure that changes don't degrade prediction accuracy, through a set of standard experiments with thresholds for the prediction metrics. For example, limiting the permutations number may cause this test to fail if it results in lower accuracy.

• **Sanity Testing:** This test is part of regression testing. It aims to check the stability of new functionality or code changes in an existing build to ensure, for example, that bugs are fixed and that no new issues are introduced as a result of these changes. The Sanity test is performed prior to rigorous regression testing. The test focuses specifically on the build to ensure that the proposed functionality works close to or as expected [30]. For the sanity test cases to fail, it indicates that the build doesn't contain the required change (hence rejected) and, therefore, doesn't need to perform regression testing. We observed Sanity test cases in only `DeepSpeech` software system. The example of the Regression and Sanity test cases in the `DeepSpeech` include the test for various Finite State Transducer (FST) algorithms[10] that allow for mapping between two sets of symbols and generates a set of relations.

• **Periodic Validation and Verification:** This test periodically or frequently checks the program under test to confirm that the requirement has been fulfilled effectively and that the program achieves its intended purpose given the objective evidence. This type of test happens during automated model retraining and continuous learning; —ensuring that models are retrained with the most recent available data based on a given frequency or other conditions. This type of test can also be viewed as one form of regression testing that happens post the rigorous regression testing with a more specific focus on validating and verifying the running ML model in the production environment. The test of this type was observed in `nupic` (Application platform), `mycroft-core` (NLP system), and `mmf` (NLP system).

• **API Testing:** API Testing aims to check the functionality, performance, reliability, and security of the programming interfaces [76]. In API Testing, the software is used to send calls to the API, the output is collected, and the system's response is analyzed. We observed test cases related to API testing in four (4) of the studied ML software system (i.e., `paperless-ng`, `deepchem`, `nupic`, and `mycroft-core`)— out of which most of the API test cases was observed in `paperless-ng` ML system. `paperless-ng` provides API for most of its endpoints related to CRUD (i.e., create, read, update, and delete) operations for documents, email processing, and assigning tags to documents—subsequently, we observed a number of test cases for these APIs [11]. Also, `deepchem` provides API for two of its hyperparameter optimization endpoints: Grid and Gaussian process hyperparameter Optimization algorithms. Similarly, `mycroft-core` a voice NLP assistant system that can run on different platforms (e.g., Raspberry-pi), provides API endpoints for managing its features (e.g., Mycroft skills)— establishing connection between the host and Mycroft server.

• **Thread Testing:** This is an incremental software testing procedure (also called thread interaction test) performed during software system integration to verify the key functional capabilities of a specific thread/task. Thread testing is usually conducted early in the Integration testing phase [79]. Thread testing operations are classified as either (i) single-thread testing, where only one application transaction is verified at a given time, or (ii) multi-thread testing, where multiple concurrently

---

[9]https://github.com/numenta/nupic/blob/master/tests/regression/run_opf_benchmarks_test.py
[10]https://github.com/mozilla/DeepSpeech/blob/master/native_client/ctcdecode/third_party/openfst-1.6.7/src/test/algo_test.h
[11]https://github.com/jonaswinkler/paperless-ng/blob/master/docs/api.rst





active transactions are verified at a time. We observed this type of test in two of the studied ML software systems: `Apollo` and `Deepspeech` software systems, accounting for an average of 0.53% of the total number of test cases. An example code of thread testing is shown in Listing 31. This example is extracted from the `Apollo` software system. In `DeepSpeech` a single threaded test case is used to verify the creation of thread safe producer-consumer queue.

```
1   TEST(TestThread, Test) {
2     MyThread my_thread;
3     EXPECT_EQ(my_thread.get_value(), 0);
4     my_thread.set_thread_name("my_thread");
5     EXPECT_EQ(my_thread.get_thread_name(), "my_thread");
6
7     my_thread.Start();
8     my_thread.Join();
9     EXPECT_EQ(my_thread.get_value(), 100);
10    EXPECT_FALSE(my_thread.IsAlive());
11    MyThread my_thread2;
12    my_thread2.Start();
13    EXPECT_TRUE(my_thread2.IsAlive());
14    my_thread2.Join();
15    EXPECT_EQ(my_thread2.tid(), 0);
16
17    MyThread my_thread3;
18    my_thread3.set_joinable(false);
19    my_thread3.Start();
20    my_thread3.set_joinable(false);
21    EXPECT_TRUE(my_thread3.IsAlive());
22    std::this_thread::sleep_for(std::chrono::milliseconds(100));
23    EXPECT_FALSE(my_thread3.IsAlive()); }
```

Listing 31: Thread Testing example code extracted from Apollo system

• ***Performance/ Blob Testing:*** Performance Testing aims to check whether the software under test meets certain performance requirements such as stability, reliability, speed, response time, scalability, and resource consumption. In Table 5, we highlighted some of the test cases related to *Performance/ Blob* test cases, such as testing for the system overhead when loading a Binary Large Object (Blob) data observed mainly in `Apollo` software system.

• ***Exploratory Testing:*** This type of test involves using human creativity to explore and hunt for unexpected behaviors in the ML software system. Typically, the test design and execution are done simultaneously, and the observed test results are used to design the next test. Example of such test cases include using the print statements at different points within the test code instead of assertions api to help manually monitor the test during runs; writing examples test cases to verify specific component of the ML software system.

The ML engineers should consider including the above newly derived testing methods in the reference Test Pyramid for ML. For the regression test, they may utilise the different regression test selection techniques summarized by Graves, T.L. et al. [61] and make it as one of the required step [12, 20] during the development and testing phases of their ML software system.

> We identified a total of 13 different types of tests, out of which only eight (8) are either included in the Test Pyramid of the delivery process of ML software system proposed by Sato et al [1] or reviewed in research [116]. The newly observed types of tests are: *Regression testing*, *Sanity testing*, *Periodic Validation and Verification*, *Thread testing*, and *Blob testing*. Also, the test type *Contract test* described in the proposed Test Pyramid were not implemented by any of our studied ML software systems; suggesting gaps in the current ML testing practices in the field.

### 4.3.1 Comparing the use of testing methods across systems

Figure 9a and Figure 9b compare the usage of identified testing methods across the studied ML





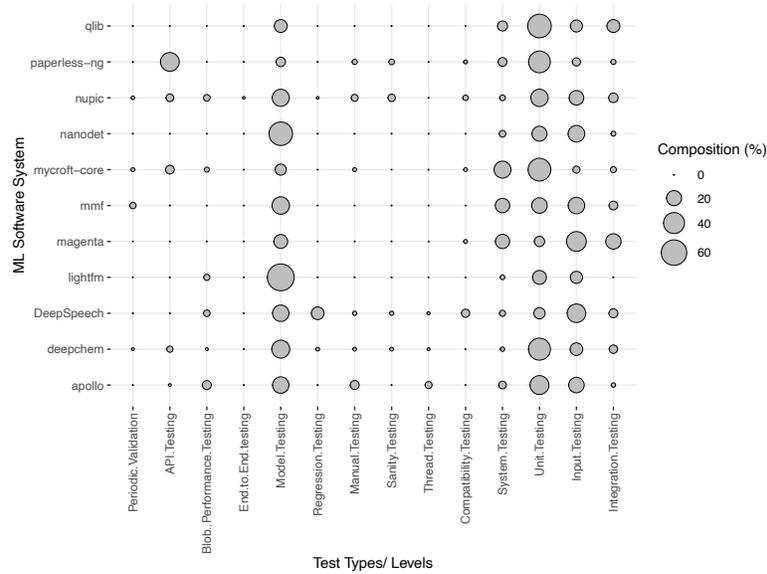

(a) The composition of types of testing across the studied ML software systems

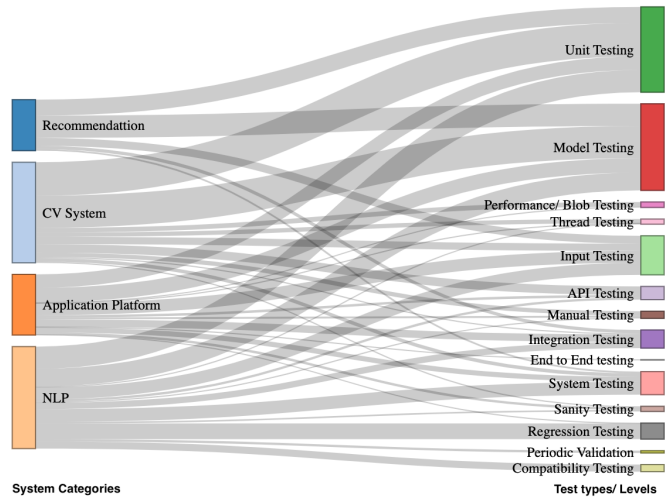

(b) Sankey diagram summarising the types of tests across ML system category

Fig. 9: Comparing the types of test across the ML Software System and their categories

software systems, and the ML domains, respectively. Figure 9a summarizes the percentage of total number of tests constituting each of the testing methods. Overall, the level of adoption of the identified testing methods varies dramatically across the studied ML software systems (ranging from zero to 32%), except for *Unit Testing*, *Model Testing*, *Input Testing*, *Integration testing*, and *System Testing* which are observed as the most used testing methods by ≥ 90% of the studied ML





software systems. *End-to-End*, *API Testing*, *Regression Test*, *Sanity Test*, *Compatibility* and *Thread Test* are implemented in fewer (< 40%) of the studied ML software systems.

According to our analysis based on the ML system domain shown in Figure 9b, the *Regression Test*, *System Testing*, *Compatibility*, and *Periodic Verification & Validation* are used more specifically in NLP system domain, while the *Model Test*, *Unit Test*, *Input Test*, *Thread Test*, and *Performance/ Blob Test* are observed more in CV system domain. On the other hand, Figure 9b show that *Integration Test* is used more in Application platform (12.1%) followed by Recommendation system (6.4%) and NLP system domains (6.1%), and fewest usage observed in CV systems (1.2%).

This can indicate that few ML software systems are thoroughly tested at different level of ML software development and deployment. Also, the non-uniform composition of the testing methods along the proposed test levels can potentially indicate that ML engineers only focus on the testing methods that related to their system's goals, while ignoring some essential requirements that might impact the ML system to be generalized. For example, according to our analysis in Figure 9a, the mean composition of the testing methods are *Unit Testing* (29.2%), *Model Testing* (27%), *Input Testing* (18.4%), *System Testing* (8.3%), *Integration Testing* (6.1%), *API Testing* (4.1%), *Regression Testing* (2.2%), *Blob/ Performance Testing* (1.6%), *Exploratory Testing* (1.3%), *Compatibility Testing* (0.9%), *Thread Testing* (0.4%), and *End-to-End Testing* (0%), representing the order of test composition in descending order. Similarly, at the software system level, `nupic` an application platform which implements most (86%) of the identified testing methods has more *Unit Testing* (26.8%), *Model Testing* (26.1%), *Input Testing* (18.6%), *Integration Testing* (7.2%), *API Testing* (4.6%), *Exploratory Testing* (3.6%), *Blob/ Performance Testing* (3.3%), *Regression Testing* (2.5%), *System Testing* (2.3%), *Compatibility Testing* (2%), *Periodic Validation* (0.7%), *End to End testing* (0.3%), *Thread Testing* (0%) .

Similarly, among three studied ML software systems that implements *Regression Test* (i.e., `DeepSpeech`, `deepchem`, and `nupic`); — the `DeepSpeech` which is an NLP systems implements the highest percentage of of test cases corresponding to *Regression Test* (13.9%), according to Figure 9a. To highlight the example of the test case corresponding to *Regression test*, in `DeepSpeech` an NLP system (a speech recognition system) ) an automated test [12] is set to re-run 25 times for every update to check that no regression is introduced for the various algorithms implementing the weighted finite-state transducers (FST) [13], such as lexical/ alphabetical ordering algorithm. The overall testing methods implemented in `DeepSpeech` system are in the order: *Input Testing* (31%), *Model Testing* (24.1%), *Regression Testing* (13.9%), *Unit Testing* (10.7%), *Integration Testing* (6.4%), *Compatibility Testing* (5.3%), *Blob/ Performance Testing* (3.2%), *System Testing* (2.7%), *Exploratory Testing* (1.1%), *Sanity Testing* (1.1%), *Thread Testing* (0.5%), *Periodic Validation* (0%) *API Testing* (0%) *End to End testing* (0%), as shown in Figure 9a. The results implies there is a non-uniform use of testing methods across the studied ML software system.

Figure 10 summarizes the usage of different testing methods in each ML systems domain, presented in the descending order of test case composition. Figure 10 indicates that certain testing methods are more dominant (with a high percentage) in some ML domains while others are less dominant. This shows variability across the studied domains. Moreover, at least half (50%) of the studied tests cases are contributed by the top 20% unique testing methods in all the studied ML domains, and the bottom 20% testing methods implements a few tests (≤ 2%). The testing method which is more used across all the studied ML system domains are: *Unit Test*, *Model Test*, *Input Test*, and *System Test*. In Recommendation system domain, the top 20% of the testing methods implements 80% of the test cases, confirming the Pareto's principle (80−20 rule) [44].

---

[12]https://github.com/mozilla/DeepSpeech/blob/master/native_client/ctcdecode/third_party/openfst-1.6.7/src/test/algo_test.cc
[13]www.openfst.org





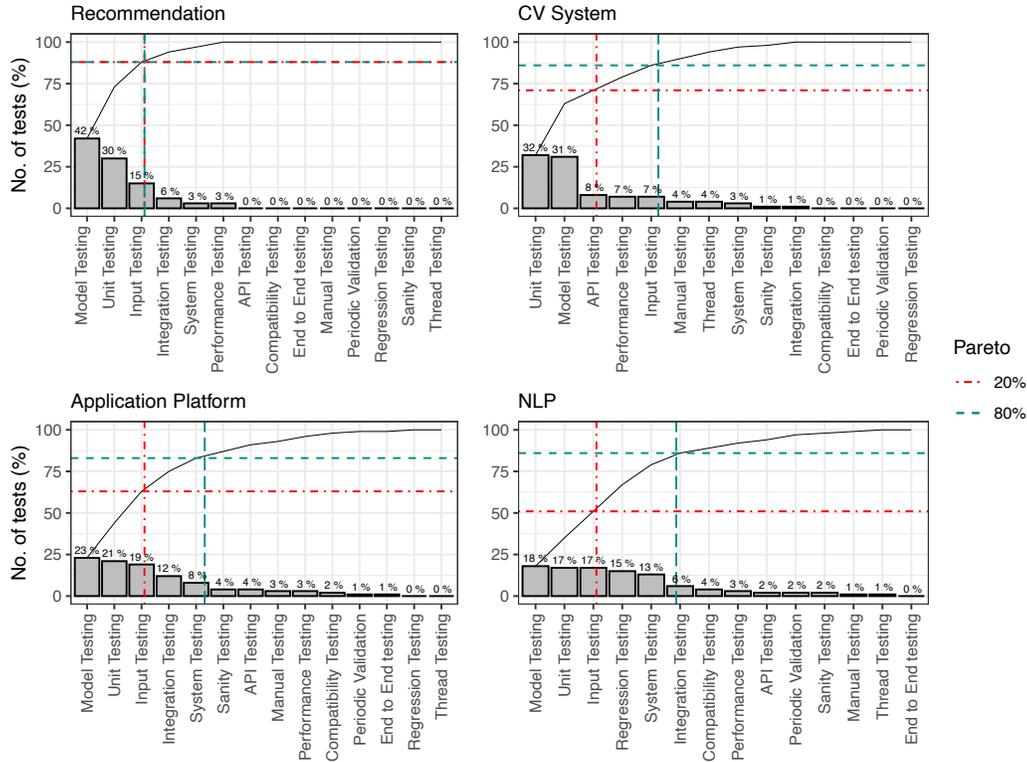

Fig. 10: Comparing Test methods across the ML system's domains, and Pareto analysis using the cumulative sum (%). The dotted red vertical line indicates the top 20% of the testing methods, while the dotted green horizontal line indicate the 80% of the test cases.

> Overall, at least half (50%) of the studied tests cases are contributed by the top 20% of testing methods in all the studied ML system's domain. On average, *Unit Testing*, *Model Testing*, *Input Testing*, *Integration testing*, and *System Testing* as on average the most used testing methods and are used by ≥ 90% of the studied ML software systems. The least implemented testing methods across the studied ML software systems (in < 40%) are: *End-to-End*, *API Test*, *Regression Test*, *Sanity Test*, *Compatibility Test* and *Thread Test*. Our results suggest that the ML software systems are not thoroughly tested during the development phases.

## 5 Discussion and Implications

### 5.1 Discussion

In the following, we discuss our findings and their implications for various stakeholders. This study presents the first fine-grained empirical analysis of testing practices for ML systems in real-world settings. We identified 16 ML properties tested using 15 major testing strategies and 13 testing methods. Our breakdown covers tested properties, testing strategies, and types of tests used in the ML workflow. This paper highlights that a significant portion of testing activities (approximately 36%) occurs during the model training and feature engineering phases. These testing activities aim





at early detection and resolution of issues, emphasizing the critical role of data in ML systems. Since most ML vulnerabilities and attacks originate from the early activities of the ML workflow, addressing these issues earlier is crucial for the overall integrity of the system [103]. Thus, focusing on these early phases of the ML lifecycle is essential as it helps identify and rectify potential problems before they propagate to later ML workflow activities. However, the uneven distribution of testing efforts across different workflow activities reveals a lack of consistent testing coverage. This inconsistency can result in overlooked issues that may emerge during deployment or in production environments. Therefore, a uniform application of testing strategies throughout the entire ML workflow is necessary. Our findings reveal that the Systems Under Tests (SUTs)—namely NLP, recommendation systems, CV systems, and application platforms—exhibit diverse distributions regarding the testing strategies employed. Specifically, the recommendation system follows a Pareto distribution, indicating that 20% of the testing strategies account for 80% of the SUTs' pipeline testing. This pattern underscores a significant concentration of testing efforts in a few key areas. For recommendation systems, model testing, unit testing, and input testing are predominant, while in CV systems, API testing replaces input testing within the top 20% test types. This shift highlights the distinct nature of CV systems, where interaction with external APIs is more critical, likely due to the frequent need for integration with various data sources and processing services.

Critically, this distribution raises important questions about the sufficiency and effectiveness of the current testing strategies. The reliance on a limited subset of test types for a significant portion of the testing process could lead to gaps in coverage, potentially overlooking nuanced issues that might arise in less frequently tested areas. This concern is particularly pertinent for complex ML systems, where the diversity of potential failure modes necessitates a comprehensive and balanced testing approach. Furthermore, the variation between SUTs, such as the emphasis on API testing in CV systems, suggests that a one-size-fits-all approach to testing may be inadequate. Instead, testing strategies should be tailored to each SUT's specific characteristics and requirements to ensure thorough validation and reliability. Effective mitigation techniques to address these disparities in ML system testing strategies include improving data quality, implementing rigorous testing practices, securing the training and deployment processes, etc. Therefore, conducting comprehensive (i.e., diversified and evenly distributed) testing across the ML workflow can help reduce risks and improve ML systems' overall reliability and performance, which is crucial for ensuring their robustness, security, and integrity. Moreover, we observed that incorporating specific properties, such as model explainability [114], data poisoning [32, 80], adversarial training [57, 134], anomaly detection [29], fairness assessments, and bias mitigation [13, 92], collectively could enhance the robustness of ML models against distribution shifts. This finding highlights the need for greater emphasis on these properties to ensure comprehensive testing and robust performance of ML systems. By integrating these techniques into existing testing frameworks, ML engineers can effectively address distribution shifts and improve their models' generalization capabilities. Future research should explore ways to encourage the adoption of these techniques across various ML applications, considering both technical feasibility and ethical considerations. This approach could enhance the reliability and performance of ML systems and contribute to building more trustworthy and ethical AI systems.

Our results suggest a significant dominance of Grey-box testing, with prevalent strategies including *Exception and Error Condition* (24.52%), *Statistical Analysis* (10.01%), and *Assertion-based Testing* (24.85%). Following this, White-box testing is also common, featuring techniques such as *Value Range Analysis* (9.24%), *Membership Testing* (8.81%), and *Decision & Logical Condition* (5.39%). This study highlights the preference for Grey-box testing due to its hybrid approach, which combines the internal insights of White-box testing with the external validation focus of Black-box testing. However, the reliance on assertion-based and negative testing methods indicates an underutilization of advanced statistical techniques. Though less commonly employed, these





sophisticated methods are crucial for thoroughly validating model performance under various conditions. We advocate for ML engineers to broaden their testing toolkit by incorporating more statistical analyses. This expansion has the potential to enhance the robustness of ML models and ensure more thorough validation, thereby improving the overall quality of ML testing practices.

**Lesson learned from the studied ML Properties and Testing Strategies**

In addition to these studied testing strategies and methods, we identified 16 distinct ML properties, emphasizing their importance in ML systems. The key focus areas include functional correctness and consistency, efficiency, security and privacy, and certainty. Five properties are frequently tested in ML workflow activities: correctness, consistency, data poisoning, efficiency, and robustness. These properties are critical for ensuring ML systems' fundamental functionality, reliability, and performance. In contrast, properties such as bias and fairness, compatibility and portability, security and privacy, data timeliness, and uncertainty are tested less frequently. This disparity highlights potential gaps in current testing practices, possibly leading to issues in these less-emphasized areas.

A comprehensive evaluation of **ML properties** and **Testing strategies** revealed key insights:

(1) **Consistency in Testing Approaches**: Grey-box and White-box testing methodologies were consistently employed to verify the various **ML properties**. This regular use underscores their effectiveness in addressing a broad spectrum of quality concerns within ML systems, demonstrating their versatility and reliability in different testing scenarios.

(2) **Focus on Core Properties**: Significant attention is given to properties such as **Correctness**, **Consistency**, and **Efficiency**, with robust and consistent testing strategies applied across these domains. This emphasis highlights their critical role in ensuring the overall reliability and performance of ML systems. The meticulous testing of these core properties ensures that the fundamental functionality and operational integrity of ML systems are maintained.

(3) **Neglect of Ethical and Security Concerns**: Properties related to **Bias & Fairness** and **Security & Privacy** are not tested as consistently, pointing to potential vulnerabilities. The less frequent testing of these properties indicates a crucial area needing greater focus, particularly given the ethical and security implications inherent in ML applications. This gap suggests a need for more rigorous and frequent evaluation to mitigate risks and uphold ethical standards in ML systems.

(4) **Evolution of Testing Methodologies**: The identification of new test types, which were not previously included in standard practices, underscores the dynamic and evolving nature of ML testing. This evolution highlights the necessity for continuous adaptation and innovation in testing methodologies to address emerging challenges and complexities in ML systems. Staying updated with and integrating these novel test types can enhance the robustness and comprehensiveness of ML testing strategies.

Below we highlight concrete use-cases of some of the identified testing strategies in practice in the following format {**Potential Application** (***Testing Strategy***: Description) } :

● **Testing for Robustness** (*Error and Exception Handling*):

Error and exception analysis is a critical process within ML systems, integral for enhancing the performance of ML models and ensuring their proper functioning in production environments. The primary objective of error and exception condition analysis in ML is to identify any errors that were inadequately handled or remained undetected during model deployment. This comprehensive process entails evaluating various ML artifacts, including data, models, features, and code, throughout the entirety of the ML workflow and fortifying their robustness. Regrettably, no one-size-fits-all





toolkit exists capable of universally addressing all ML problems. Distinct domains, such as computer vision and natural language processing (NLP), often demand tailored error and exception conditions specific to their respective challenges. For instance, in computer vision tasks, ensuring the quality of input data, such as its validity, and confirming adherence to the model's anticipated conditions for features are imperative. Conversely, the intricate and opaque nature of deep learning models complicates the understanding and interpreting of features, thereby presenting challenges in elucidating their behavior. It is critical to customize the approach based on domain-specific expertise to develop reliable and resilient ML systems. Practitioners should proactively detect and rectify potential errors in the development process to maintain system effectiveness in production environments.

- **Testing for interpretable model** *(Decision and Logical Condition)*:

Ensuring transparency and accountability of predictive models is essential in ML system testing. However, achieving model interpretability is a multifaceted endeavor influenced by contextual factors [50, 71]. Interpretability often hinges on domain-specific structural knowledge, encompassing generative constraints, additivity principles [87], monotonicity considerations [64], as well as causal and physical constraints rooted in domain expertise. This context-dependency underscores the potential applicability of the Decision and Logical Condition testing strategy in testing models that adhere to such properties. By leveraging Decision and Logical Condition testing, tailored test cases can be designed to account for semantic characteristics and structural intricacies inherent in the model architecture. This strategy enables a thorough evaluation of model behavior against domain-specific constraints, thereby enhancing the comprehensiveness and efficacy of ML testing practices within software engineering.

- **Towards building robust and reliable ML system** (*Decision and Logical Condition*):

Testing based on decision-making and logical conditions offers a versatile utility extending beyond mere error and execution handling. These tests empower software engineers/testers to scrutinize and manage specific program conditions at various levels, thereby enhancing the resilience and adaptability of the system. For instance, when diverse error-handling strategies are deployed across different stages of program execution, leveraging decision-making and logical conditions enables seamless navigation through active handlers in the event of errors. This capability contributes significantly to the construction of ML systems that are both robust and reliable.

- **Regret analysis and minimization (***Thresholding***)**:

In uncertain environments, decision thresholds often necessitate updating or modification due to shifts in the distribution of new data and changes in business decisions as the ML system evolves. Consequently, there arises a demand for robust and automated thresholding techniques capable of minimizing worst-case regret. This need becomes particularly evident in scenarios such as online machine learning or during the training of machine learning models that must adhere to various constraints, such as fairness, without compromising the accuracy of the model's performance.

- **Image classification and Object detection (***Thresholding***)**:

Object detection combines the principles of image classification and object localization. Image classification involves identifying the class of objects present in an image, while object localization precisely pinpoints the location of object(s)/ classes in an image, and by drawing abounding boxes around them. Object detection is commonly used in safety-critical domains where high confidence scores and model robustness are essential, due to legal requirements or the potential consequences of incorrect decision. It is crucial to employ proper thresholding techniques to ensure that the system operates with the highest degree of accuracy and reliability. This involves meticulously





testing the various threshold parameters used to maximize the system's confidence levels prior to deployment. By doing so, the system can meet the stringent demands of safety-critical applications.

• **Improving and Optimizing the performance of predictive models (*Back-testing*):**
Back-testing strategy plays a significant role in developing and validating predictive models. It is an essential tool to improve model reliability, identify technical or theoretical weaknesses, and build confidence in its performance before deploying it in real-world scenarios. For example, by testing a model against historical data from a selected period and evaluating its performance across different time frames or out-of-sample datasets, practitioners can refine and optimize the model, ensuring its effectiveness across diverse market conditions.

• **Models performance validation:**

• **Validate online model performance *(Back-testing)*:**
Back-testing, a common practice in historical data analysis, has recently been adapted for online performance testing of Machine Learning (ML) systems. This adaptation involves creating a back-testing service to validate ML model performance in real-time using incoming data streams. The service evaluates model predictions against live data at regular intervals, allowing for daily, weekly, or monthly assessments as needed. Using a sliding scale approach, the back-testing service analyzes model performance across multiple time windows, identifying potential fluctuations and accuracy drifts. Insights from back-testing inform model training optimization. Upon detecting significant performance shifts, the service triggers ML model retraining using new data windows, ensuring adaptability to evolving data dynamics. Integration of back-testing into online performance testing not only strengthens ML system reliability but also enables proactive adjustments to maintain model effectiveness.

• **Validate online model performance (*Oracle Approximation*):**
The Oracle Approximation testing technique is widely used in ML and DL models due to several factors such as randomness and numerical errors, which can lead to slight variations between actual and estimated outcomes. Although it does not ensure an optimal solution, this testing strategy offers numerous advantages, especially in model evaluation scenarios where outcomes from cross-validation techniques may vary across multiple runs. The primary objective of the Oracle Approximation testing technique is to closely approximate the optimal value within a reasonable computational time frame, typically polynomial time complexity. Similar to Back-testing in finance, the concept of Oracle Approximation can be used to validate the real-time performance of ML models. In this context, the model's decisions on new data are compared against the closest best-fixed decision, providing insights into its performance under dynamic conditions. Furthermore, the Oracle Approximation technique holds promise for refinement over time, especially in online learning environments where continuous updates and adaptations are essential for model improvement and adaptation to evolving data patterns.

• **Model construction e.g., structure learning of Bayesian networks (*Swarm testing*):**
Bayesian networks are probabilistic graphical models that depict a set of random variables and their conditional dependencies in a directed acyclic graph. The presence or absence of edges between nodes in the network signifies the relationships of dependence among variables. Dynamic Bayesian networks extend this concept to model sequences of variables, allowing for the representation of temporal dependencies. Wu et al. proposed a methodology for generating Bayesian networks using the swarm testing approach, as outlined in [144]. This approach leverages Ant Colony Optimization (ACO) to construct Bayesian networks in two steps: ChainACO and K2ACO. In the ChainACO step,





the node order is determined based on dependencies, prioritizing computational efficiency over structural complexity. Conversely, the K2ACO step employs the K2 algorithm to explore richer structural configurations and identify the optimal variable ordering. Experimental evaluations demonstrated that ACO-based Bayesian network learning algorithms outperform greedy search and simulated annealing methods. In addition, as discussed in Section 4.1, the Nupic ML system utilizes a Swarming algorithm for model selection and hyperparameter optimization. This approach involves creating and evaluating multiple models on a dataset and selecting the parameters that yield the lowest error scores.

● **Image Segmentation (*Swarm testing*):**

Clustering algorithms, optimized with swarming techniques, have emerged as promising methods for aggregating data into distinct groups while ensuring intra-group homogeneity and inter-group distinctiveness [33, 46, 113]. Drawing inspiration from natural systems' behaviors [108], swarming techniques have been effectively employed to enhance traditional clustering algorithms such as K-means and Simple Competitive Learning (SCL). Specifically, the Ant Colony Optimization for K-means (ACO-K-means)[113] and Particle Swarm Optimization for K-means (PSO-K-means)[33, 46] algorithms have been proposed to overcome limitations associated with traditional K-means clustering. These techniques facilitate the discovery of global optima while reducing dependency on initial seed selection, a common challenge in traditional K-means clustering. Comparative studies have demonstrated the superior performance of these optimized algorithms over conventional K-means, which often converge to local optima and are highly sensitive to initial cluster center selection. Moreover, swarming techniques have shown promise in enhancing the capabilities of SCL algorithms. By leveraging swarming optimization, SCL algorithms can effectively identify global optima using comparable parameter sets and learning rates. Thus, enabling SCL algorithms to identify meaningful clusters even in scenarios where traditional SCL approaches may fail.

● **Test case generation and test optimization (*Swarm testing*)**:

Swarming techniques have potential applications in generating test cases, which is a crucial activity in producing a set of test data that satisfies selected testing criteria. By leveraging swarming techniques, the process of generating test cases can be achieved through several key steps. Firstly, the testing problem can be transformed into a graph representation, facilitating a structured approach to test case generation. Subsequently, defining a scoring technique becomes essential for measuring the goodness of paths within the graph, enabling the evaluation of potential test case solutions. Moreover, identifying suitable and effective solutions, along with establishing stopping criteria, are vital aspects of the test case generation process. Additionally, adopting suitable techniques for updating criteria ensures adaptability and optimization throughout the generation process. Furthermore, optimization strategies have been extensively employed to address the optimization problem associated with test cases. Techniques such as test case prioritization (e.g., [9, 10]), test case selection (e.g., [148]), and test case reduction (e.g., [10]) have been instrumental in identifying the most critical test cases, thereby enhancing testing efficiency. These optimization strategies not only aid in finding optimal solutions but also contribute to cost-effective testing practices. For instance, in regression testing scenarios, the application of optimization techniques ([10, 148]) has led to significant improvements in testing efficacy and resource utilization.

● **Decision making and Planning under uncertainty (Swarm testing):**

When addressing unknown and unstructured environments, the Swarm technique has emerged as a valuable approach for planning amidst uncertainty, drawing inspiration from the collective behavior of natural swarms. Notably, the swarm technique has found application in tackling complex, so-called wicked problems [43], such as climate change [120, 121], real-time distribution





planning [146], and path planning [117, 130], offering more adaptable and flexible solutions. In the context of software testing of ML models, understanding the landscape of tested ML properties is crucial for ensuring the reliability and effectiveness of ML-based systems.

## 5.2 Implications

Based on the analysis presented in this paper, we suggest the following to researchers/academics and ML practitioners.

**ML Engineers:**

– **Testing Strategy Implementation:** ML engineers should consider adopting a diverse range of testing strategies, particularly focusing on Grey-box and White-box approaches, which have shown dominance in identifying software bugs across various ML workflow activities.
– **Property Testing:** There's a need to broaden the scope of tested ML properties beyond the commonly studied ones. ML engineers should prioritize testing for a wider range of properties such as Bias & Fairness, Explainability & Interpretability, Compatibility & Portability, Security & Privacy, Data Timeliness, and Uncertainty.
– **Test Methodologies:** The study highlights the non-uniform implementation of different testing methods across ML software systems. ML engineers should ensure thorough testing throughout the development phases, covering a comprehensive set of testing methods to enhance software reliability.

**Testers:**

– **Diversified Testing:** Testers should employ a variety of testing strategies, particularly focusing on Oracle Approximation, Instance and Type Checks, Statistical testing, State Transition, and Value Range Analysis, which are consistently used across a significant proportion of ML software systems.
– **Understanding Domain-Specific Testing Needs:** Testers need to recognize that the effectiveness of certain testing strategies varies across ML domains. For instance, Statistical Testing is more prevalent in CV systems and NLP systems but less utilized in Recommendation Systems, necessitating domain-specific testing approaches.

**Tool Builders:**

– **Tool Development:** Tool builders can develop tools that facilitate the implementation of diverse testing strategies identified in the study. This includes tools for conducting Grey-box, White-box, and Black-box testing efficiently across various activities of the ML workflow.
– **Domain-Specific Support:** Tools should offer support for domain-specific testing needs, considering the varying requirements across different ML domains such as CV, NLP, and Recommendation Systems.

**Researchers:**

– **Research Focus:** Future research should delve deeper into evaluating the effectiveness of identified testing strategies, particularly those that have been less explored in previous studies. This includes testing strategies like Heuristic-based testing and Boundary Value Analysis, which show uniqueness in test objectives.
– **Property Testing Enhancement:** Researchers should extend their focus beyond the commonly studied ML properties and explore testing methodologies for properties like Bias & Fairness, Explainability & Interpretability, Compatibility & Portability, Security & Privacy, Data Timeliness, and Uncertainty.





**Academics:**

- **Curriculum Enhancement:** Academics can incorporate insights from the study into ML curriculum, emphasizing the importance of testing practices and strategies in ML software development. This includes covering a comprehensive range of testing methods and strategies across different ML domains.
- **Encouraging Interdisciplinary Research:** Encouragement should be given for interdisciplinary research collaborations between ML practitioners, software engineers, and testers to enhance the understanding and application of effective testing practices in ML.

### 5.3 Future direction

This paper highlights critical areas for future research/ innovation in ML testing practices. Future studies should prioritize evaluating the effectiveness of the identified testing strategies and investigating the underlying reasons for the non-uniform application of these methods across different ML software systems. Developing standardized best practices for ML testing is also essential. Furthermore, interdisciplinary collaborations are crucial to bridging the gap between theoretical insights and practical applications, ultimately advancing the field of ML and software engineering.

## 6 Threats to validity

In this section, we discuss the threats that could affect the validity of our results.

**Internal Validity threats** concern our selection of subject systems and analysis method. We have selected ML-based software systems where tests are written in either Python or C/C++ programming languages. We followed an iterative process to extract the relevant test cases, test functions, and assertions. First, we analyze the code base of all the test files while referring to the official documentation of the studied ML software systems. This step was performed manually by researchers with extensive ML expertise. Yet, it is still possible that we may have missed some test cases and–or assertions that are rarely used in ML-based systems and hence harder to recognize. However, we believe that this threat should have a minor impact on our analysis and the results presented in this study.

**External Validity threats** concern the possibility to generalize our results. First, we studied only nine open-source ML software systems hosted in GitHub. Although these systems are selected from different domains such as AutoML frameworks, ML systems, and autonomous systems, it is still possible that we may have missed some important aspects of the testing practices of ML software systems. Also, the selected systems do not cover all domains of ML software systems. In the future, we plan to expand our study to cover more ML software systems to further validate our results. Second, the focus of this study is ML software systems programmed in either Python and or C/C++. Therefore our findings may not generalize to ML software systems written in other programming languages. We also plan to expand our study in the future to include other programming languages, such as Java or GO. Also, when analyzing the different testing strategies in ML systems, we limited our analysis to fewer cases, such as categorizing the assertions and error-handling techniques, while mainly relying on the documentation and the relevant literature. Generally, identifying all the different test strategies would require more manual analysis efforts and a complete understanding of the relevant implementation of the studied systems. We plan to expand the scope of this study to cover all these different aspects in the future. We also plan to conduct some qualitative studies involving the original developers of the studied systems. We recognize that robustness to distribution shifts is crucial for generalizing ML models. To mitigate this, we have identified several properties that can enhance model robustness, including model explainability, data poisoning, adversarial training, anomaly detection, fairness assessments, and bias mitigation. Integrating these





properties helps ensure improved generalization and reliable performance across diverse real-world applications, addressing both technical and ethical considerations. We recommend future work to build on our taxonomy and explore additional ML properties not covered in this study.

**Reliability validity threats** concern the possibility of replicating this study. Every result obtained through empirical studies is threatened by potential bias from data sets. To mitigate these threats we chose to conduct manual analysis in this study, leverage up to five different participants with extensive ML expertise. We also provide in the paper, all the necessary details required to replicate our study. The source code repositories of the studied systems are publicly available to obtain the same data. In addition, we provide a replication package in [3], containing the list of the studied ML software systems and their source codes for the selected system's versions, and the data containing the analysis for each of our four research questions (both in raw and processed form).

**Construct to validity:** We based our categorization on the international standards like the ISO 25010 Software and Data Quality standard[14] and the ISO/IEC TR 29119-11:2020 Software and systems engineering — Software testing standard (Part 11: Guidelines on the testing of AI-based systems)[15], while we categorize and track with the hope of standardize things, and how to count and measure say one test, or one strategy, or ML property and double check with the peers.

## 7    Conclusion

This paper presents the first fine-grained empirical study of ML testing strategies implemented, the specific ML properties that are tested, and the types of testing used in the ML workflow. Our results highlight many interesting findings: 1) Although we uncover a broad range of 15 testing strategies used by ML engineers to verify 17 ML properties across the ML workflow, in line with the previous work [143], our results suggest that ML software systems may not be currently tested thoroughly at different test levels of ML development life cycle. This is true due to the inconsistency in the usage of the testing strategies, the verified ML properties, and the types of tests used across the ML workflow activities. For instance, few of the testing strategies ($< 40\%$) are consistently used across multiple ML systems and similarly 20% to 30% of the ML properties are tested across all the studied ML software systems. 2) Grey-Box testing are the most dominated testing strategies, specifically Assertion-based (e.g, Oracle Approximation) and negative testing (e..g., Error and Exception condition, Data quality assessment). These testing strategies are predominantly implemented during the model training and feature engineering. 3) Some of the ML critical properties like *Bias/ Fairness*, *Security & Privacy*, and *Uncertainly* are verified in specific ML software systems such as CV and Recommendations systems as compared to the less usage in other domains like NLP - instead, NLP systems tend to focus on verifying the commonly tested ML properties: Correctness and Consistency.

We encourage researchers, to build on our finding of 16 common ML properties to develop novel testing techniques and better tool support to help ML engineers tests for these identified properties. We also invite more studies on the evaluation of the effectiveness of the identified ML testing strategies in future. ML engineers can use our presented taxonomy, to learn about the existing ML testing strategies, and implement them in their ML workflow. Also, ML maintenance teams can test for our identified ML properties in their systems in order to ensure their systems' trustworthiness.

## Acknowledgement

This work is supported by the DEEL Project CRDPJ 537462-18 funded by the National Science and Engineering Research Council of Canada (NSERC) and the Consortium for Research and Innovation

---







in Aerospace in Québec (CRIAQ), together with its industrial partners Thales Canada inc, Bell Textron Canada Limited, CAE inc and Bombardier inc.[16]

## References


[1] 2019. Continuous Delivery for Machine Learning. https://martinfowler.com/articles/cd4ml.html.

[2] 2021. *GitHub REST API.* Retrieved January 5, 2021 from https://developer.github.com/v3/

[3] 2021. TOSEM-2021-Replication; "Studying the Practices of Testing Machine Learning Software in the Wild". https://github.com/SWATLab-git/TOSEM-2021-Replication.

[4] Aniya Aggarwal, Pranay Lohia, Seema Nagar, Kuntal Dey, and Diptikalyan Saha. 2019. Black box fairness testing of machine learning models. In *Proceedings of the 2019 27th ACM Joint Meeting on European Software Engineering Conference and Symposium on the Foundations of Software Engineering.* 625–635.

[5] Saleema Amershi, Andrew Begel, Christian Bird, Robert DeLine, Harald Gall, Ece Kamar, Nachiappan Nagappan, Besmira Nushi, and Thomas Zimmermann. 2019. Software Engineering for Machine Learning: A Case Study. In *2019 IEEE/ACM 41st International Conference on Software Engineering: Software Engineering in Practice (ICSE-SEIP).* 291–300. https://doi.org/10.1109/ICSE-SEIP.2019.00042

[6] Maryam Ashoori and Justin D Weisz. 2019. In AI we trust? Factors that influence trustworthiness of AI-infused decision-making processes. *arXiv preprint arXiv:1912.02675* (2019).

[7] auto-sklearn automl. 2021. test automl. https://github.com/automl/auto-sklearn/blob/master/test/test_automl/test_automl.py.

[8] auto-sklearn automl. 2021. test metrics. https://github.com/automl/auto-sklearn/blob/master/test/test_metric/test_metrics.py.

[9] Anu Bajaj, Ajith Abraham, Saroj Ratnoo, and Lubna Abdelkareim Gabralla. 2022. Test case prioritization, selection, and reduction using improved quantum-behaved particle swarm optimization. *Sensors* 22, 12 (2022), 4374.

[10] Anu Bajaj and Om Prakash Sangwan. 2021. Tri-level regression testing using nature-inspired algorithms. *Innovations in Systems and Software Engineering* 17 (2021), 1–16.

[11] Gagan Bansal, Besmira Nushi, Ece Kamar, Daniel S Weld, Walter S Lasecki, and Eric Horvitz. 2019. Updates in human-ai teams: Understanding and addressing the performance/compatibility tradeoff. In *Proceedings of the AAAI Conference on Artificial Intelligence,* Vol. 33. 2429–2437.

[12] Luciano Baresi and Mauro Pezze. 2006. An introduction to software testing. *Electronic Notes in Theoretical Computer Science* 148, 1 (2006), 89–111.

[13] Solon Barocas, Moritz Hardt, and Arvind Narayanan. 2023. *Fairness and machine learning: Limitations and opportunities.* MIT Press.

[14] Carlo Batini, Cinzia Cappiello, Chiara Francalanci, and Andrea Maurino. 2009. Methodologies for data quality assessment and improvement. *ACM computing surveys (CSUR)* 41, 3 (2009), 1–52.

[15] Yonatan Belinkov and Yonatan Bisk. 2017. Synthetic and natural noise both break neural machine translation. *arXiv preprint arXiv:1711.02173* (2017).

[16] Barry W. Boehm. 1984. Verifying and validating software requirements and design specifications. *IEEE software* 1, 1 (1984), 75.

[17] Matthew Wamsley Bovee. 2004. *Information quality: A conceptual framework and empirical validation.* Ph.D. Dissertation. University of Kansas.

[18] Houssem Ben Braiek and Foutse Khomh. 2019. TFCheck: A TensorFlow Library for Detecting Training Issues in Neural Network Programs. In *2019 IEEE 19th International Conference on Software Quality, Reliability and Security (QRS).* IEEE, 426–433.

[19] Houssem Ben Braiek and Foutse Khomh. 2020. On testing machine learning programs. *Journal of Systems and Software* 164 (2020), 110542. https://doi.org/10.1016/j.jss.2020.110542

[20] Eric Breck, Shanqing Cai, Eric Nielsen, Michael Salib, and D Sculley. 2017. The ML test score: A rubric for ML production readiness and technical debt reduction. In *2017 IEEE International Conference on Big Data (Big Data).* IEEE, 1123–1132.

[21] J. Businge, M. Openja, D. Kavaler, E. Bainomugisha, F. Khomh, and V. Filkov. 2019. Studying Android App Popularity by Cross-Linking GitHub and Google Play Store. In *2019 IEEE 26th International Conference on Software Analysis, Evolution and Reengineering (SANER).* 287–297. https://doi.org/10.1109/SANER.2019.8667998

[22] J. Businge, M. Openja, S. Nadi, E. Bainomugisha, and T. Berger. 2018. Clone-Based Variability Management in the Android Ecosystem. In *2018 IEEE International Conference on Software Maintenance and Evolution (ICSME).* 625–634. https://doi.org/10.1109/ICSME.2018.00072


---

[16]https://deel.quebec





[23] Bruno Cabral and Paulo Marques. 2007. Exception handling: A field study in java and. net. In *ECOOP 2007–Object-Oriented Programming: 21st European Conference, Berlin, Germany, July 30-August 3, 2007. Proceedings 21*. Springer, 151–175.

[24] Sean D Campbell. 2005. A review of backtesting and backtesting procedures. (2005).

[25] Yulong Cao, Chaowei Xiao, Benjamin Cyr, Yimeng Zhou, Won Park, Sara Rampazzi, Qi Alfred Chen, Kevin Fu, and Z Morley Mao. 2019. Adversarial sensor attack on lidar-based perception in autonomous driving. In *Proceedings of the 2019 ACM SIGSAC conference on computer and communications security*. 2267–2281.

[26] Nicholas Carlini and David Wagner. 2017. Towards evaluating the robustness of neural networks. In *2017 ieee symposium on security and privacy (sp)*. Ieee, 39–57.

[27] Adnan Causevic, Rakesh Shukla, Sasikumar Punnekkat, and Daniel Sundmark. 2013. Effects of Negative Testing on TDD: An Industrial Experiment. In *Agile Processes in Software Engineering and Extreme Programming*, Hubert Baumeister and Barbara Weber (Eds.). Springer Berlin Heidelberg, Berlin, Heidelberg, 91–105.

[28] Joymallya Chakraborty, Suvodeep Majumder, and Tim Menzies. 2021. Bias in machine learning software: why? how? what to do?. In *Proceedings of the 29th ACM Joint Meeting on European Software Engineering Conference and Symposium on the Foundations of Software Engineering*. 429–440.

[29] Varun Chandola, Arindam Banerjee, and Vipin Kumar. 2009. Anomaly detection: A survey. *ACM computing surveys (CSUR)* 41, 3 (2009), 1–58.

[30] Vinod Kumar Chauhan. 2014. Smoke testing. *Int. J. Sci. Res. Publ* 4, 1 (2014), 2250–3153.

[31] Chenyi Chen, Ari Seff, Alain Kornhauser, and Jianxiong Xiao. 2015. Deepdriving: Learning affordance for direct perception in autonomous driving. In *Proceedings of the IEEE international conference on computer vision*. 2722–2730.

[32] Xinyun Chen, Chang Liu, Bo Li, Kimberly Lu, and Dawn Song. 2017. Targeted backdoor attacks on deep learning systems using data poisoning. *arXiv preprint arXiv:1712.05526* (2017).

[33] Xuexin Chen, Pu Miao, and Qingkai Bu. 2019. Image segmentation algorithm based on particle swarm optimization with k-means optimization. In *2019 IEEE International Conference on Power, Intelligent Computing and Systems (ICPICS)*. IEEE, 156–159.

[34] Dawei Cheng, Chun Cao, Chang Xu, and Xiaoxing Ma. 2018. Manifesting bugs in machine learning code: An explorative study with mutation testing. In *2018 IEEE International Conference on Software Quality, Reliability and Security (QRS)*. IEEE, 313–324.

[35] John Joseph Chilenski and Steven P Miller. 1994. Applicability of modified condition/decision coverage to software testing. *Software Engineering Journal* 9, 5 (1994), 193–200.

[36] Voskoglou Christina. 2017. What is the best programming language for Machine Learning? https://towardsdatascience.com/what-is-the-best-programming-language-for-machine-learning-a745c156d6b7.

[37] Peter Christoffersen and Denis Pelletier. 2004. Backtesting value-at-risk: A duration-based approach. *Journal of Financial Econometrics* 2, 1 (2004), 84–108.

[38] Israel Cohen, Yiteng Huang, Jingdong Chen, Jacob Benesty, Jacob Benesty, Jingdong Chen, Yiteng Huang, and Israel Cohen. 2009. Pearson correlation coefficient. *Noise reduction in speech processing* (2009), 1–4.

[39] Darren Cook. 2016. *Practical machine learning with H2O: powerful, scalable techniques for deep learning and AI.* " O'Reilly Media, Inc.".

[40] cpputest. 2021. CppUTest: CppUTest unit testing and mocking framework for C/C++. https://cpputest.github.io/.

[41] Brian d'Alessandro, Cathy O'Neil, and Tom LaGatta. 2017. Conscientious classification: A data scientist's guide to discrimination-aware classification. *Big data* 5, 2 (2017), 120–134.

[42] Alexander D'Amour, Katherine Heller, Dan Moldovan, Ben Adlam, Babak Alipanahi, Alex Beutel, Christina Chen, Jonathan Deaton, Jacob Eisenstein, Matthew D Hoffman, et al. 2020. Underspecification presents challenges for credibility in modern machine learning. *arXiv preprint arXiv:2011.03395* (2020).

[43] Ruth DeFries and Harini Nagendra. 2017. Ecosystem management as a wicked problem. *Science* 356, 6335 (2017), 265–270.

[44] Rosie Dunford, Quanrong Su, and Ekraj Tamang. 2014. The pareto principle. (2014).

[45] Anna Fariha, Ashish Tiwari, Arjun Radhakrishna, Sumit Gulwani, and Alexandra Meliou. 2020. Data invariants: On trust in data-driven systems. *arXiv preprint arXiv:2003.01289* (2020).

[46] Taymaz Rahkar Farshi, John H Drake, and Ender Özcan. 2020. A multimodal particle swarm optimization-based approach for image segmentation. *Expert Systems with Applications* 149 (2020), 113233.

[47] Giuseppe Fenza, Mariacristina Gallo, Vincenzo Loia, Francesco Orciuoli, and Enrique Herrera-Viedma. 2021. Data set quality in Machine Learning: Consistency measure based on Group Decision Making. *Applied Soft Computing* 106 (2021), 107366. https://doi.org/10.1016/j.asoc.2021.107366

[48] Melanie Fink. 2021. The EU Artificial Intelligence Act and Access to Justice. *EU Law Live* (2021).

[49] A. Foundjem, E. E. Eghan, and B. Adams. 2023. A Grounded Theory of Cross-Community SECOs: Feedback Diversity Versus Synchronization. *IEEE Transactions on Software Engineering* 49, 10 (oct 2023), 4731–4750. https://doi.org/10.





1109/TSE.2023.3313875

[50] Alex A Freitas. 2014. Comprehensible classification models: a position paper. *ACM SIGKDD explorations newsletter* 15, 1 (2014), 1–10.

[51] Sainyam Galhotra, Yuriy Brun, and Alexandra Meliou. 2017. Fairness testing: testing software for discrimination. In *Proceedings of the 2017 11th Joint Meeting on Foundations of Software Engineering*. 498–510.

[52] Fernando Gama, Elvin Isufi, Geert Leus, and Alejandro Ribeiro. 2020. Graphs, Convolutions, and Neural Networks: From Graph Filters to Graph Neural Networks. *IEEE Signal Processing Magazine* 37, 6 (2020), 128–138. https://doi.org/10.1109/MSP.2020.3016143

[53] Rozental Gennadiy and Enficiaud Raffi. 2020. Boost C++ Libraries. https://www.boost.org/.

[54] S. Gheisari and M.R. Meybodi. 2016. BNC-PSO: structure learning of Bayesian networks by Particle Swarm Optimization. *Information Sciences* 348 (2016), 272–289. https://doi.org/10.1016/j.ins.2016.01.090

[55] Inc GitHub. 2021. The GitHub Search API lets you to search for the specific item efficiently. https://docs.github.com/en/rest/reference/search.

[56] Martin Glinz. 2000. Improving the quality of requirements with scenarios. In *Proceedings of the second world congress on software quality*, Vol. 9. 55–60.

[57] Ian J Goodfellow, Jonathon Shlens, and Christian Szegedy. 2014. Explaining and harnessing adversarial examples. *arXiv preprint arXiv:1412.6572* (2014).

[58] Cloud Google. 2021. MLOps: Continuous delivery and automation pipelines in machine learning. https://cloud.google.com/architecture/mlops-continuous-delivery-and-automation-pipelines-in-machine-learning.

[59] Googletest. 2021. Googletest Primer. http://google.github.io/googletest/primer.html.

[60] Palash Goyal and Emilio Ferrara. 2018. Graph embedding techniques, applications, and performance: A survey. *Knowledge-Based Systems* 151 (2018), 78–94. https://doi.org/10.1016/j.knosys.2018.03.022

[61] Todd L Graves, Mary Jean Harrold, Jung-Min Kim, Adam Porter, and Gregg Rothermel. 2001. An empirical study of regression test selection techniques. *ACM Transactions on Software Engineering and Methodology (TOSEM)* 10, 2 (2001), 184–208.

[62] Alex Groce, Chaoqiang Zhang, Eric Eide, Yang Chen, and John Regehr. 2012. Swarm Testing. In *Proceedings of the 2012 International Symposium on Software Testing and Analysis* (Minneapolis, MN, USA) *(ISSTA 2012)*. Association for Computing Machinery, New York, NY, USA, 78–88. https://doi.org/10.1145/2338965.2336763

[63] Riccardo Guidotti, Anna Monreale, Salvatore Ruggieri, Franco Turini, Fosca Giannotti, and Dino Pedreschi. 2018. A survey of methods for explaining black box models. *ACM computing surveys (CSUR)* 51, 5 (2018), 1–42.

[64] Maya Gupta, Andrew Cotter, Jan Pfeifer, Konstantin Voevodski, Kevin Canini, Alexander Mangylov, Wojciech Moczydlowski, and Alexander Van Esbroeck. 2016. Monotonic calibrated interpolated look-up tables. *The Journal of Machine Learning Research* 17, 1 (2016), 3790–3836.

[65] Naveen Gv. 2019. Memory errors in C++. https://www.cprogramming.com/tutorial/memory_debugging_parallel_inspector.html.

[66] Patrick Hall, Navdeep Gill, and Nicholas Schmidt. 2019. Proposed guidelines for the responsible use of explainable machine learning. *arXiv preprint arXiv:1906.03533* (2019).

[67] W. H. Harrison. 1977. Compiler Analysis of the Value Ranges for Variables. *IEEE Trans. Softw. Eng.* 3, 3 (May 1977), 243–250. https://doi.org/10.1109/TSE.1977.231133

[68] Campbell R Harvey and Yan Liu. 2015. Backtesting. *The Journal of Portfolio Management* 42, 1 (2015), 13–28.

[69] Kenneth Heafield. 2011. KenLM: Faster and Smaller Language Model Queries. In *Proceedings of the Sixth Workshop on Statistical Machine Translation*. Association for Computational Linguistics, Edinburgh, Scotland, 187–197. https://www.aclweb.org/anthology/W11-2123

[70] Eyke Hüllermeier, Thomas Fober, and Marco Mernberger. 2013. *Inductive Bias*. Springer New York, New York, NY, 1018–1018. https://doi.org/10.1007/978-1-4419-9863-7_927

[71] Johan Huysmans, Karel Dejaeger, Christophe Mues, Jan Vanthienen, and Bart Baesens. 2011. An empirical evaluation of the comprehensibility of decision table, tree and rule based predictive models. *Decision Support Systems* 51, 1 (2011), 141–154.

[72] IBM. 2020. The Machine Learning Development and Operations. https://ibm-cloud-architecture.github.io/refarch-data-ai-analytics/methodology/MLops/.

[73] ICS-33. 2021. Complexity of Python Operations. https://www.ics.uci.edu/~pattis/ICS-33/lectures/complexitypython.txt.

[74] K. Ilgun, R.A. Kemmerer, and P.A. Porras. 1995. State transition analysis: a rule-based intrusion detection approach. *IEEE Transactions on Software Engineering* 21, 3 (1995), 181–199. https://doi.org/10.1109/32.372146

[75] Tatjana Ille and Natasa Milic. 2008. *Statistical tests*. Springer Netherlands, Dordrecht, 1341–1344. https://doi.org/10.1007/978-1-4020-5614-7_3349





[76] Isha, Abhinav Sharma, and M. Revathi. 2018. Automated API Testing. In *2018 3rd International Conference on Inventive Computation Technologies (ICICT)*. 788–791. https://doi.org/10.1109/ICICT43934.2018.9034254

[77] Nataliya V Ivankova, John W Creswell, and Sheldon L Stick. 2006. Using mixed-methods sequential explanatory design: From theory to practice. *Field methods* 18, 1 (2006), 3–20.

[78] Mohit Iyyer, John Wieting, Kevin Gimpel, and Luke Zettlemoyer. 2018. Adversarial example generation with syntactically controlled paraphrase networks. *arXiv preprint arXiv:1804.06059* (2018).

[79] Paul C Jorgensen. 2013. *Software testing: a craftsman's approach*. Auerbach Publications.

[80] Pang Wei Koh and Percy Liang. 2017. Understanding black-box predictions via influence functions. In *International conference on machine learning*. PMLR, 1885–1894.

[81] Maciej Kula. 2015. Metadata Embeddings for User and Item Cold-start Recommendations. In *Proceedings of the 2nd Workshop on New Trends on Content-Based Recommender Systems co-located with 9th ACM Conference on Recommender Systems (RecSys 2015), Vienna, Austria, September 16-20, 2015. (CEUR Workshop Proceedings, Vol. 1448)*, Toine Bogers and Marijn Koolen (Eds.). CEUR-WS.org, 14–21. http://ceur-ws.org/Vol-1448/paper4.pdf

[82] J Lawrence, Steven Clarke, Margaret Burnett, and Gregg Rothermel. 2005. How well do professional developers test with code coverage visualizations? An empirical study. In *2005 IEEE Symposium on Visual Languages and Human-Centric Computing (VL/HCC'05)*. IEEE, 53–60.

[83] Xiang Li, Wenhai Wang, Lijun Wu, Shuo Chen, Xiaolin Hu, Jun Li, Jinhui Tang, and Jian Yang. 2020. Generalized focal loss: Learning qualified and distributed bounding boxes for dense object detection. *Advances in Neural Information Processing Systems* 33 (2020), 21002–21012.

[84] Geert Litjens, Thijs Kooi, Babak Ehteshami Bejnordi, Arnaud Arindra Adiyoso Setio, Francesco Ciompi, Mohsen Ghafoorian, Jeroen Awm Van Der Laak, Bram Van Ginneken, and Clara I Sánchez. 2017. A survey on deep learning in medical image analysis. *Medical image analysis* 42 (2017), 60–88.

[85] Liping Liu and Lauren Chi. 2002. Evolutional Data Quality: A Theory-Specific View.. In *ICIQ*. 292–304.

[86] Zhuang Liu, Jianguo Li, Zhiqiang Shen, Gao Huang, Shoumeng Yan, and Changshui Zhang. 2017. Learning efficient convolutional networks through network slimming. In *Proceedings of the IEEE International Conference on Computer Vision*. 2736–2744.

[87] Yin Lou, Rich Caruana, Johannes Gehrke, and Giles Hooker. 2013. Accurate intelligible models with pairwise interactions. In *Proceedings of the 19th ACM SIGKDD international conference on Knowledge discovery and data mining*. 623–631.

[88] Lei Ma, Fuyuan Zhang, Jiyuan Sun, Minhui Xue, Bo Li, Felix Juefei-Xu, Chao Xie, Li Li, Yang Liu, Jianjun Zhao, et al. 2018. Deepmutation: Mutation testing of deep learning systems. In *2018 IEEE 29th International Symposium on Software Reliability Engineering (ISSRE)*. IEEE, 100–111.

[89] Forough Majidi, Moses Openja, Foutse Khomh, and Heng Li. 2022. An Empirical Study on the Usage of Automated Machine Learning Tools. In *2022 IEEE International Conference on Software Maintenance and Evolution (ICSME)*. 59–70. https://doi.org/10.1109/ICSME55016.2022.00014

[90] VanderVoord Mark, Karlesky Mike, and Williams Greg. 2015. Unity Unit Testing for C (especially embedded software). http://www.throwtheswitch.org/unity.

[91] Lindsey Fiona Masson, Geraldine McNeill, JO Tomany, JA Simpson, Heather Sinclair Peace, L Wei, DA Grubb, and C Bolton-Smith. 2003. Statistical approaches for assessing the relative validity of a food-frequency questionnaire: use of correlation coefficients and the kappa statistic. *Public health nutrition* 6, 3 (2003), 313–321.

[92] Ninareh Mehrabi, Fred Morstatter, Nripsuta Saxena, Kristina Lerman, and Aram Galstyan. 2021. A survey on bias and fairness in machine learning. *ACM computing surveys (CSUR)* 54, 6 (2021), 1–35.

[93] Cohn Mike. 2009. The Forgotten Layer of the Test Automation Pyramid. https://www.mountaingoatsoftware.com/blog/the-forgotten-layer-of-the-test-automation-pyramid.

[94] Pavlo Molchanov, Arun Mallya, Stephen Tyree, Iuri Frosio, and Jan Kautz. 2019. Importance estimation for neural network pruning. In *Proceedings of the IEEE/CVF Conference on Computer Vision and Pattern Recognition*. 11264–11272.

[95] Ramon E Moore. 1966. *Interval analysis*. Vol. 4. Prentice-Hall Englewood Cliffs.

[96] Nuthan Munaiah, Steven Kroh, Craig Cabrey, and Meiyappan Nagappan. 2017. Curating github for engineered software projects. *Empirical Software Engineering* 22, 6 (2017), 3219–3253.

[97] W James Murdoch, Chandan Singh, Karl Kumbier, Reza Abbasi-Asl, and Bin Yu. 2019. Interpretable machine learning: definitions, methods, and applications. *arXiv preprint arXiv:1901.04592* (2019).

[98] Mahdi Nejadgholi and Jinqiu Yang. 2019. A study of oracle approximations in testing deep learning libraries. In *2019 34th IEEE/ACM International Conference on Automated Software Engineering (ASE)*. IEEE, 785–796.

[99] NumPy. 2021. NumPy: The fundamental package for scientific computing with Python. https://numpy.org/.

[100] Moses Openja, Gabriel Laberge, and Foutse Khomh. 2023. Detection and Evaluation of bias-inducing Features in Machine learning. *arXiv preprint arXiv:2310.12805* (2023).





[101] Moses Openja, Forough Majidi, Foutse Khomh, Bhagya Chembakottu, and Heng Li. 2022. Studying the Practices of Deploying Machine Learning Projects on Docker. In *Proceedings of the International Conference on Evaluation and Assessment in Software Engineering 2022* (Gothenburg, Sweden) *(EASE '22)*. Association for Computing Machinery, New York, NY, USA, 190–200. https://doi.org/10.1145/3530019.3530039

[102] Moses Openja, Amin Nikanjam, Ahmed Haj Yahmed, Foutse Khomh, and Zhen Ming Jack Jiang. 2022. An empirical study of challenges in converting deep learning models. In *2022 IEEE International Conference on Software Maintenance and Evolution (ICSME)*. IEEE, 13–23.

[103] Nicolas Papernot, Patrick McDaniel, Ian Goodfellow, Somesh Jha, Z. Berkay Celik, and Ananthram Swami. 2017. Practical Black-Box Attacks against Machine Learning. In *Proceedings of the 2017 ACM on Asia Conference on Computer and Communications Security* (Abu Dhabi, United Arab Emirates) *(ASIA CCS '17)*. Association for Computing Machinery, New York, NY, USA, 506–519. https://doi.org/10.1145/3052973.3053009

[104] Kexin Pei, Yinzhi Cao, Junfeng Yang, and Suman Jana. 2017. Deepxplore: Automated whitebox testing of deep learning systems. In *proceedings of the 26th Symposium on Operating Systems Principles*. 1–18.

[105] Anjana Perera, Aldeida Aleti, Chakkrit Tantithamthavorn, Jirayus Jiarpakdee, Burak Turhan, Lisa Kuhn, and Katie Walker. 2022. Search-based fairness testing for regression-based machine learning systems. *Empirical Software Engineering* 27, 3 (2022), 1–36.

[106] Anthony Peruma, Khalid Almalki, Christian D. Newman, Mohamed Wiem Mkaouer, Ali Ouni, and Fabio Palomba. 2019. On the Distribution of Test Smells in Open Source Android Applications: An Exploratory Study. In *Proceedings of the 29th Annual International Conference on Computer Science and Software Engineering* (Toronto, Ontario, Canada) *(CASCON '19)*. IBM Corp., USA, 193–202.

[107] Fortunato Pesarin and Luigi Salmaso. 2010. The permutation testing approach: a review. *Statistica* 70, 4 (2010), 481–509.

[108] DT Pham, Dervis Karaboga, DT Pham, and D Karaboga. 2000. Genetic Algorithms. *Intelligent Optimisation Techniques: Genetic Algorithms, Tabu Search, Simulated Annealing and Neural Networks* (2000), 51–147.

[109] Vinodkumar Prabhakaran, Ben Hutchinson, and Margaret Mitchell. 2019. Perturbation sensitivity analysis to detect unintended model biases. *arXiv preprint arXiv:1910.04210* (2019).

[110] Pytest. 2021. Pytest, About fixtures. https://docs.pytest.org/en/latest/explanation/fixtures.html.

[111] Sivaramakrishnan Rajaraman, Prasanth Ganesan, and Sameer Antani. 2022. Deep learning model calibration for improving performance in class-imbalanced medical image classification tasks. *PloS one* 17, 1 (2022), e0262838.

[112] Bharath Ramsundar, Peter Eastman, Patrick Walters, Vijay Pande, Karl Leswing, and Zhenqin Wu. 2019. *Deep Learning for the Life Sciences*. O'Reilly Media. https://www.amazon.com/Deep-Learning-Life-Sciences-Microscopy/dp/1492039837.

[113] T Namratha Reddy and KP Supreethi. 2017. Optimization of K-means algorithm: Ant colony optimization. In *2017 International Conference on Computing Methodologies and Communication (ICCMC)*. IEEE, 530–535.

[114] Marco Tulio Ribeiro, Sameer Singh, and Carlos Guestrin. 2016. " Why should i trust you?" Explaining the predictions of any classifier. In *Proceedings of the 22nd ACM SIGKDD international conference on knowledge discovery and data mining*. 1135–1144.

[115] Marco Tulio Ribeiro, Sameer Singh, and Carlos Guestrin. 2018. Semantically equivalent adversarial rules for debugging nlp models. In *Proceedings of the 56th Annual Meeting of the Association for Computational Linguistics (Volume 1: Long Papers)*. 856–865.

[116] Vincenzo Riccio, Gunel Jahangirova, Andrea Stocco, Nargiz Humbatova, Michael Weiss, and Paolo Tonella. 2020. Testing machine learning based systems: a systematic mapping. *Empirical Software Engineering* 25 (2020), 5193–5254.

[117] Vincent Roberge, Mohammed Tarbouchi, and Gilles Labonté. 2012. Comparison of parallel genetic algorithm and particle swarm optimization for real-time UAV path planning. *IEEE Transactions on industrial informatics* 9, 1 (2012), 132–141.

[118] Suzanne Robertson and James Robertson. 2012. *Mastering the requirements process: Getting requirements right*. Addison-wesley.

[119] Emmanuel Roche and Yves Schabes. 1997. *Finite-state language processing*. MIT press.

[120] Rob Roggema. 2013. *Swarm planning: The development of a planning methodology to deal with climate adaptation*. Springer Science & Business Media.

[121] Rob Roggema and Andy Van den Dobbelsteen. 2012. Swarm planning for climate change: an alternative pathway for resilience. *Building research & information* 40, 5 (2012), 606–624.

[122] Barbara Rychalska, Dominika Basaj, Alicja Gosiewska, and Przemysław Biecek. 2019. Models in the wild: On corruption robustness of neural nlp systems. In *International Conference on Neural Information Processing*. Springer, 235–247.

[123] Roshni Sahoo, Shengjia Zhao, Alyssa Chen, and Stefano Ermon. 2021. Reliable decisions with threshold calibration. *Advances in Neural Information Processing Systems* 34 (2021), 1831–1844.





[124] Pedro Sandoval-Segura, Vasu Singla, Jonas Geiping, Micah Goldblum, Tom Goldstein, and David Jacobs. 2022. Autoregressive Perturbations for Data Poisoning. In *Advances in Neural Information Processing Systems*, S. Koyejo, S. Mohamed, A. Agarwal, D. Belgrave, K. Cho, and A. Oh (Eds.), Vol. 35. Curran Associates, Inc., 27374–27386. https://proceedings.neurips.cc/paper_files/paper/2022/file/af66ac99716a64476c07ae8b089d59f8-Paper-Conference.pdf

[125] Monica Scannapieco and Tiziana Catarci. 2002. Data quality under a computer science perspective. *Archivi & Computer* 2 (2002), 1–15.

[126] Carolyn B. Seaman. 1999. Qualitative methods in empirical studies of software engineering. *IEEE Transactions on software engineering* 25, 4 (1999), 557–572.

[127] Witowski Sebastian. 2021. Membership Testing. https://switowski.com/blog/membership-testing.

[128] Jasmine Sekhon and Cody Fleming. 2019. Towards improved testing for deep learning. In *2019 IEEE/ACM 41st International Conference on Software Engineering: New Ideas and Emerging Results (ICSE-NIER)*. IEEE, 85–88.

[129] Sina Shafaei, Stefan Kugele, Mohd Hafeez Osman, and Alois Knoll. 2018. Uncertainty in machine learning: A safety perspective on autonomous driving. In *Computer Safety, Reliability, and Security: SAFECOMP 2018 Workshops, ASSURE, DECSoS, SASSUR, STRIVE, and WAISE, Västerås, Sweden, September 18, 2018, Proceedings 37*. Springer, 458–464.

[130] Abhishek Sharma, Shraga Shoval, Abhinav Sharma, and Jitendra Kumar Pandey. 2022. Path planning for multiple targets interception by the swarm of UAVs based on swarm intelligence algorithms: A review. *IETE Technical Review* 39, 3 (2022), 675–697.

[131] Hocheol Shin, Dohyun Kim, Yujin Kwon, and Yongdae Kim. 2017. Illusion and dazzle: Adversarial optical channel exploits against lidars for automotive applications. In *International Conference on Cryptographic Hardware and Embedded Systems*. Springer, 445–467.

[132] Julien Siebert, Lisa Joeckel, Jens Heidrich, Adam Trendowicz, Koji Nakamichi, Kyoko Ohashi, Isao Namba, Rieko Yamamoto, and Mikio Aoyama. 2021. Construction of a quality model for machine learning systems. *Software Quality Journal* (2021), 1–29.

[133] Amanpreet Singh, Vedanuj Goswami, Vivek Natarajan, Yu Jiang, Xinlei Chen, Meet Shah, Marcus Rohrbach, Dhruv Batra, and Devi Parikh. 2020. Mmf: A multimodal framework for vision and language research.

[134] Aman Sinha, Hongseok Namkoong, Riccardo Volpi, and John Duchi. 2017. Certifying some distributional robustness with principled adversarial training. *arXiv preprint arXiv:1710.10571* (2017).

[135] British Computer Society. 1998. Glossary of terms used in software testing (Version 6.3). http://www.testingstandards.co.uk/bs_7925-1_online.htm.

[136] Iain Sommerville and Peter Sawyer. 1997. *Requirements engineering: a good practice guide.* John Wiley & Sons, Inc.

[137] Megha Srivastava, Besmira Nushi, Ece Kamar, Shital Shah, and Eric Horvitz. 2020. An Empirical Analysis of Backward Compatibility in Machine Learning Systems. In *Proceedings of the 26th ACM SIGKDD International Conference on Knowledge Discovery & Data Mining* (Virtual Event, CA, USA) *(KDD '20)*. Association for Computing Machinery, New York, NY, USA, 3272–3280. https://doi.org/10.1145/3394486.3403379

[138] StackExchange. 2017. "GitHub Stars" is a very useful metric. But for "what"? https://opensource.stackexchange.com/questions/5110/github-stars-is-a-very-useful-metric-but-for-what.

[139] Hu Su, Yonghao He, Rui Jiang, Jiabin Zhang, Wei Zou, and Bin Fan. 2022. DSLA: Dynamic smooth label assignment for efficient anchor-free object detection. *Pattern Recognition* 131 (2022), 108868.

[140] Youcheng Sun, Xiaowei Huang, Daniel Kroening, James Sharp, Matthew Hill, and Rob Ashmore. 2018. Testing deep neural networks. *arXiv preprint arXiv:1803.04792* (2018).

[141] Vincent Tjeng, Kai Xiao, and Russ Tedrake. 2017. Evaluating robustness of neural networks with mixed integer programming. *arXiv preprint arXiv:1711.07356* (2017).

[142] William J Vetter. 1973. Matrix calculus operations and Taylor expansions. *SIAM review* 15, 2 (1973), 352–369.

[143] Song Wang, Nishtha Shrestha, Abarna Kucheri Subburaman, Junjie Wang, Moshi Wei, and Nachiappan Nagappan. 2021. Automatic unit test generation for machine learning libraries: How far are we?. In *2021 IEEE/ACM 43rd International Conference on Software Engineering (ICSE)*. IEEE, 1548–1560.

[144] Yanghui Wu, John McCall, and David Corne. 2010. Two novel ant colony optimization approaches for Bayesian network structure learning. In *IEEE Congress on Evolutionary Computation*. IEEE, 1–7.

[145] Zhicong Yan, Gaolei Li, Yuan Tlan, Jun Wu, Shenghong Li, Mingzhe Chen, and H Vincent Poor. 2021. Dehib: Deep hidden backdoor attack on semi-supervised learning via adversarial perturbation. In *Proceedings of the AAAI conference on artificial intelligence*, Vol. 35. 10585–10593.

[146] Lidong Yang, Jialin Jiang, Xiaojie Gao, Qinglong Wang, Qi Dou, and Li Zhang. 2022. Autonomous environment-adaptive microrobot swarm navigation enabled by deep learning-based real-time distribution planning. *Nature Machine Intelligence* 4, 5 (2022), 480–493.

[147] Xiao Yang, Weiqing Liu, Dong Zhou, Jiang Bian, and Tie-Yan Liu. 2020. Qlib: An AI-oriented Quantitative Investment Platform. *arXiv preprint arXiv:2009.11189* (2020).






[148] Shin Yoo and Mark Harman. 2012. Regression testing minimization, selection and prioritization: a survey. *Software testing, verification and reliability* 22, 2 (2012), 67–120.

[149] Andy Zaidman, Bart Van Rompaey, Serge Demeyer, and Arie van Deursen. 2008. Mining Software Repositories to Study Co-Evolution of Production amp; Test Code. In *2008 1st International Conference on Software Testing, Verification, and Validation.* 220–229. https://doi.org/10.1109/ICST.2008.47

[150] Jie M Zhang, Mark Harman, Lei Ma, and Yang Liu. 2020. Machine learning testing: Survey, landscapes and horizons. *IEEE Transactions on Software Engineering* (2020).

[151] Quan Zou, Sifa Xie, Ziyu Lin, Meihong Wu, and Ying Ju. 2016. Finding the best classification threshold in imbalanced classification. *Big Data Research* 5 (2016), 2–8.


## A Assertion API and example code representing the Test strategies

Table 6: API code example for expressing *Value Range analysis* and *Decision & Logical Condition Test*

| Strategy | Test scenarios and Assertion | Framework |
|---|---|---|
| Value Range Analysis | Interval Analysis (i.e., $[a, b] = \{x \in \mathbb{R} : a \leqslant x \leqslant b\}$) | *** |
| | assertTrue(0 ≤ a and b ≤ n) | unnittest |
| | assertTrue(0 ≤ b4 ≤ n) | unittest |
| | assertion inside loop statement | *** |
| | value-range datatype assertion eg assert*list*, assert*set*, assert*dict*, assert*turple* | unittest |
| Decision & Logical Condition | Control flows (statement, branch or path) | *** |
| | Conditional assertions eg EXPECT_TRUE(a exp b && x exp y) | *** |
| | Check IsValidRowCol e.g row ≥ 0 && row < rows && col ≥ 0 && col < cols | deepspeech |





## Table 7: Assertion API example code expressing the selected testing strategies

(a) The API code example for expressing Negative Test, Instance Checks and Sub component test strategies

| Strategy | | Assertion API | Framework |
|---|---|---|---|
| | Negative Test | assertNot** | python |
| | | assertFalse(statement), assert not (statement) | unittest, pytest |
| | | assert_not_equal(statement) | numpy |
| | | assertTrue(a != b), assert a != b | unittest, pytest |
| | | ASSERT (a != b) | C/C++ |
| | | CHECK(!statement), CHECK (a != b) | gLog |
| | | RAY_CHECK(a != b) | Ray |
| | | ACHECK(!statement) | C/C++ |
| | | **_NE(a, b) | gTest |
| | | **_FALSE(statement) | gTest |
| | Instance and Type Checks | assertIsInstance(a,b) | python |
| | | **equal(a.astype(datatype), typeid(a)) | numpy |
| | | **equal(a.dtype, typeid(a)) | unittest |
| | | **equal(type(a), typeid(a)) | unittest |
| | | assert a.dstype == typeid(a) | pytest |
| | | assert **_type(a) == typeid(a) | nupic, apollo |
| | | **_EQ(a**::Instance(), instance(a)) | apollo |
| | | **_EQ(a**::Type(), typeid(a)) | gTest |
| | | EXPECT_**(→type, typeid(a)) | apollo |
| | | CHECK(a**::Type() == typeid(a)) | gLog |
| | Membership Testing | assertIn(a, A) | unittest |
| | | assert a in A | pytest |
| | | CHECK(Subset(A, B)); | gLog |
| | | EXPECT_TRUE(A.has**()); | apollo |
| | | **TRUE(a, indexof(a)); | gtest, unittest |
| | | assert a == indexof(a); | pytest |
| | | Checking Sub Component using loop statement (e.g., in a List, Set, dictionary) | * |

(b) The API code example for expressing Oracle approximation

| Strategy | | | Assertion API | Framework |
|---|---|---|---|---|
| | Oracle Approximation | Absolute Relative Tolerance/ Absolute Tolerance | assert_allclose(result, expect, rVal,aVal) | numpy |
| | | | assert allclose(result, expect, rVal,aVal) | numpy |
| | | | assertTrue(isclose(result, expect, rVal,aVal)) | numpy |
| | | | assert isclose(result, expect, rVal,aVal) | numpy |
| | | | assert torch.all(torch.isclose(result, expect, rVal)) | torch |
| | | | assert torch.allclose(result, expect, rVal) | torch |
| | | | assert math.isclose(result, expect, aVal) | python |
| | | | assertAllClose(result, expect, rVal,aVal) | numpy |
| | | | assertAllCloseAccordingToType(result, expect, aVal) | numpy |
| | | | SLOPPY_CHECK_CLOSE(result, expect, rVal,aVal) | deepspeech |
| | | | BOOST_CHECK_CLOSE((result, expect, rVal,aVal) | Boost |
| | | | EXPECT_NEAR(result, expect, aVal) | gTest |
| | | | ASSERT_NEAR(result, expect, aVal) | gTest |
| | | Rounding Tolerance | assert_almost_equal(result, expect, dp) | numpy |
| | | | assertAlmostEquals(result, expect, dp) | unittest |
| | | | assertAlmostEqual(result, expect, dp) | unittest |
| | | | assertListAlmostEqual(result, expect) | unittest |
| | | | assert_array_almost_equal(result, expect, dp) | numpy |
| | | | ApproxEqual(result, expect, dp) | deepspeech |
| | | | CHECK(ApproxEqual(result, expect, dp)) | deepspeech |
| | | | EXPECT_TRUE(almost_equal(result, expect, dp)) | gTest |
| | | Error Bounding | assertTrue(x < y) | numpy |
| | | | assertLess(x,y) | numpy |
| | | | assertLessEqual(x,y) | numpy |
| | | | assert_array_less(x,y) | numpy |
| | | | EXPECT_LT(x,y), ASSERT_LT(x,y) | gTest |
| | | | EXPECT_TRUE(EXPECT_LT(x,y)), ASSERT_TRUE(ASSERT_LT(x,y)) | gTest |